\documentclass[
floatfix,
twocolumn,
showpacs,showkeys,preprintnumbers,aps,prd]
{revtex4}

\usepackage{amsmath}
\usepackage{amssymb}
\usepackage{revsymb}
\usepackage{graphicx}
\usepackage{dcolumn}
\begin{document}

\title{Quark-Model Identification of Baryon Ground and Resonant States}
\author{T. Melde, W. Plessas, and B. Sengl}
\affiliation{Theoretische Physik, Institut f\"ur Physik, Karl-Franzens-Universit\"at,
Universit\"atsplatz 5, A-8010 Graz, Austria}

\begin{abstract}
We present a new classification scheme of baryon ground states and resonances into
$SU(3)$ flavor multiplets. The scheme is worked out along a covariant formalism with
relativistic constituent quark models and it relies on detailed investigations of the
baryon spectra, the spin-flavor structure of the baryon eigenstates, the behaviour
of their probability density distributions as well as covariant predictions for
mesonic decay widths.
The results are found to be quite independent of the specific types of relativistic constituent quark models employed. It turns out that a consistent classification
requires to include also resonances that are presently reported from experiment
with only two-star status.
\end{abstract}


\maketitle

\section{Introduction}
Recently a comprehensive study of hadronic decays of baryon resonances along
relativistic constituent quark models (RCQMs) has become available. In particular,
one considered the $\pi$, $\eta$, and $K$ decay modes in the light and strange 
baryon sectors~\cite{Melde:2005hy,Melde:2006yw,Sengl:2007yq}. The calculations
were done within the framework of relativistic quantum mechanics, specifically
in the point form. The covariant
quark-model predictions were found to be drastically different from results of
previous nonrelativistic or relativized calculations as, e.g., in
refs.~\cite{Stancu:1989iu,Capstick:1993th,Geiger:1994kr,Krassnigg:1999ky,%
Plessas:1999nb,Theussl:2000sj}. The relativistic results for partial decay widths
systematically underestimate the available experimental data. Similar results for
hadronic decay widths with practically the same characteristics were also seen in 
another relativistic study~\cite{Metsch:2004qk,Migura:2007}. This is remarkable,
since the latter investigation was carried out along a completely different
approach using the Bethe-Salpeter
equation~\cite{Merten:2002nz,Metsch:2003ix}.

The observed systematics of the relativistic results for decay widths has
suggested to revisit the identification of quark-model eigenstates with
established baryon resonances~\cite{Melde:2007iu}. For instance, it was found that
in the $\Sigma$ spectrum the lowest $J^P=\frac{1}{2}^-$ state should not be
identified with the $\Sigma(1750)$ resonance, which is established at
three-star status, but rather with the $\Sigma(1560)$
resonance~\cite{Melde:2006yw}. The latter
is often neglected because it is reported only as a two-star resonance
without a $J^P$ assignment~\cite{PDBook}. 

Evidently, it is not enough to consider only energy levels to reach a conclusive
classification of baryon resonances. Rather one should take into account also properties
relating to the resonance structure such as decay widths. Theoretically one can
examine the detailed spin and flavor contents as well as the spatial symmetries
of all resonance states, since their wave functions are directly accessible.
This leads to a consistent identification of states
along $SU(3)$ flavor multiplets. The present work is essentially devoted to a
comprehensive investigation of the multiplet classification within a covariant
formalism along RCQMs. In several cases this yields results different from
what is understood as the usual quark-model classification. In addition, the
inclusion of low-lying resonances with less than three-star status turns
out to be necessary.

It is mandatory to work within a relativistic framework. Obviously, the
constituent quarks confined to a finite volume carry high momenta. Large
relativistic effects are present in all aspects of baryon states, as
was found in covariant studies of electroweak nucleon form
factors~\cite{Cardarelli:1995dc,Wagenbrunn:2000es,Glozman:2001zc,%
Boffi:2001zb,Merten:2002nz,Melde:2007zz}, electric radii as well as magnetic
moments~\cite{Berger:2004yi,Metsch:2004qk}, and mesonic
decays~\cite{Metsch:2004qk,Melde:2005hy,Melde:2006yw,Migura:2007,Sengl:2007yq}.
For the relativistic description we use the formalism of 
Poincar\'e-invariant quantum mechanics~\cite{Keister:1991sb}. The theory relies on a relativistically invariant mass operator, and it allows to
incorporate all symmetries required by special relativity. 
Different forms of relativistic quantum mechanics are characterized 
by different kinematical subgroups of the Poincar\'e group. The most common
ones are the instant, front, and point forms~\cite{Dirac:1949,Leutwyler:1977vy}.
Here we adhere to the point form.

In the following section we shortly outline the theoretical framework and
describe the specific RCQMs used. The third section deals with the baryon
excitation spectra as produced by the RCQMs. Then we discuss the
mesonic decays of the established baryon resonances in comparison to the
available experimental data. In the fifth section we identify the theoretical
eigenstates with the phenomenologically known resonances and propose a
new classification scheme for baryon ground and resonant states. There, also
the spin-flavor structures as well as spatial density distributions of the
various flavor-multiplet members are detailed. Subsequently we discuss their
decay properties and end with our conclusions.

\section{Relativistic Quark Models
\label{sec:theory}}
Here we formulate the Poincar\'e-invariant description of baryon eigenstates.
Following the point form of relativistic quantum mechanics we outline the
eigenvalue problem of the invariant mass operator and specify the dynamics
of two different types of RCQMs.

\subsection{Eigenvalue problem of the invariant mass operator
\label{subsec:massoperator}}

Starting out from the free mass operator ${\hat M}_{\rm free}$ the interactions
are introduced according to the Bakamjian-Thomas
construction~\cite{Bakamjian:1953}. Thus the free mass operator
is replaced by a full mass operator ${\hat M}$
containing an interacting term ${\hat M}_{\rm int}$
\begin{equation}
\label{eq:masses}
\hat M_{\rm free} \rightarrow \hat M = \hat M_{\rm free} + \hat M_{\rm int}.
\end{equation}
For a given baryon state of mass $M$ and total angular momentum (intrinsic spin)
$J$ with z-projection $\Sigma$ the eigenvalue problem of the mass operator reads
\begin{equation}
    {\hat M}\left|V,M,J,\Sigma\right>
    =M\left|V,M,J,\Sigma\right>.
    \label{eq:mass}
\end{equation}
Here we have written the eigenstates in obvious notation as
$\left|V,M,J,\Sigma\right>$, where $V$ indicates the four eigenvalues
of the velocity operator $\hat V^{\mu}$, of which only three are independent. 

The four-momentum operator is then defined by multiplying the mass 
operator ${\hat M}$ by the four-velocity operator
\begin{equation}
    \hat P^{\mu}={\hat M}\hat V^{\mu} \, ,
\end{equation}
and it becomes interaction-dependent.
Alternatively we can thus express the baryon eigenstates also as
\begin{equation}
\left|V,M,J,\Sigma\right> \equiv \left|P,J,\Sigma\right>,
\label{eq:eigenstates}
\end{equation}
where $P$ represents the four eigenvalues of $\hat P^{\mu}$, whose 
square gives the invariant mass operator.

The ground and resonance state wave functions are defined through
the velocity-state representations of the mass-operator eigenstates
\begin{eqnarray}
    \left\langle v;\vec{k}_1,\vec{k}_2,\vec{k}_3;\mu_1,\mu_2,\mu_3
    |V,M,J,\Sigma\right\rangle=
\phantom{0000000}&& \nonumber\\ 
     \frac{\sqrt{2}}{M} v_{0}\delta^{3}\left(\vec{v}-\vec{V}\right)
  \sqrt{\frac{2\omega_{1}2\omega_{2}2\omega_{3}}
    {\left(\omega_{1}+\omega_{2}+\omega_{3}\right)^{3}}
    }
   \Psi_{MJ\Sigma}\left(
   \vec{k}_i;\mu_i
   \right) \, ,
   \phantom{00}
&&
\end{eqnarray}
and they are normalized to unity as
\begin{eqnarray}
   && \delta_{MM'}\delta_{JJ'}\delta_{\Sigma\Sigma'}
    =
    \sum_{\mu_{1}\mu_{2}\mu_{3}}
    \int{d^{3}k_{2}d^{3}k_{3}}
\nonumber\\
&&
    \times
   \Psi^{\star}_{M'J'\Sigma'}\left( 
    \vec{k}_i;\mu_i
   \right)
   \Psi_{MJ\Sigma}\left(
   \vec{k}_i;\mu_i
   \right) \, .
\end{eqnarray}
The velocity states build a specific basis of free three-body states defined by
\begin{eqnarray}
&&\left|v;\vec{k}_1,\vec{k}_2,\vec{k}_3;\mu_1,\mu_2,\mu_3\right\rangle
=U_{B(v)}
\left|k_1,k_2,k_3;\mu_1,\mu_2,\mu_3\right\rangle
\nonumber \\
&&=\sum_{\sigma_1,\sigma_2,\sigma_3}
\prod\limits_{i=1}^3D^{\frac{1}{2}}_{\sigma_i\mu_i}[R_W(k_i,B(v))]
\left|p_1,p_2,p_3;\sigma_1,\sigma_2,\sigma_3\right\rangle .
\nonumber \\
& &
\label{eq:velstates}
\end{eqnarray}
Here $B\left(v\right)$, with unitary representation 
$U_{B\left(v\right)}$,
is a boost with four-velocity $v$ on the free three-body states
$\left|k_1,k_2,k_3;\mu_1,\mu_2,\mu_3\right\rangle$ in the 
centre-of-momentum system, i.e., for which the individual quark momenta
sum up as $\sum{\vec k_{i}}=0$.
The second line in Eq.~(\ref{eq:velstates}) expresses the 
corresponding Lorentz transformation as 
acting on general three-body states 
$\left|p_1,p_2,p_3;\sigma_1,\sigma_2,\sigma_3\right\rangle$. 
The quark momenta $p_{i}$ and $k_{i}$ are related by 
$p_{i}=B\left(v\right)k_{i}$, where
$k_{i}=\left(\omega_{i},{\vec k}_{i}\right)$.
The $D^{\frac{1}{2}}$ are the spin-$\frac{1}{2}$ representation matrices
of Wigner rotations $R_{W}\left(k_{i},B\left(v\right)\right)$.
The velocity-state representation allows to separate
the motion of the system as a whole and the internal 
motion. The latter is described by the wave function
$\Psi_{MJ\Sigma}\left(\vec{k}_i;\mu_i\right)$, which is also the rest-frame
wave function. It contains the whole information on the flavor, spin, and spatial
structure of a baryon state.

In the rest frame the invariant mass operator coincides with the Hamiltonian
\begin{equation}
\hat H=\hat H_{\rm free}+\hat H_{\rm int}=
\hat H_{\rm free}+\sum_{i<j=1}^3{\hat V_{ij}} \, ,
\end{equation}
where the quark-quark dynamics is decomposed into a confinement and a hyperfine
interaction
\begin{equation}
\hat H_{\rm int}=\sum_{i<j=1}^3{\left(\hat V^{\rm conf}_{ij}+\hat V^{\rm hyper}_{ij}
\right)} \, .
\end{equation}
While the confinement interaction is nowadays usually taken as a potential linearly
rising towards longer distances (as suggested from quantum chromodynamics), the
hyperfine interaction is qualitatively distinct among
different constituent quark models.

\subsection{Goldstone-boson-exchange RCQM}
The Goldstone-boson-exchange (GBE) RCQM~\cite{Glozman:1998ag,Glozman:1998fs}
relies on a linear confinement potential
and a hyperfine interaction that is motivated by the spontaneous breaking
of chiral symmetry. The different parts in the Hamiltonian are thus represented by
\begin{equation}
H_{\rm free}=\sum_{i=1}^3{\sqrt{m_i^2+\vec k^2_i}} \, ,
\end{equation}
\begin{equation}
V^{\rm conf}_{ij}=V_0+Cr_{ij} \, ,
\end{equation}
and
\begin{widetext}
\begin{equation}
V^{\rm hf}_{ij}=\left[
\sum_{a=1}^3{V^\pi_{ij}\lambda_i^a\lambda_j^a}+
\sum_{a=4}^7{V^K_{ij}\lambda_i^a\lambda_j^a}+
V^\eta_{ij}\lambda_i^8\lambda_j^8+
\frac{2}{3}V^{\eta'}_{ij}
\right]\vec \sigma\left(i\right)\cdot\vec \sigma\left(j\right) \, .
\label{eq:gbehyper}
\end{equation}
\end{widetext}
The hyperfine interaction is furnished only by the spin-spin part of the
GBE (identified with the exchange of pseudoscalar mesons) and it comes up with
an explicit flavor dependence reflected by the $SU\left(3\right)$
Gell-Mann flavour matrices $\lambda^a_i$ in Eq.~(\ref{eq:gbehyper}).
This specific property is favorable for a unified description of all
light and strange baryon ground and resonances states. In particular,
it provides for the correct level orderings of positive- and negative-parity
states in both the nucleon and $\Lambda$ excitation spectra.

The terms $V^\gamma_{ij}$, with $\gamma=\pi,K,\eta,\eta'$, assume the form of
instantaneous meson-exchange potentials. The parameters were determined by
fitting the established baryon resonances (with at least three-star status) 
below 2 GeV. In total, the GBE RCQM involves four open parameters. The detailed
parametrization can be found in ref.~\cite{Glozman:1998ag}. 

\subsection{One-gluon-exchange RCQM}
Another type of RCQM consists in constructing the hyperfine interaction from
one-gluon exchange (OGE). Here, we consider in particular the relativistic variant
of the Bhaduri-Cohler-Nogami (BCN) OGE CQM~\cite{Bhaduri:1981pn} in the parametrization
of ref.~\cite{Theussl:2000sj}. The confinement interaction is given by
\begin{equation}
V^{\rm conf}_{ij}=V_0+Cr_{ij}-\frac{2b}{3r_{ij}} \, ,
\end{equation}
and the hyperfine potential relies on the (flavor-independent) color-magnetic
spin-spin interaction
\begin{equation}
V^{\rm hf}_{ij}=\frac{\alpha_S}{9m_i m_j}
\Lambda^2\frac{e^{-\Lambda r_{ij}}}{r_{ij}}
\vec \sigma\left(i\right)\cdot\vec \sigma\left(j\right)\, .
\end{equation}
The OGE RCQM also has four open parameters that were obtained through a fit to the
baryon spectrum. The detailed parametrization is given in ref.~\cite{Theussl:2000sj}.

\section{Baryon Spectra}
The solution of the eigenvalue problem of the mass operator $\hat M$ in
Eq.~(\ref{eq:mass}) has been
performed with the stochastic variational method (SVM)~\cite{Suzuki:1998bn}.
Thereby we have produced the invariant mass spectra of all the light and strange
baryons
\footnote{An alternative approach consists in solving the three-quark problem by
Faddeev-type integral equations, which leads to identical results for the mass
spectra~\cite{Papp:2000kp}.}. 
In addition we have obtained the rest-frame wave functions of all
ground and resonance states.  

\begin{figure*}
\includegraphics
[height=8.2cm]{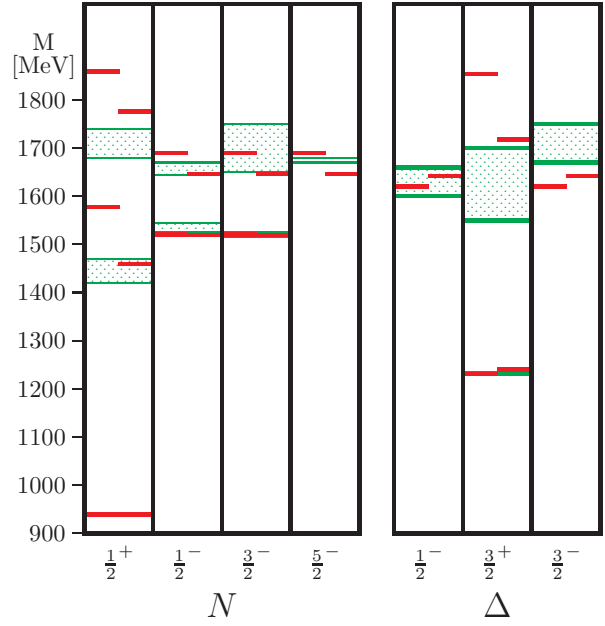}
\caption{Energy levels (red solid lines) of the lowest $N$ 
and $\Delta$ states with total angular momentum and parity $J^P$ for the OGE
(left levels) and GBE (right levels) RCQMs in comparison to
experimental values with uncertainties~\cite{PDBook}, represented as (green)
shadowed boxes.}
\label{fig:NDel}
\end{figure*}

\begin{figure*}
\includegraphics
[height=8.2cm]{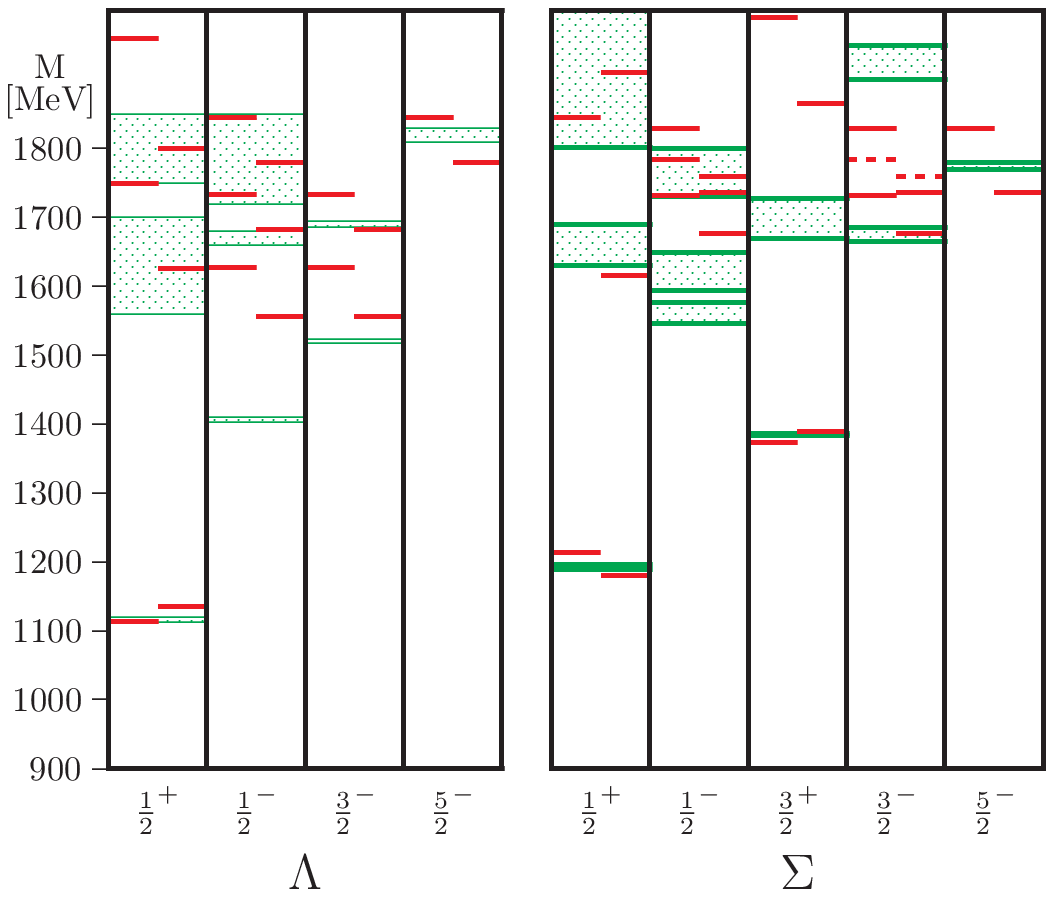}
\hspace{0.05cm}
\includegraphics
[height=8.2cm]{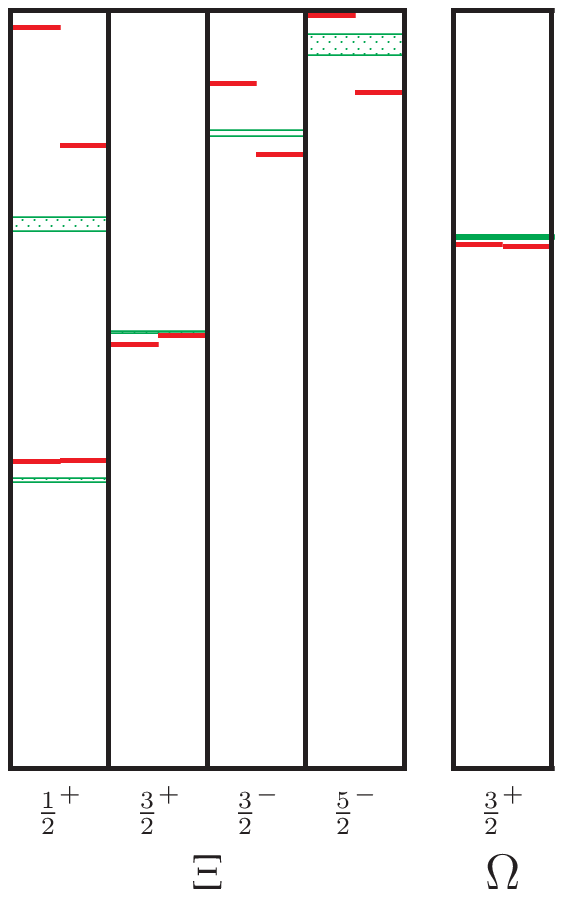}
\caption{Sames as in Fig.~\ref{fig:NDel} for the lowest $\Lambda$, $\Sigma$, $\Xi$, 
and $\Omega$ states. The dashed lines in the $J^P=\frac{3}{2}^-$ $\Sigma$ spectrum
represent (decuplet) eigenstates, for which there is no experimental counterpart yet.
\label{fig:LaSig}
}
\end{figure*}

The energy eigenvalues of the ground states and resonances below $\approx 2$ GeV
as resulting with the GBE and OGE RCQMs are quoted in Table~\ref{tab:masses}.
Here, only eigenstates with (rest-frame) total orbital angular momentum $L<2$ are
considered, as the theoretical decay properties necessary in the present work are
available only for such states~\cite{Melde:2005hy,Melde:2006yw,Sengl:2007yq}. 
For the complete spectra of the GBE and OGE RCQMs we refer to
refs.~\cite{Glozman:1998ag,Glozman:1998fs} and~\cite{Theussl:2000sj}, respectively.
The same results are depicted also in
Figs.~\ref{fig:NDel} and~\ref{fig:LaSig} in comparison to the phenomenological
data by the PDG~\cite{PDBook}. The left (red) lines in each one of the
$J^P$ columns denote the energy levels produced by the OGE RCQM, while the right
(red) lines belong to the GBE RCQM. With respect to the number of states below
about 2 GeV one immediately observes a one-to-one correspondence
between theory and experiment in each one of the $J^P$ sets in the light-flavor
sector (Fig.~\ref{fig:NDel}). In the strange sector (Fig.~\ref{fig:LaSig}),
however, there would be more theoretical
levels than experimental ones, if only established resonances (with at least
three-star status) were taken into account. For instance, in the
$J^P=\frac{1}{2}^-$ column of the $\Sigma$ excitation spectrum the RCQMs produce
three states, while there is only one established resonance reported. This
problem is remedied, if one includes also the two lower-lying resonances
$\Sigma(1560)$ and $\Sigma(1620)$, as is done in Fig.~\ref{fig:LaSig}.
Of course, these two resonances have only a two-star confidence status and the
$J^P$ value is only known for the latter
\footnote{We note that these two states have not been considered in
refs.~\cite{Glozman:1998ag,Theussl:2000sj} neither for fitting the parameters
nor for the comparison of the theoretical results to the experimental data.}.
As will become clear in the subsequent sections, these additional states are
required and can/should be accommodated in a consistent classification into
$SU(3)$ flavor multiplets. A similar situation occurs
in the $J^P=\frac{3}{2}^-$ $\Sigma$ spectrum, where the RCQMs again produce three
levels, while the PDG reports only two established resonances. In this case,
however, there is essentially no further candidate seen in experiment, even not with
one- or two-star status. Therefore the RCQM state that misses a corresponding
experimental counterpart (the last $\Sigma$ entry in Table~\ref{tab:masses}) is
indicated by dashed lines in Fig.~\ref{fig:LaSig}.
Nevertheless, all of these three theoretical states fit into the
$SU(3)$ multiplet classification we propose in the following sections.

The PDG~\cite{PDBook} gives assignments of baryon states (for $L<2$) in terms of
flavor multiplets as summarized in Table~\ref{tab:multiplet_PDG}. In this
context not only established (three- and four-star) resonances are included but
also some two-star states. Besides the octet and decuplet of ground states only one
more multiplet is complete, namely, the octet of the lowest $J^P=\frac{3}{2}^-$
excitations (involving the $N(1520)$ resonance). All other multiplets miss at least
a $\Xi$, the decuplets in addition also a $\Sigma$. The assignments of some
states, especially of $\Lambda(1810)$ and $\Xi(1820)$, are merely based on
educated guesses~\cite{PDBook}. The resulting scheme is mostly in line with the one by
Samios et al.~\cite{Samios:1974tw} proposed back in 1974, when many of the
resonances known today have not yet been found from pehenomenology.

To a large extent the PDG
classification also coincides with a more recent one by Guzey and
Polyakov (GP)~\cite{Guzey:2005vz}. There occur only differences with regard to the
identification of the $\Sigma(1620)$ in the $J^P=\frac{1}{2}^-$ octet, involving
the $N(1535)$, and the $\Sigma(1750)$ in the $J^P=\frac{1}{2}^-$ octet, involving
the $N(1650)$; according to GP the $\Sigma(1750)$ falls into the
$J^P=\frac{1}{2}^-$ decuplet involving the $\Delta(1620)$. In addition, some further
states without assignment by the PDG were included by GP, such as the
three-star resonances $\Xi(1690)$ and $\Xi(1950)$ as well as the two-star resonances
$\Sigma(1560)$ and $\Sigma(1690)$. All of the latter will also be considered
(and needed) in the classification we elaborate and propose below. 

\section{Mesonic Decays of Established Baryon Resonances}

The mesonic decays of the baryon resonances from Figs.~\ref{fig:NDel}
and~\ref{fig:LaSig} were comprehensively studied with regard to their
$\pi$, $\eta$, and $K$ decay modes within a relativistic
framework~\cite{Melde:2005hy,Melde:2006yw,Sengl:2007yq}. In these works the
decay operator was constructed along the point-form spectator model
(PFSM)~\cite{Melde:2004qu}. It consists in the simplifying assumption that
the meson is emitted from one quark while the other two act as spectators.
The PFSM decay operator is manifestly covariant and it preserves its
spectator-model character in all reference frames. Though it formally looks
like a one-body operator, it nevertheless includes many-body
effects~\cite{Melde:2007zz}. In particular, the recoil effect on the residual
baryon state is naturally taken into account. The nonrelativistic limit of
the PFSM leads to the familiar elementary emission model (EEM).

From the covariant PFSM calculations a new pattern of partial decay widths
as predicted by modern RCQMs has emerged. It has turned out to be rather different
from what had been known from previous nonrelativistic or relativized studies.
Notably, quite large relativistic effects have been detected, and the results
generally underestimate the experimental data. In this context it is remarkable
that quite similar findings have been obtained by the Bonn group from a completely
different relativistic investigation along the Bethe-Salpeter
formalism~\cite{Metsch:2004qk,Migura:2007}.

\begin{figure*}
\includegraphics[angle=0,clip=,width=0.99\textwidth]{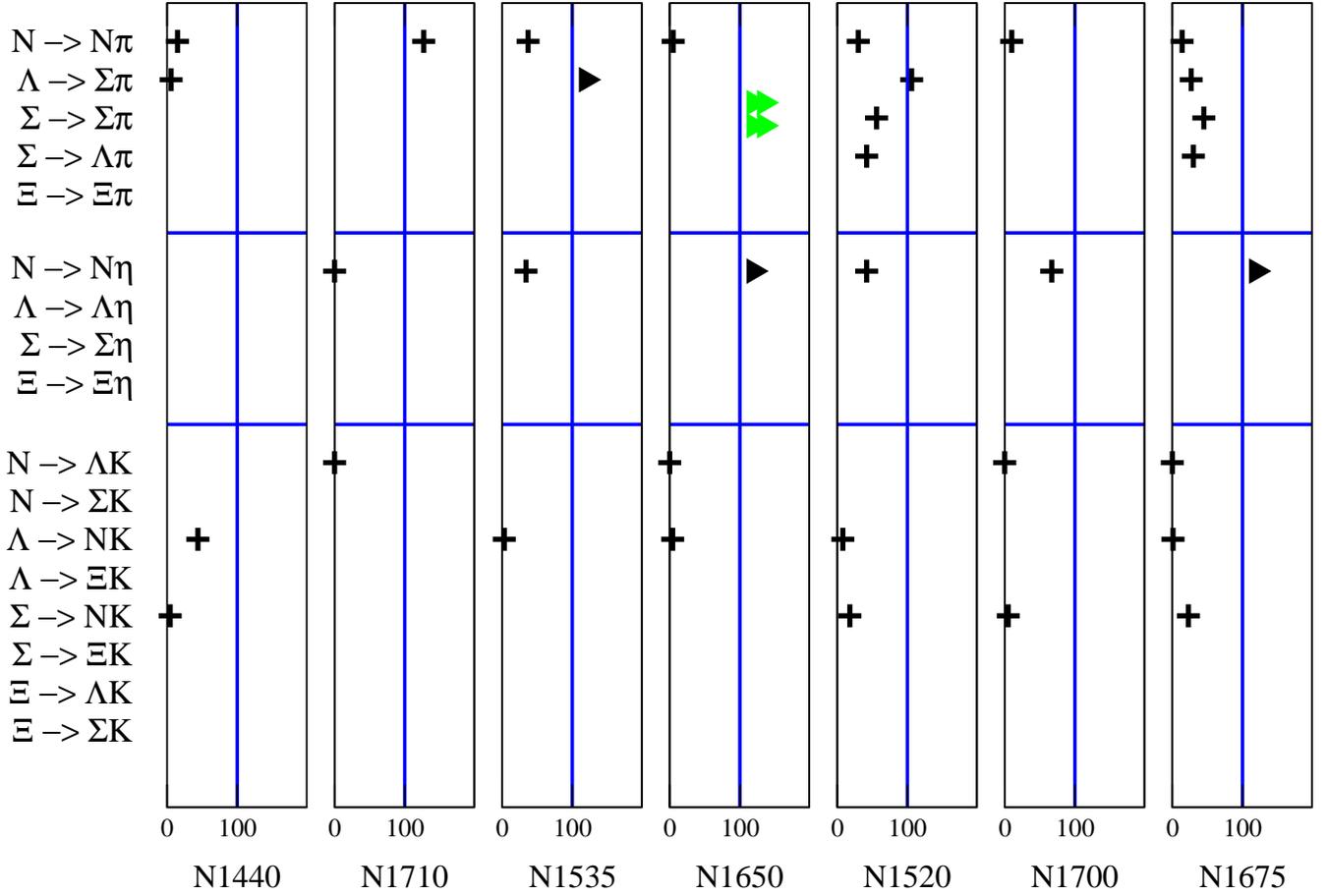}
\caption{Predictions for partial $\pi$, $\eta$, and $K$ decay widths of the GBE CQM
from the PFSM calculation for the lowest octets according to
refs.~\cite{Melde:2005hy,Melde:2006yw,Sengl:2007yq}. The results (crosses)
are presented as percentages of the best estimates for experimental data reported by
the PDG~\cite{PDBook}. The various resonances are grouped according to the octet
assignments in Table~\ref{tab:multiplet_PDG}. The (double) triangles point to results
(far) outside the plotted range. For the latter see also the discussion in the text.}
\label{fig:decay_graph_est}
\end{figure*}

In Fig.~\ref{fig:decay_graph_est} the situation is exemplified with regard to the
low-lying octet baryon resonances for which partial decay widths are reported by
the PDG. The theoretical decay widths as resulting with the GBE RCQM are depicted as
percentages of the (best estimates of the) experimental data. In each
octet one observes a clear pattern: The magnitudes of the theoretical
widths remain far below the experimental measurements, i.e. they lie to the left of
the vertical 100\% lines. There are only a few
exceptions. In the octet involving the $N(1710)$ the decay width for the $N\pi$
mode appears to be unusually large. The same is true for the
$\Lambda \rightarrow \Sigma \pi$ decays in the $N(1535)$ and $N(1520)$ octets.
Regarding the $\eta$ decays the relatively large percentages
should not be taken too serious in the cases of $N(1700)$ and $N(1675)$ as the
magnitudes of the experimental widths are very small, practically identical to
zero. In the $\eta$ channel only two partial widths are sizable, namely the ones
of $N(1535)$ and $N(1650)$. The latter one comes out considerably larger than
experimentally measured. This must be considered as a notorious problem of
constituent quark models and may point to an unidentified deficiency in the
decay operator and/or resonance wave function.

Of particular interest in the context of the present study are the $\pi$ decays
of the $J^P=\frac{1}{2}^-$ $\Sigma$ resonances. Out of the three eigenstates
appearing in Fig.~\ref{fig:LaSig} two of them are octets, as can be clearly
determined from the RCQM calculations. If one considers
only resonances with at least three-star status and known phenomenological
partial decay widths, these two octet states should be related to the $\Sigma(1750)$
resonance. In Fig.~\ref{fig:decay_graph_est} their $\Sigma \rightarrow \Sigma \pi$
decay widths are represented relative to the experimental width measured for
$\Sigma(1750)$. They are denoted by the
double triangles, since for both octet states when interpreted in this manner
the theoretical results grossly overshoot the data. However, we must consider
that there are two further $\Sigma$ resonances observed in experiments that
may be taken as candidates for the identification of the $J^P=\frac{1}{2}^-$
eigenstates, namely the unassigned $\Sigma(1560)$ and the $J^P=\frac{1}{2}^-$
$\Sigma(1620)$, both with two-star status. If the $J^P=\frac{1}{2}^-$ $\Sigma$
octet states are related to these resonances, then the third eigenstate, found
to be a decuplet state in the RCQM calculations, has to be interpreted as
$\Sigma(1750)$. In this way the
decay widths of all three $J^P=\frac{1}{2}^-$ $\Sigma$ fit into the general
pattern of relativistic mesonic decay widths~\cite{Melde:2006yw} (cf. the results
presented in section~\ref{sec:mesdecay} below).

One may ask for the causes of the deficiencies of the relativistic spectator-model
calculations, especially also in view of results existing in the literature
with apparently better agreement with phenomenology. Past studies have revealed
that a nonrelativistic spectator-model decay operator, such as the EEM, is
not sufficiently sophisticated to yield a reasonable description of the mesonic
decays. A more elaborate decay mechanism is provided, for example, by the relativized
pair-creation model (PCM)~\cite{LeYaouanc:1988aa}. Studies along this line were
performed among others in
refs.~\cite{Stancu:1989iu,Capstick:1993th,Geiger:1994kr,Theussl:2000sj}. All of
these works have in common that some additional parametrizations were introduced
on top of the direct predictions of the underlying constituent quark models. In
this way one adjusted the extensions of the meson-creation vertices and/or the
coupling strengths. Furthermore, different phase-space factors were employed in
the decay amplitude. As a result the different approaches can hardly be compared
to each other and the degree of agreement with experimental data does not really
allow to judge on the appropriateness of the approach followed.
So, it could well be that missing relativistic
effects or other shortcomings either of the quark-model wave functions or the
decay operator were compensated by introducing ad-hoc parameters.

For the relativistic results of decay widths reported in
refs.~\cite{Melde:2005hy,Melde:2006yw,Sengl:2007yq} as well as in
refs.~\cite{Metsch:2004qk,Migura:2007} one refrained from applying
any additional parametrization to the predictions of the RCQMs. Even though
in these works by the Bonn and Graz groups different RCQMs were employed and
distinct relativistic approaches were followed, the predicted decay widths
came out surprisingly similar. This might be a consequence of implementing
full Poincar\'e invariance, which is strictly observed in our point-form
approach and similarly in the Bethe-Salpeter formalism followed by the Bonn
group. In both types of relativistic studies one lacks substantial contributions
to the mesonic decay widths. One is left with a systematic underestimation of the
experimental data with only a few exceptional cases. In view of the present insight
it is difficult to decide from where the defects come. However, one must bear in
mind that in present-day constituent quark models the baryon ground and resonance
states are all described as bound three-quark eigenstates of the invariant mass
operator. In addition, the decay operators used so far might miss important
contributions from explicit many-body parts. Still, the covariant results
achieved so far can serve as benchmarks and provide a solid basis for further
explorations.

In the following section we present a classification of the singlet, octet,
and decuplet baryon resonances based on the evidences of their properties from
the mass spectra, the mesonic decay widths as well as the spin-flavor contents
and the spatial structures of the wave functions.
It will turn out that for a comprehensive classification
of the various states we shall need to consider the decay widths of additional
RCQM eigenstates beyond the results already published in
refs.~\cite{Melde:2005hy,Melde:2006yw,Sengl:2007yq}. The corresponding predictions
are collected in Table~\ref{tab:multi3}.

\section{Classification of Baryon Resonances}
\label{sec:class}

In order to arrive at a conclusive assignment of RCQM eigenstates to
$SU(3)$ multiplets we now analyze the spin-flavor contents and spatial structures
of the baryon wave functions. Through the results from the solution of the
mass-operator eigenvalue problem with the SVM we have access to their detailed
dependences on flavor, spin, and spatial variables. The SVM uses a completely
general basis of flavor, spin, and spatial test functions. In the process of
stochastic variation they are selected by varying all flavor, spin, orbital
angular momentum, and radial dependences so as to couple to baryon states
characterized by definite intrinsic spin $J$, $z$-component $\Sigma$,
hypercharge $Y$, total isospin $T$ as well as isospin $z$-component $M_T$
(for details see ref.~\cite{Sengl:2006}). In practice the eigenvalue problem is
solved in the rest frame of the corresponding state, where the point-form
version of relativistic quantum mechanics is employed. We note, however,
that the solution for the mass spectra is relativistically invariant and thus
independent of the specific form of relativistic quantum mechanics.

\subsection{Spin-flavor content of baryons}
In our solution of the mass-operator eigenvalue problem (in the rest frame)
utilizing the SVM, the total orbital angular momentum is restricted to $L<2$.
Even higher angular-momentum components have turned out to be rather small
and may be neglected for the ground and resonance states considered here.
For any eigenstate the total spin $S$ is definitely determined. $L$ and $S$
uniquely produce the intrinsic spin $J$ and parity $P$.

With regard to the flavor content, the SVM may pick up basis states from
different $SU(3)$ flavor multiplets. In particular, mixtures of flavor contributions
from singlet and octet as well as octet and decuplet can occur. This happens
specifically for the $\Lambda$, $\Sigma$, and $\Xi$ hyperons. For their
mass-operator eigenstates we can easily determine the singlet, octet, and
decuplet contents, respectively. For this purpose we employ the appropriate
flavor projection operators. For the singlet it reads
\begin{multline}
P^1_F=\frac{1}{6}\left(|uds\rangle-|usd\rangle+|dsu\rangle-
|dus\rangle+|sud\rangle-|sdu\rangle\right)
\nonumber
\\\times\left(\langle uds|-\langle usd|+\langle dsu|-\langle dus|+\langle 
sud|-\langle sdu|\right)\, .
\end{multline}
Evidently it sorts out the completely antisymmetric singlet component of a
certain baryon state whose probability is then given by
\begin{equation}
\alpha^1=\left<V,J,M,\Sigma\right|P^1_F\left|V,J,M,\Sigma\right>\, .
\label{eq:alpha1}
\end{equation}
For the special case of singlet-octet mixing, such as $\Lambda$, the octet
probability is then simply
\begin{equation}
\alpha^{8}=1-\alpha^1\, .
\end{equation}
Similarly, for the $\Sigma$ and $\Xi$ baryons, mixtures between 
flavor octet and decuplet can occur. Here, we determine the decuplet content 
by employing the decuplet projection operator $P^{10}_F$. In case of $\Sigma$
it suffices to consider the corresponding projection operator for $\Sigma^0$
\begin{multline}
P^{10}_F\left(\Sigma^0\right)
\nonumber\\
=\frac{1}{6}\left(|uds\rangle+|usd\rangle+|dsu\rangle+
|dus\rangle+|sud\rangle+|sdu\rangle\right)
\nonumber
\\
\times\left(\langle uds|+\langle usd|+\langle dsu|+\langle dus|+\langle 
sud|+\langle sdu|\right) \, ,
\end{multline}
since the decuplet content is the same for $\Sigma^+$ and $\Sigma^-$ due to
isospin symmetry. Analogous considerations hold for the $\Xi$ states, and we
may use the decuplet projection operator
\begin{multline}
P^{10}_F\left(\Xi^0\right)
\nonumber\\
=\frac{1}{3}\left(|ssu\rangle+|sus\rangle+
|uss\rangle\right)\left(\langle ssu|+\langle sus|+\langle uss|\right) \, .
\nonumber
\end{multline}
Again, the octet content is obtained by
\begin{equation}
\alpha^{8}=1-\alpha^{10}\, .
\end{equation}

\subsection{Spatial structure of baryons}
For the characterization of the spatial structure of any baryon ground state or
resonance we consider the angle-integrated spatial probability density distribution
\begin{multline}
\rho(\xi,\eta) = \xi^2 \eta^2 \int d\Omega_{\xi}d\Omega_{\eta}\,
\\
\Psi_{M\Sigma M_\Sigma TM_T}^{\star}(\xi,\Omega_\xi,\eta,\Omega_\eta)
\Psi_{M\Sigma M_\Sigma TM_T} 
(\xi,\Omega_\xi,\eta,\Omega_\eta) \, ,
\label{normwave1}
\end{multline}
where $\vec \xi=(\xi,\Omega_\xi)$ and $\vec \eta=(\eta,\Omega_\eta)$ are
the Jacobi coordinates.
It provides an idea of the matter distribution in a baryon. In
Fig.~\ref{fig:GBE-OGE-wf} probability density distributions are exemplified
for the nucleon ground state and the Roper resonance. Whereas the nucleon
shows a rather symmetric shape with the density distribution peaked at a
root-mean-square radius of about
0.3 fm, the $N(1440)$ exhibits the typical behaviour of a first 
radial excitation, i.e. with a nodal line in the wave function. One also observes
that $\rho(\xi,\eta)$ is very similar for both the GBE and OGE RCQMs. The
latter is slightly more localized, as its confinement interaction is relatively
stronger. In the following subsections we therefore restrict ourselves to showing
only the probability density distributions for the GBE RCQM.

\subsection{Baryon multiplets}

For the relativistic mass-operator eigenstates we obtain the flavor multiplet
classifications as given in
Tables~\ref{tab:multiplet_oct},~\ref{tab:multiplet_decu}, and~\ref{tab:multiplet_singl}.
We group the states according to their $(LS)$ values which determine the $J^P$. 
Evidently not all states have a pure singlet, octet, or decuplet content. In particular, 
considerable admixtures can occur for the $\Lambda$ singlet and octet states.

\subsection*{Octet: $\bf N(939)$, $\bf \Lambda(1116)$,  
$\bf \Sigma(1193)$, $\bf \Xi(1318)$}
\begin{figure*}
\includegraphics[width=3.9cm]{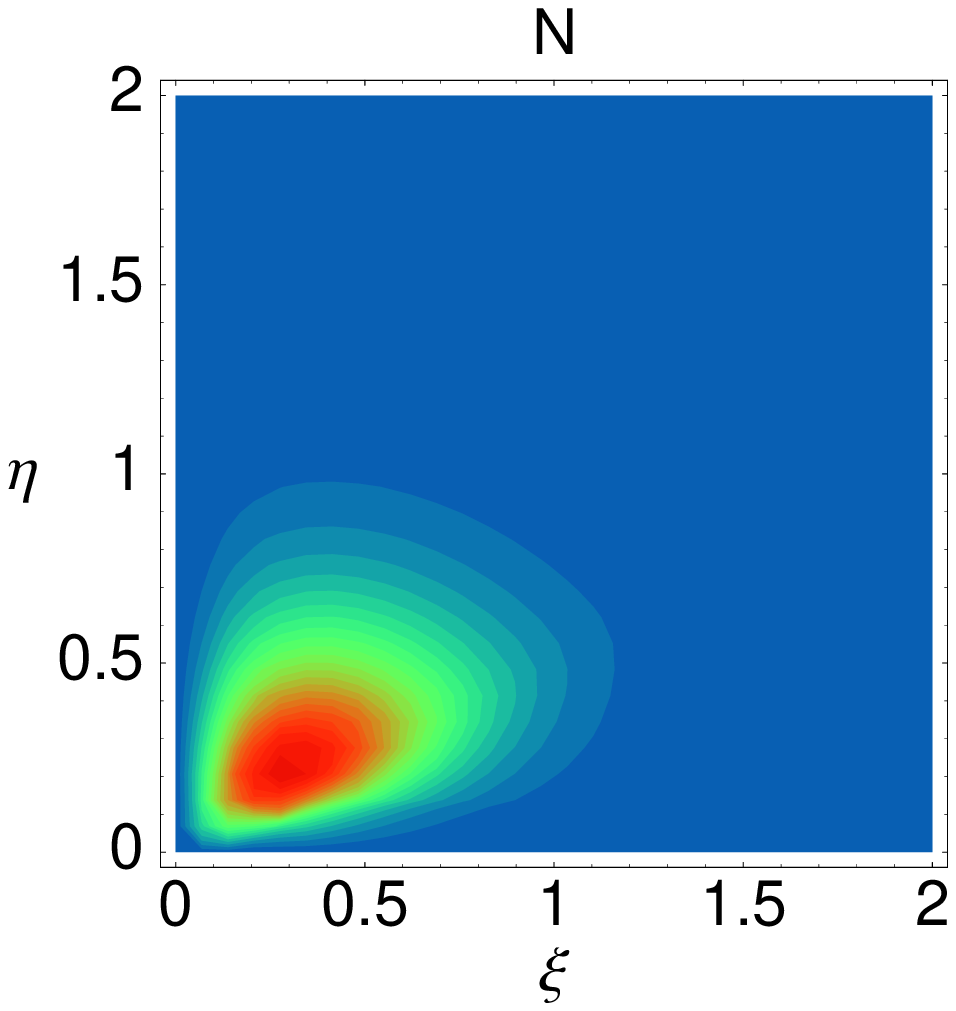}
\includegraphics[width=3.9cm]{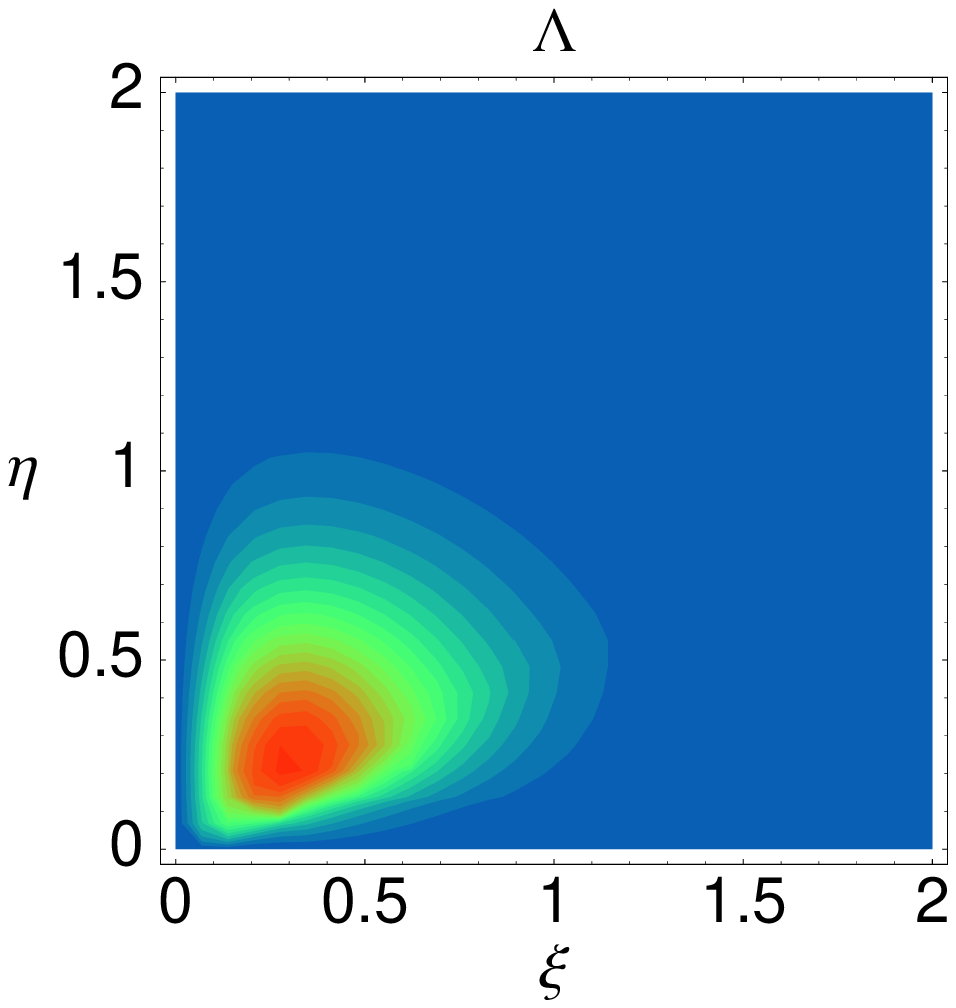}
\includegraphics[width=3.9cm]{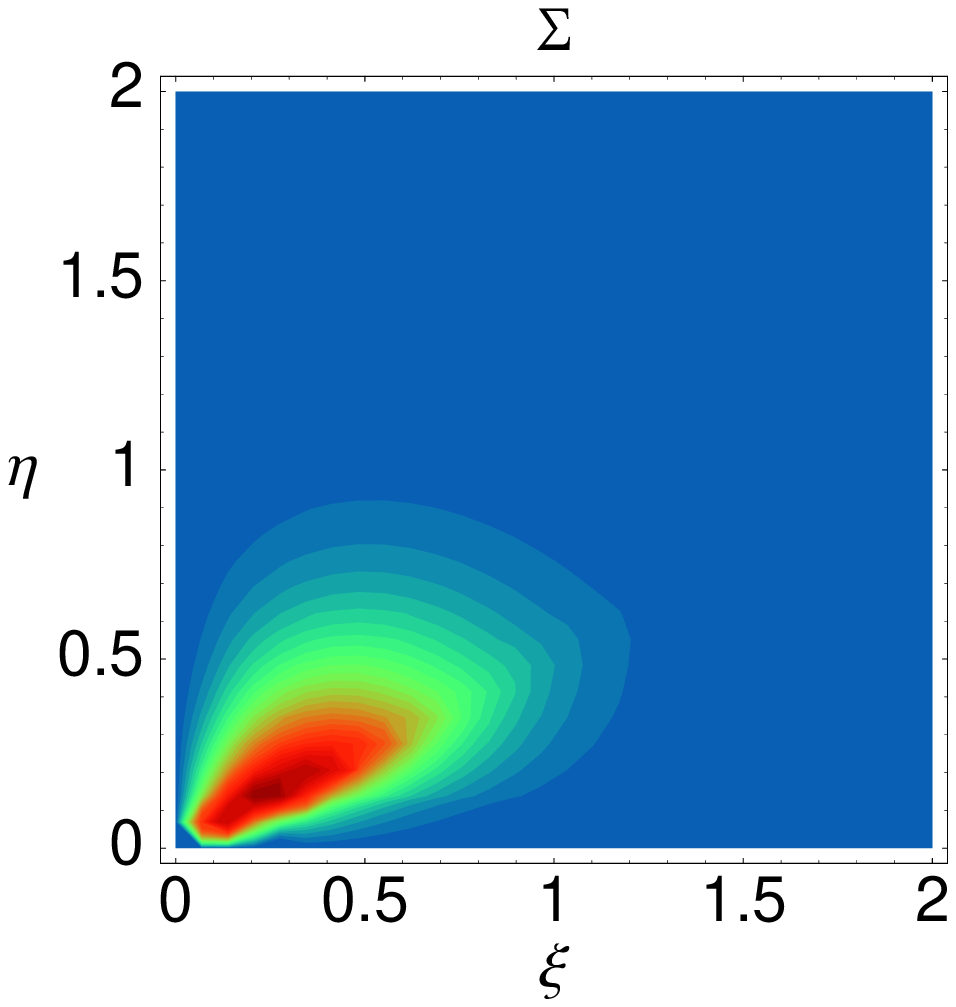}
\includegraphics[width=3.9cm]{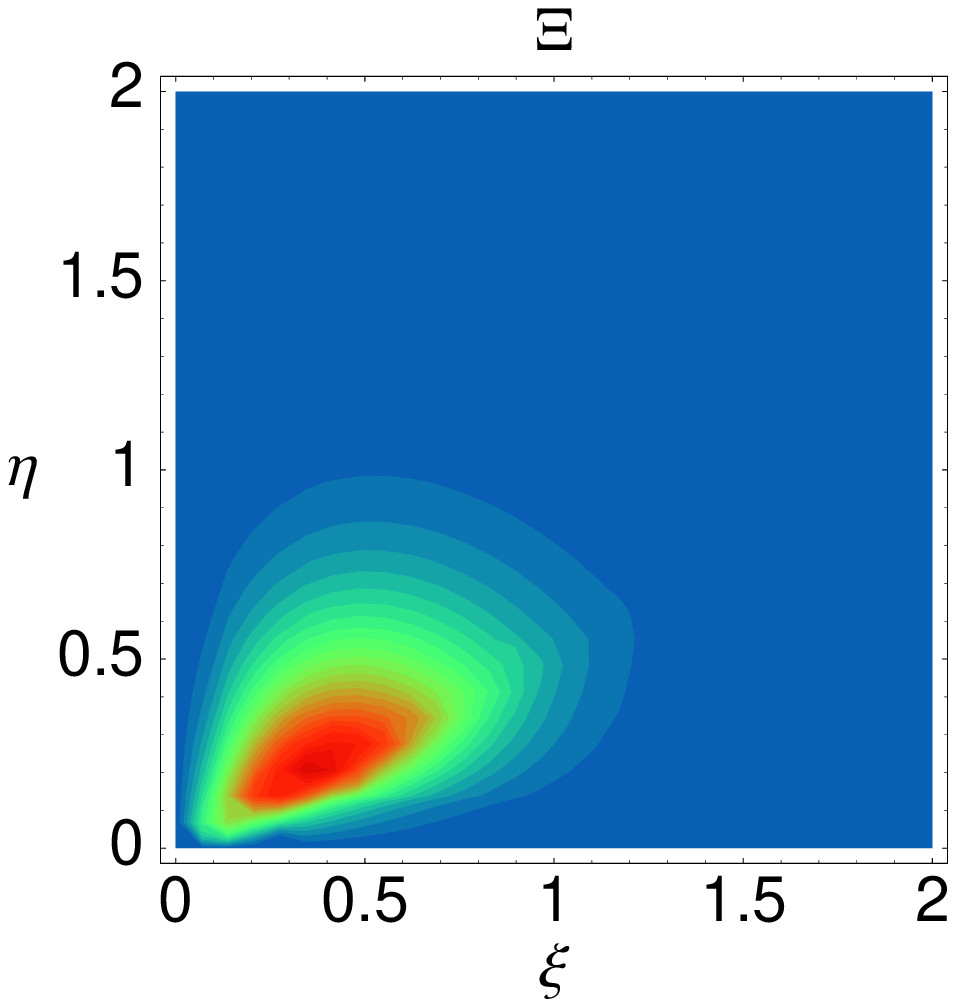}
\caption{Spatial probability density distributions $\rho(\xi,\eta)$ of the
$\frac{1}{2}^+$ octet baryon ground states
$N(939)$, $\Lambda(1116)$, $\Sigma(1193)$, $\Xi(1318)$.}
\label{fig:multi_1}
\end{figure*}

The identification of the octet ground states is natural, all bear
total intrinsic spin and parity $J^P=\frac{1}{2}^+$, and they have pure octet flavor
content. The corresponding spatial probability densities are plotted in
Fig.~\ref{fig:multi_1}. They all exhibit the typical behaviour
of ground states with no nodal lines. It should be noted that with these eigenstates
all electroweak properties of the nucleons as well as the electric radii and magnetic
moments of the other ground states are predicted in good agreement with 
experiment~\cite{Wagenbrunn:2000es,Glozman:2001zc,Boffi:2001zb,Berger:2004yi,
Melde:2007zz}.

\subsection*{Decuplet: $\bf \Delta(1232)$,  
$\bf \Sigma(1385)$, $\bf \Xi(1530)$, $\bf \Omega(1672)$}
\begin{figure*}
\includegraphics[width=3.9cm]{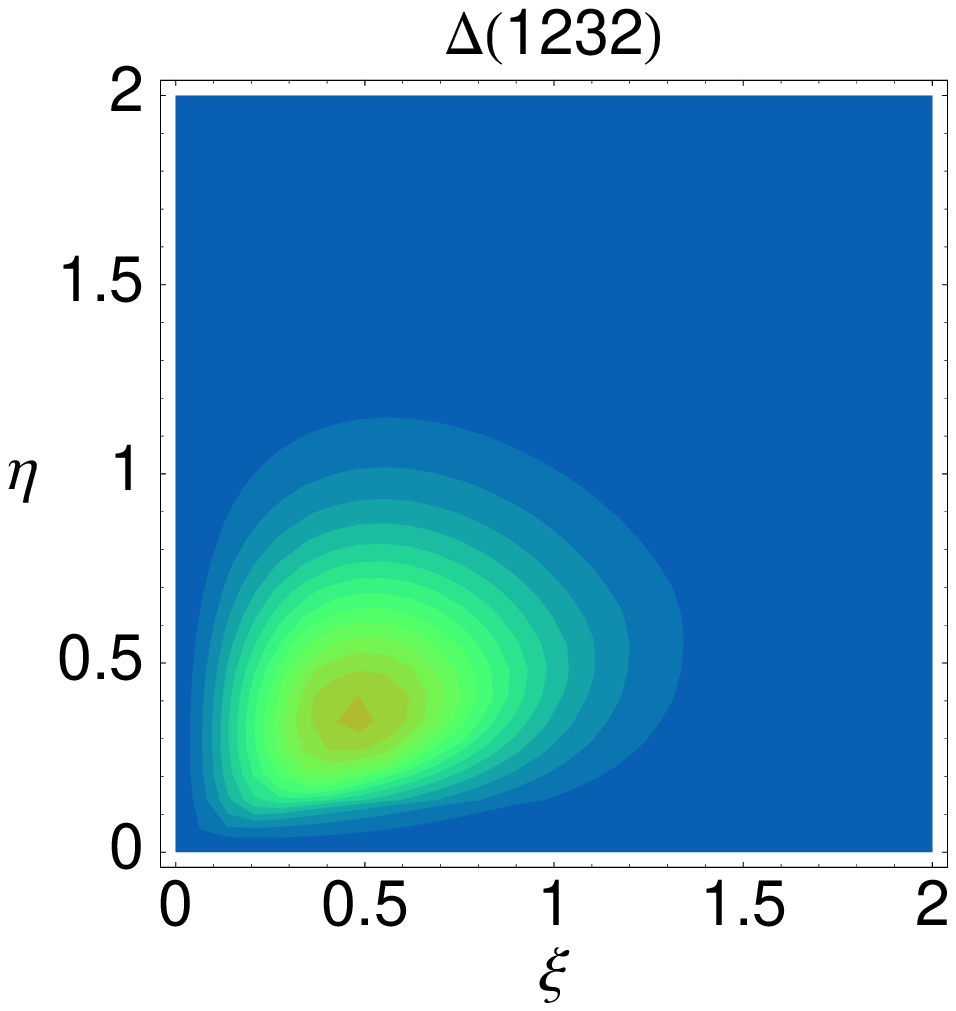}
\includegraphics[width=3.9cm]{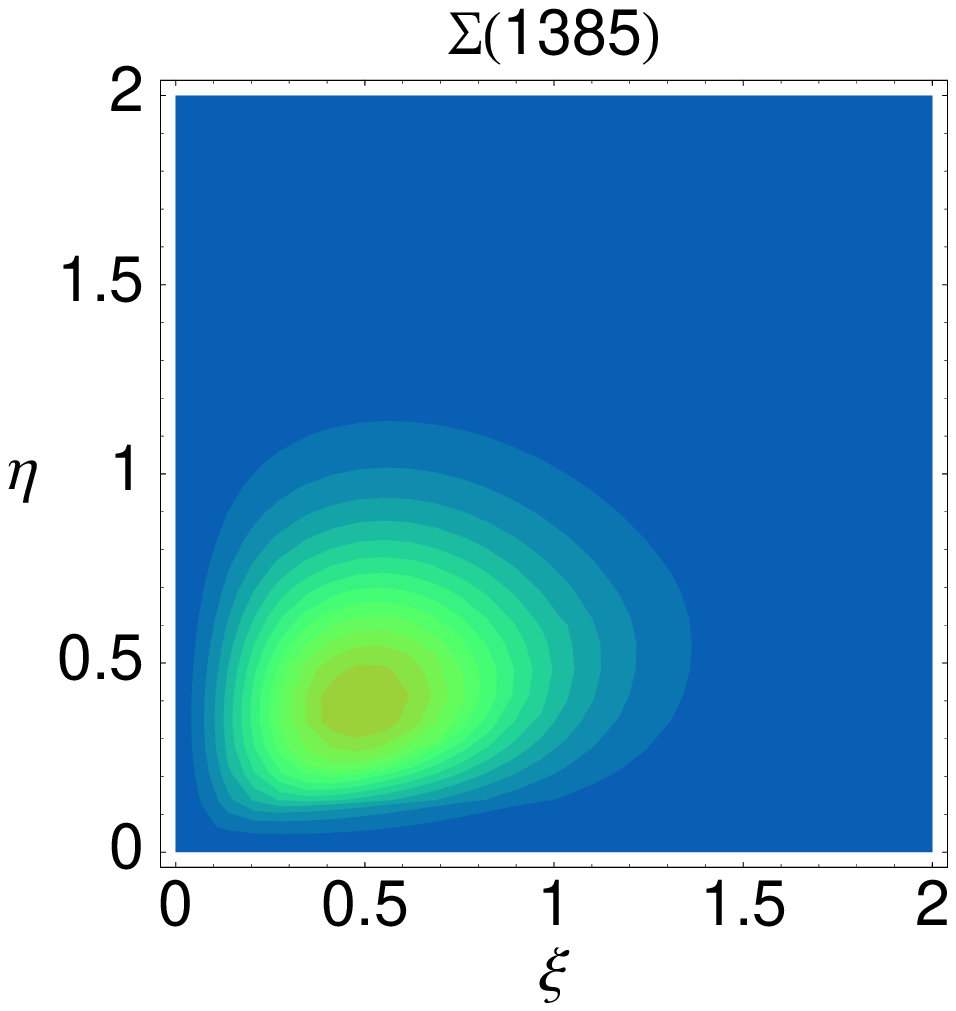}
\includegraphics[width=3.9cm]{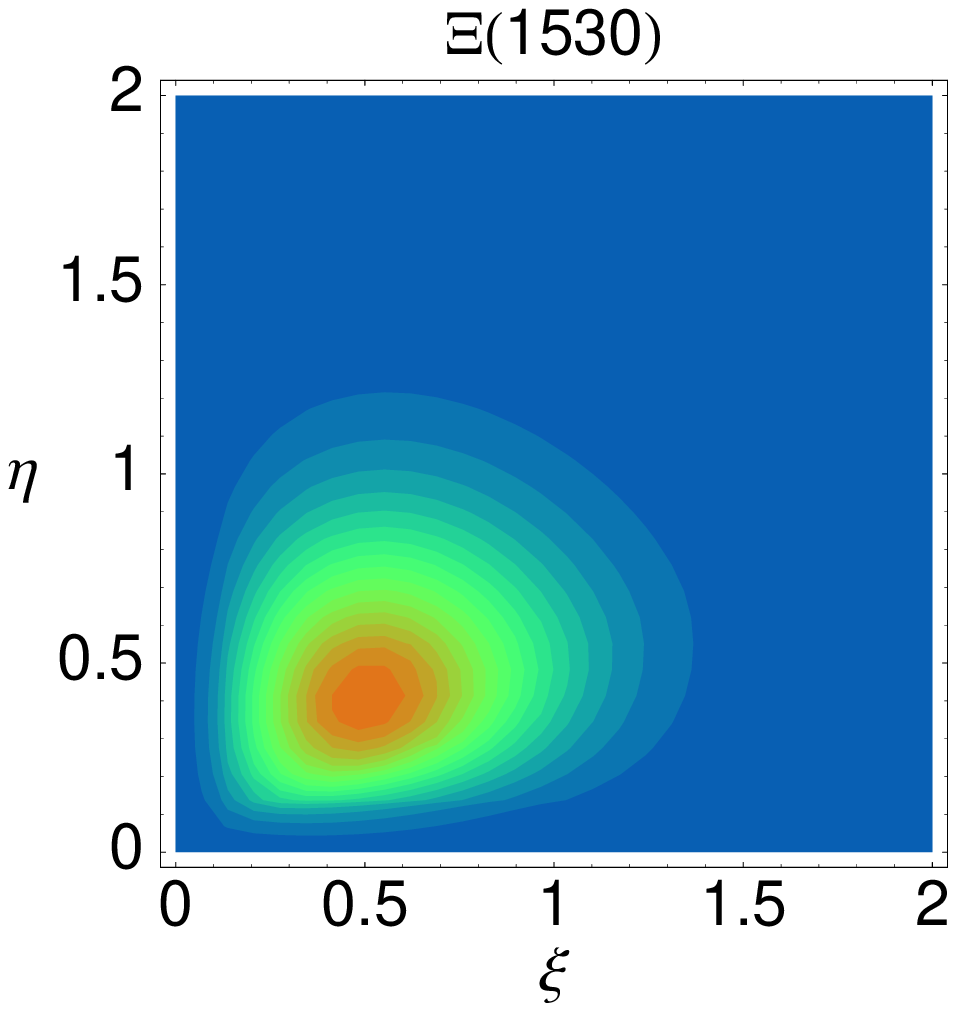}
\includegraphics[width=3.9cm]{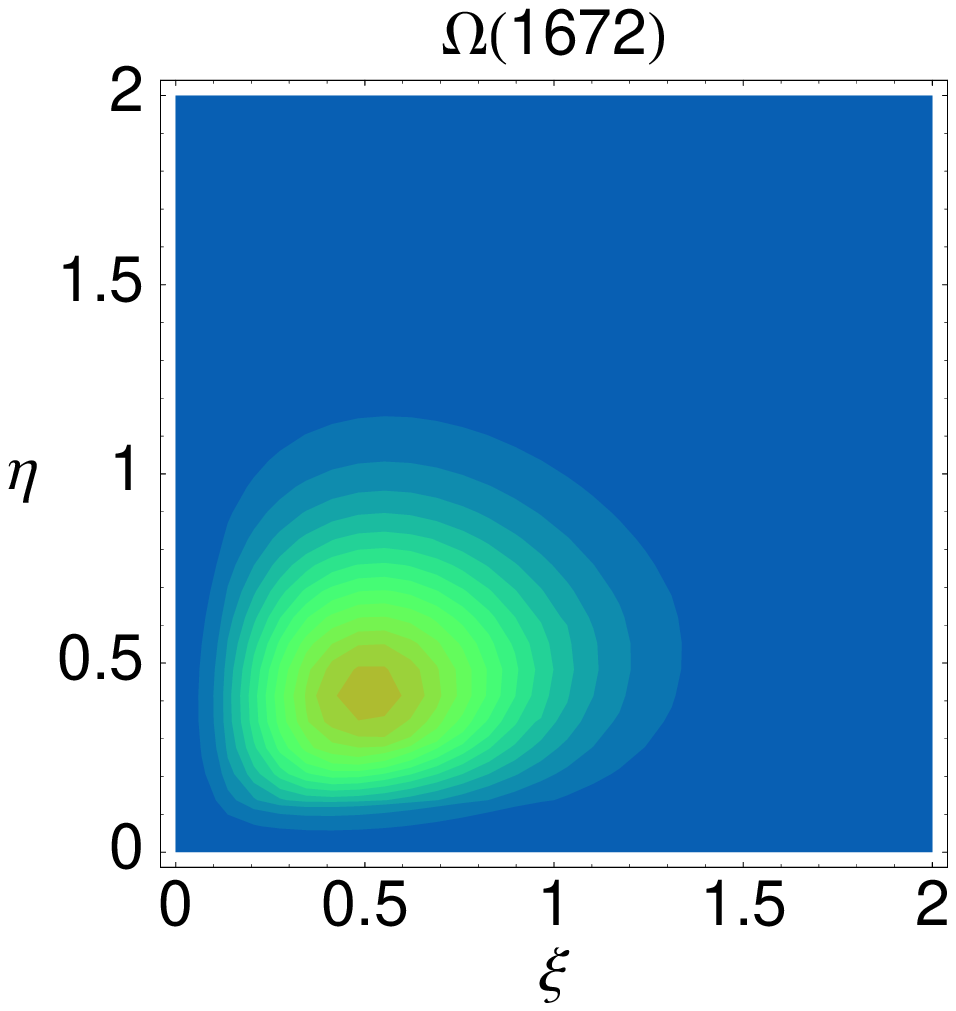}
\caption{Same as Fig.~\ref{fig:multi_1} for the $\frac{3}{2}^+$ decuplet baryon
states $\Delta$(1232), $\Sigma$(1385), $\Xi$(1530), $\Omega$(1672).}
\label{fig:multi_2}
\end{figure*}

The classification of the lowest decuplet states $\Delta(1232)$,  
$\Sigma(1385)$, $\Xi(1530)$, and $\Omega(1672)$ is also straightforward. All are
characterized by $J^P=\frac{3}{2}^+$, where
$L=0$ and $S=\frac{3}{2}$. The eigenstates are totally symmetric with respect
to spin and flavour. This leads to similar spatial probability densities as for
the octet ground states but with a larger extension (see Fig.~\ref{fig:multi_2}).

\subsection*{Octet: $\bf N(1440)$,  $\bf \Lambda(1600)$,
$\bf \Sigma(1660)$, $\bf \Xi(1690)$}
\begin{figure*}
\includegraphics[width=3.9cm]{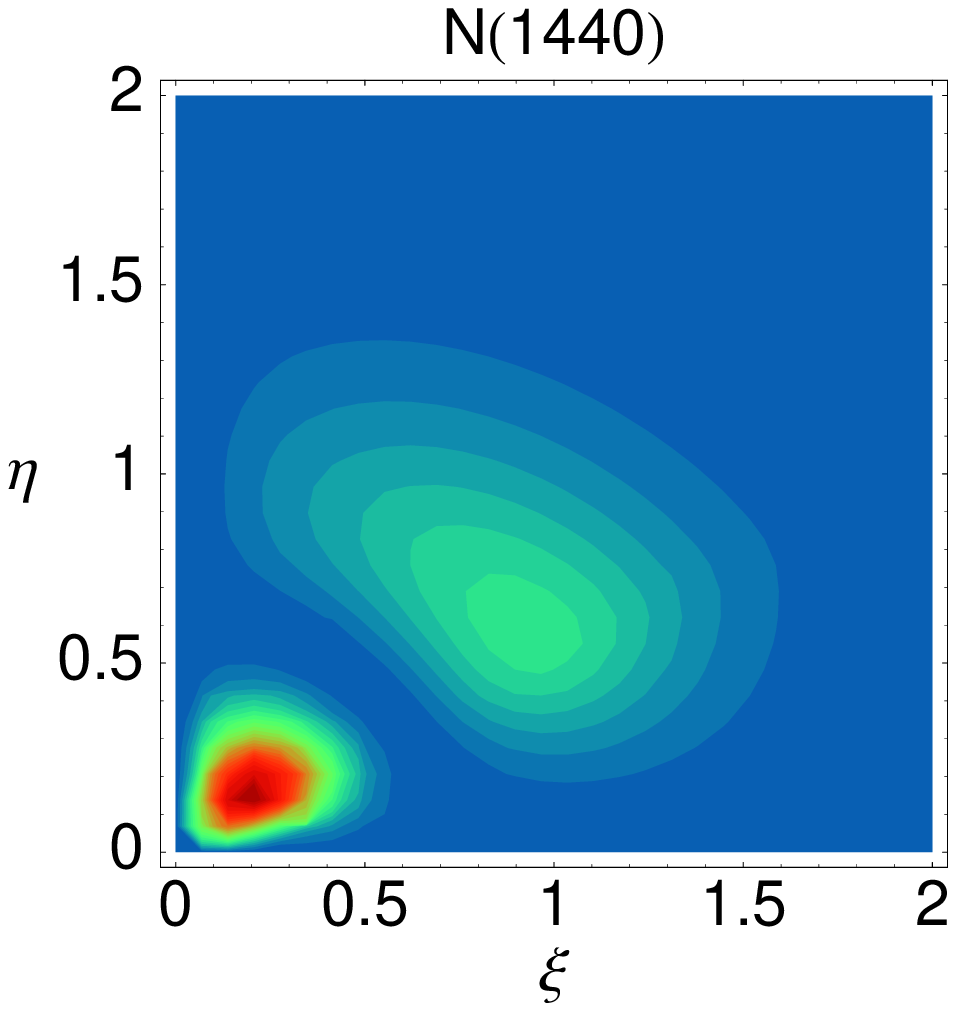}
\includegraphics[width=3.9cm]{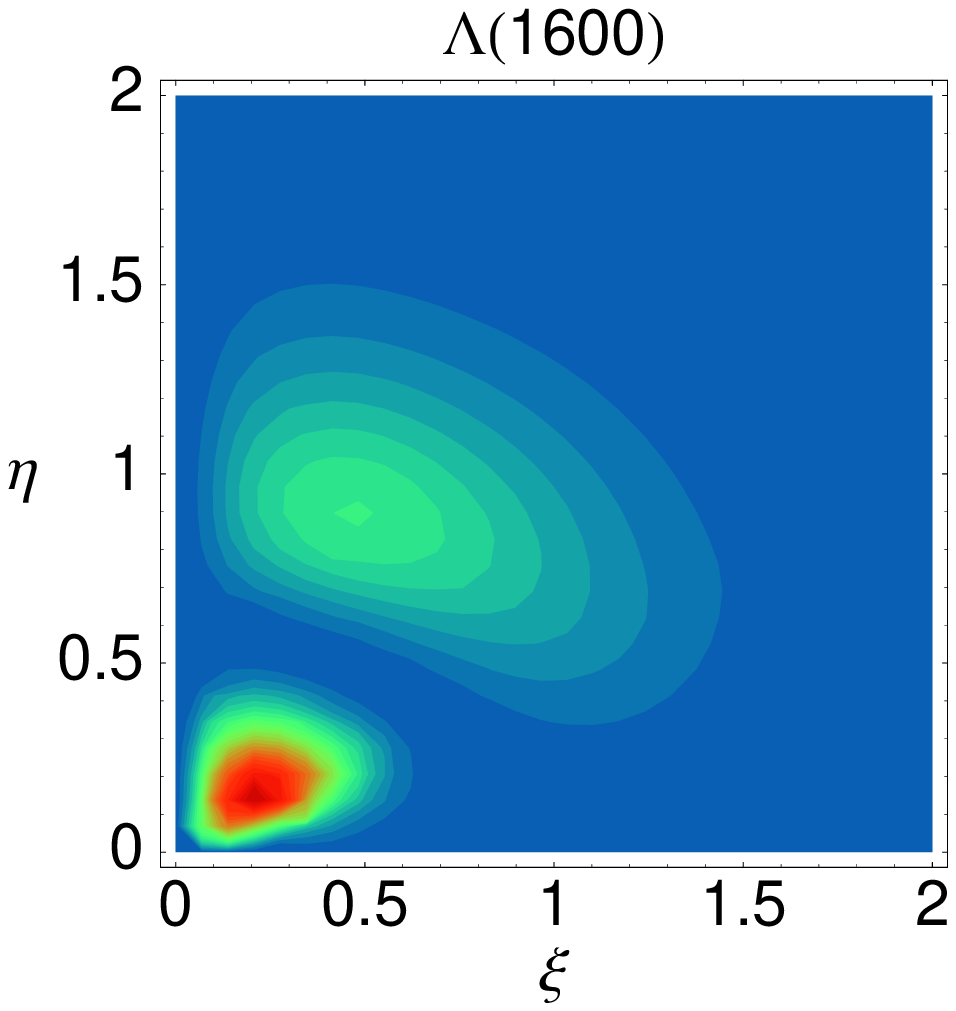}
\includegraphics[width=3.9cm]{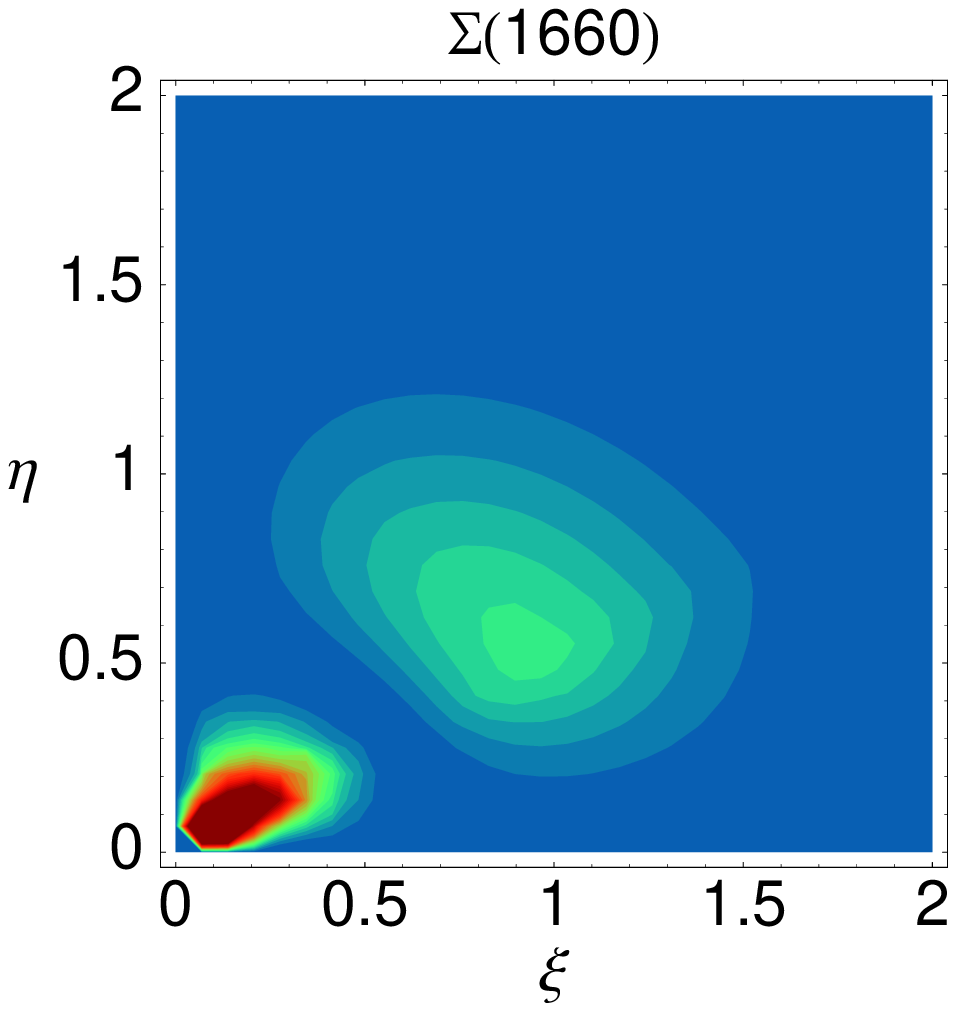}
\includegraphics[width=3.9cm]{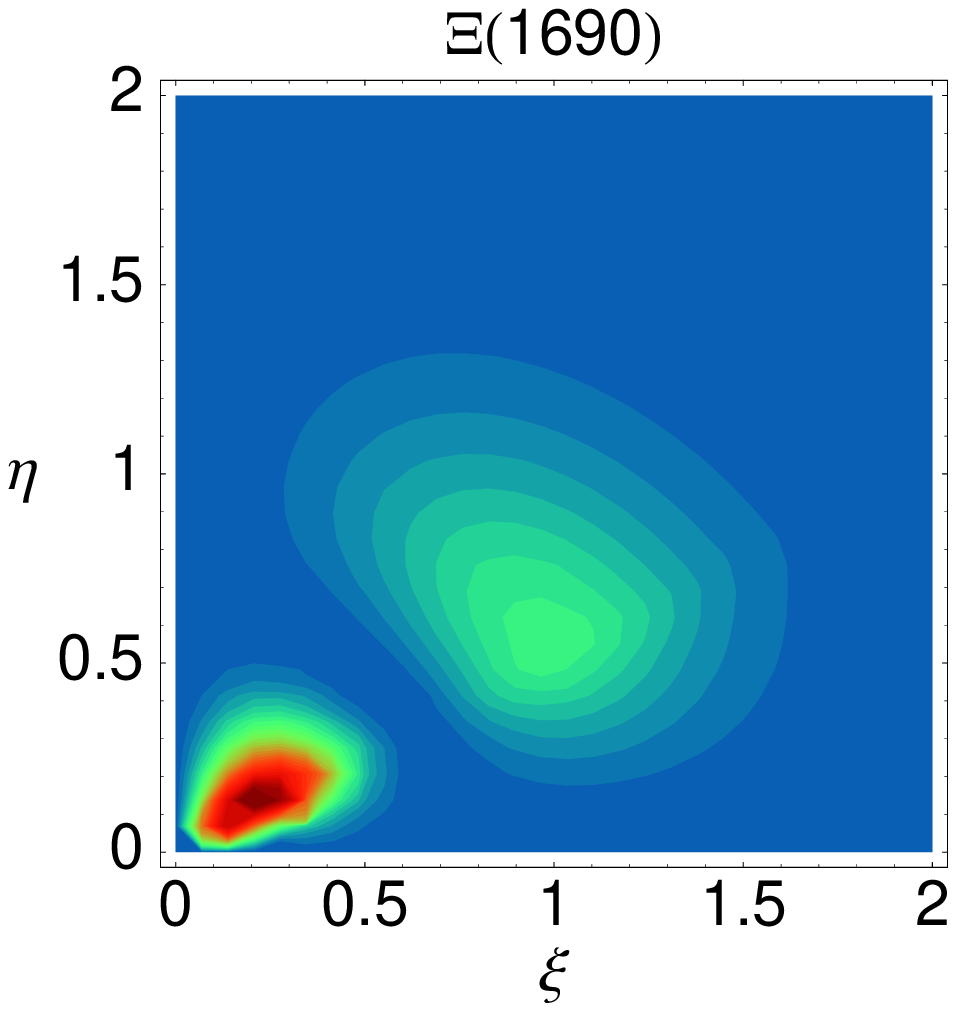}
\caption{Same as Fig.~\ref{fig:multi_1} for the $\frac{1}{2}^+$ octet baryon
states $N(1440)$,  $\Lambda(1600)$, $\Sigma(1660)$, $\Xi(1690)$.}
\label{fig:multi_3}
\end{figure*}

The states $N(1440)$, $\Lambda(1600)$, $\Sigma(1660)$, and $\Xi(1690)$ represent the
first radial excitations with $J^P=\frac{1}{2}^+$ above the octet ground states. 
Their spatial probability densities are shown in Fig.~\ref{fig:multi_3}, with all
of them exhibiting the typical nodal structures.
The RCQM predicts all of these states to be pure flavor octets with the
exception of the $\Lambda(1600)$, which has a singlet contribution of 4\%.
For the
$\Xi$ member of this multiplet the GBE RCQM produces an eigenstate with a theoretical
mass of 1805 MeV. We may identify it with the $\Xi(1690)$ resonance, which appears
as a three-star resonance with no $J^P$ value in the listings of the PDG and is not
included in their multiplet assignments (cf. Table~\ref{tab:multiplet_PDG}).
Similarly, GP classify the $\Xi(1690)$ into the $J^P=\frac{1}{2}^+$
octet~\cite{Guzey:2005vz}. The $\Xi(1690)$ is tentatively a spin $\frac{1}{2}$
state~\cite{Aubert:2006ux}. As we shall see in the next section, the
decay width of the theoretical $J^P=\frac{1}{2}^+$ $\Xi$ state at least
fits into the general pattern observed for the PFSM results and the
magnitude of the total width is in line with the value measured by the BaBar
Collaboration~\cite{Aubert:2006ux}. On the other hand, in a recent study Pervin
and Roberts~\cite{Pervin:2007wa} classify the $\Xi(1690)$ as a $J^P=\frac{1}{2}^-$
octet resonance. More conclusive experimental data on this state would be highly
welcome in order to clarify the situation with respect to its parity.

\subsection*{Octet: $\bf N(1710)$, $\bf \Sigma(1880)$}
\begin{figure*}
\includegraphics[width=3.9cm]{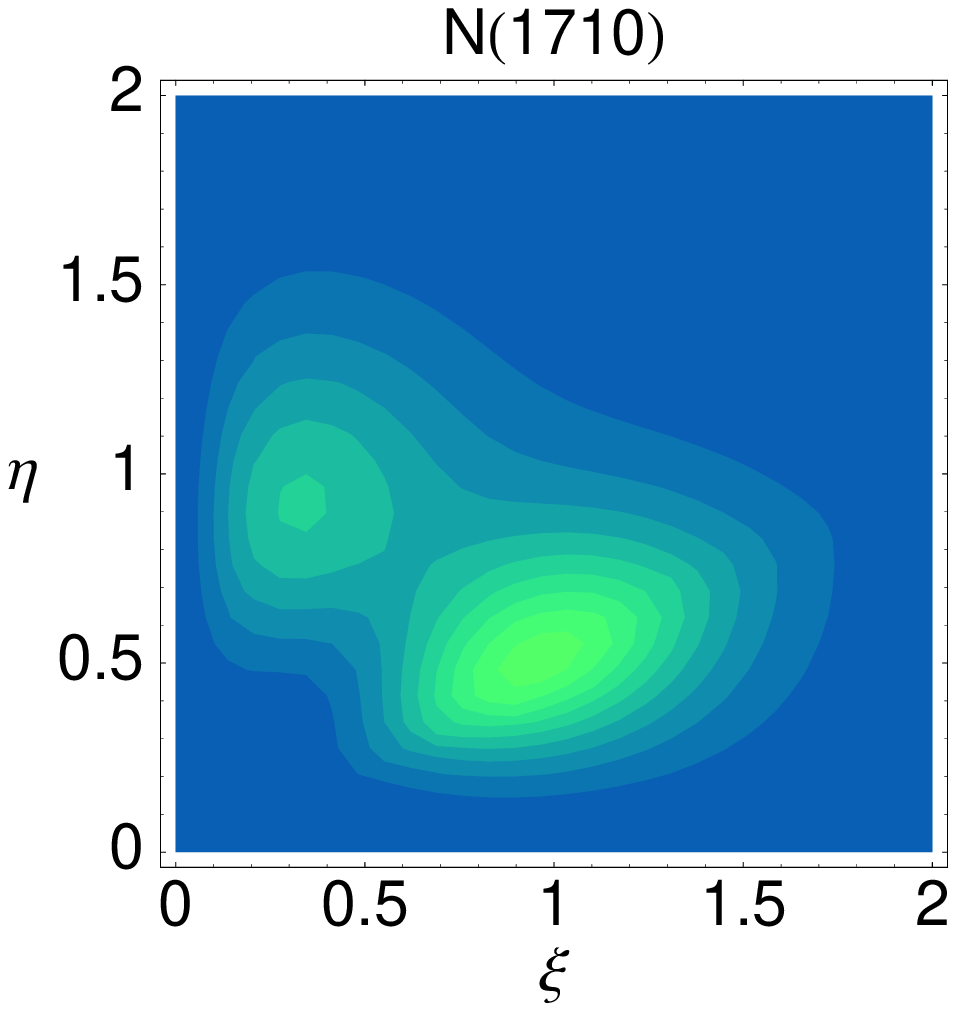}
\includegraphics[width=3.9cm]{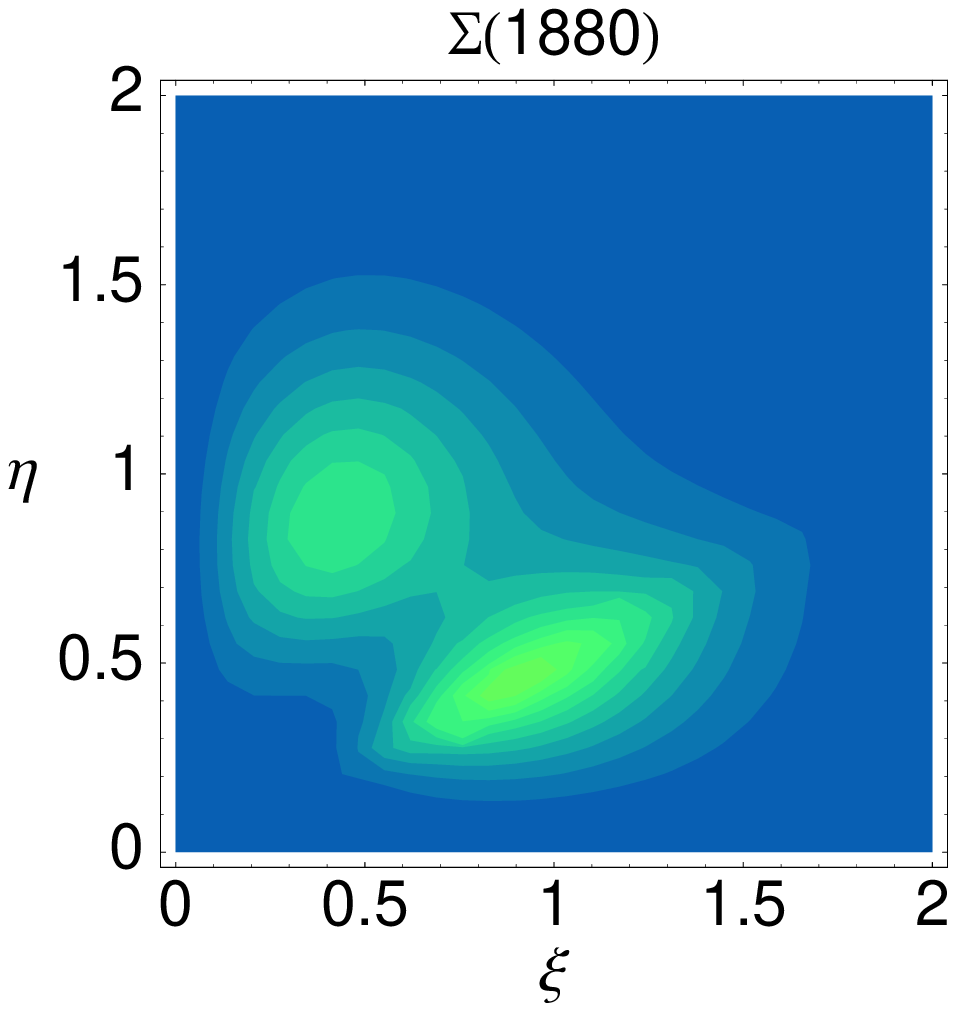}
\caption{Same as Fig.~\ref{fig:multi_1} for the $\frac{1}{2}^+$ octet baryon
states $N(1710)$, $\Sigma(1880)$.}
\label{fig:multi_4}
\end{figure*}

For the second octet of excited states with $J^P=\frac{1}{2}^+$ we can only classify
the $N(1710)$ and the $\Sigma(1880)$. They both have $L=0$ and $S=\frac{1}{2}$
and are predominantly of mixed flavour-spin symmetry. These resonances are
characterized by the typical spatial probability density distributions of second
radial excitations as shown in Fig.~\ref{fig:multi_4}, with a
dip following a straight line through the origin of the ($\xi ,\eta$) plane.
The flavor content of these two resonances is practically pure octet.
The $\Sigma(1880)$ resonance has only two-star status, while its $J^P$ is experimentally confirmed to be $\frac{1}{2}^+$. Besides the $\Sigma(1620)$ it is the only two-star
resonance that is taken into account by the PDG, and they classify it into the same
octet as we do. 

The PDG classifies also the $\Lambda(1810)$ to be a member of this octet.
However, in the RCQMs no flavor octet $\Lambda$ state with $J^P=\frac{1}{2}^+$
is found below 2 GeV, rather one obtains a flavor singlet. Thus the classification
of the $\Lambda$ and likewise the $\Xi$ members of this octet must be left open,
as the possible candidates $\Lambda(2000)$ and $\Xi(2120)$, or even higher $\Xi$'s,
are not well enough established experimentally.

\subsection*{Singlet: $\bf \Lambda(1810)$}
\begin{figure*}
\includegraphics[width=3.9cm]{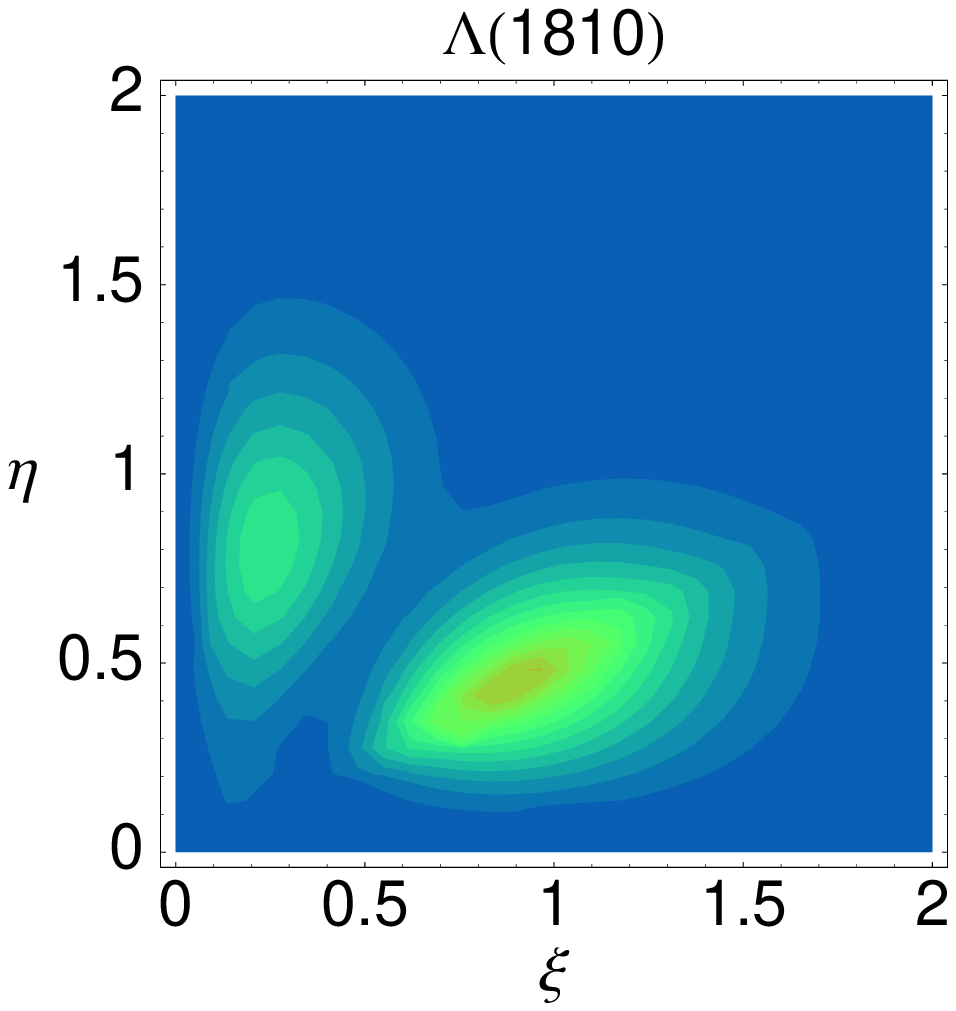}
\includegraphics[width=3.9cm]{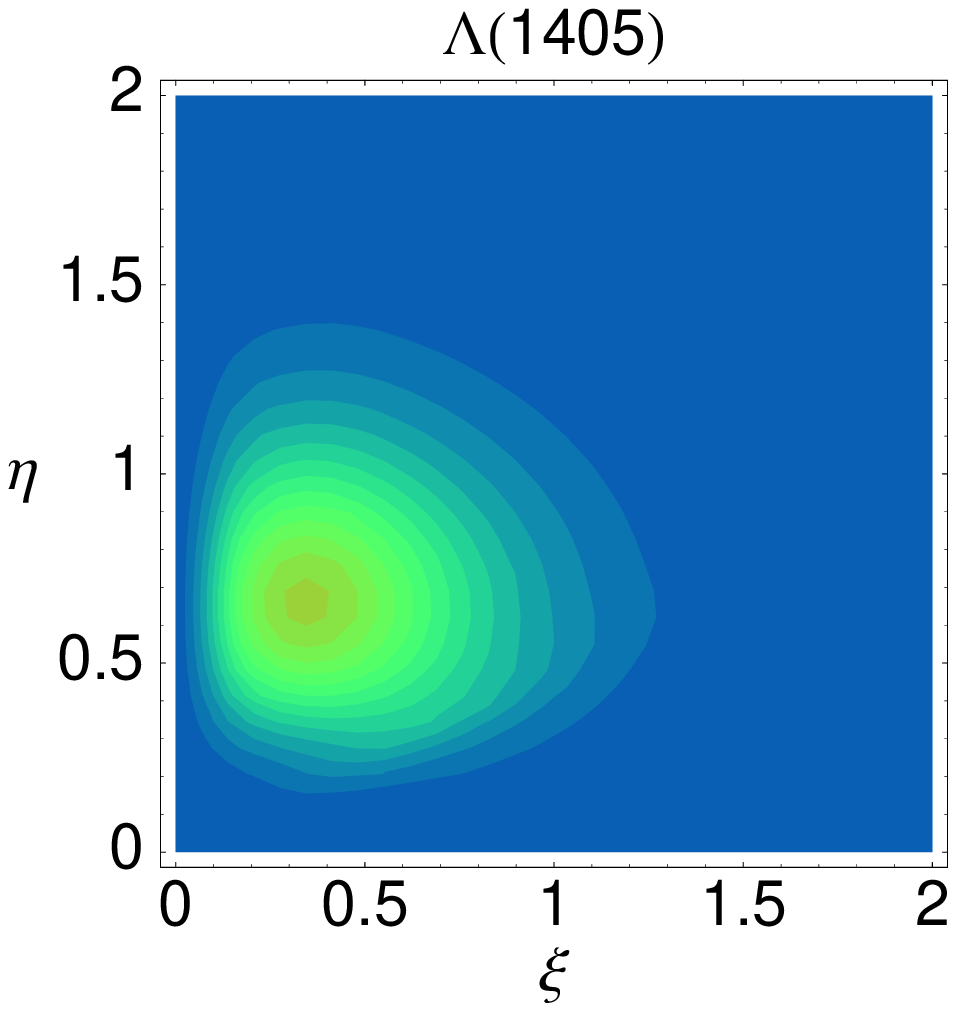}
\includegraphics[width=3.9cm]{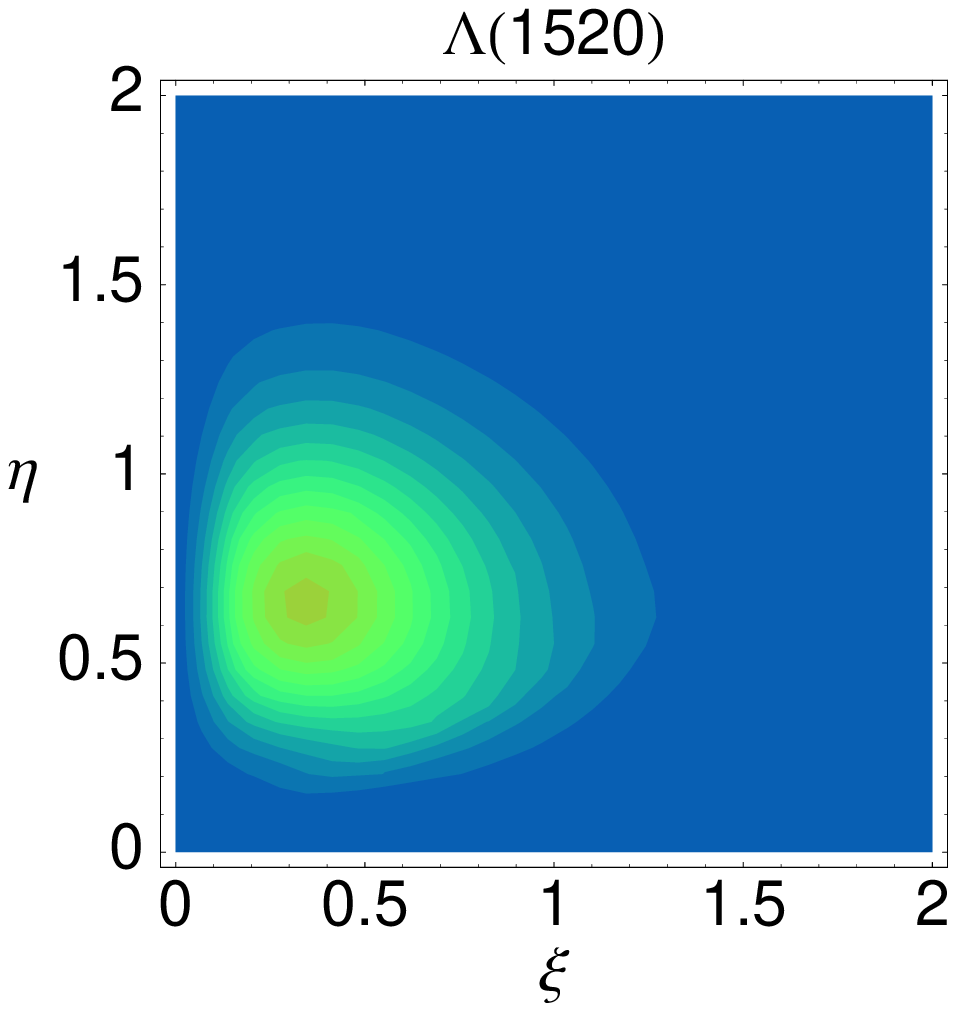}
\caption{Same as Fig.~\ref{fig:multi_1} for the $\frac{1}{2}^+$ singlet baryon
state $\Lambda(1810)$, the $\frac{1}{2}^-$ singlet baryon state $\Lambda(1405)$,
and the $\frac{3}{2}^-$ singlet baryon state $\Lambda$(1520).}
\label{fig:multi_6}
\end{figure*}

The RCQMs produce a $J^P=\frac{1}{2}^+$ with $L=0$ and $S=\frac{1}{2}$ at an
energy of 1799 and 1957 MeV for the GBE an OGE hyperfine interactions, respectively.
They are definitely flavor singlets with only a few percents of octet admixture.
No other suitable $\Lambda$ resonances are found below 2 GeV. Therefore we identify
this $J^P=\frac{1}{2}^+$ eigenstate with the $\Lambda(1810)$.
While this classification differs from the one by the PDG, the same
identification as ours is also suggested by Matagne and
Stancu~\cite{Matagne:2006zf}. The probability density distribution of the
$\Lambda(1810)$ is depicted in Fig.~\ref{fig:multi_6}.

\subsection*{Singlet: $\bf \Lambda(1405)$}
The state $\Lambda(1405)$ is predominantly a flavour singlet with $J^P=\frac{1}{2}^-$,
and it is constructed from $L=1$ and $S=\frac{1}{2}$. The octet admixture is
about one third, by far larger as in the case of $\Lambda(1810)$. The corresponding
probability density distribution is also shown in Fig.~\ref{fig:multi_6}. It should
be mentioned that the $\Lambda(1405)$ represents a notorious difficulty in
reproducing its mass for all constituent quark models relying on \{QQQ\} configurations,
very probably because the mass value happens to lie so close to the $NK$ threshold.

The next higher $\Lambda$ eigenstate is an octet and it should thus be identified
with the $\Lambda(1670)$ (see below). This reflects the typical behaviour of the
flavor singlet always lying lower than the octet as it is also found with the
pairs of $\Lambda(1520)$ and $\Lambda(1690)$ and tentatively in the case of
$\Lambda(1810)$.

\subsection*{Singlet: $\bf \Lambda(1520)$}
For the RCQMs used in this work the singlet states $\Lambda(1405)$
and $\Lambda(1520)$ are degenerate. Consequently, there is no difference between
$J^P=\frac{1}{2}^-$ and $J^P=\frac{3}{2}^-$ with respect to baryon spectroscopy and the wave functions (see Fig.~\ref{fig:multi_6}). However, when considering the decays, the distinct
total angular momenta lead to other couplings and thus produce different results
for the decay widths (cf. Fig.~\ref{fig:decay_graph_OGE_GBE_singl} in the next section).

\subsection*{Octet: $\bf N(1535)$, $\bf \Lambda(1670)$, $\bf \Sigma(1560)$} 
\begin{figure*}
\includegraphics[width=3.9cm]{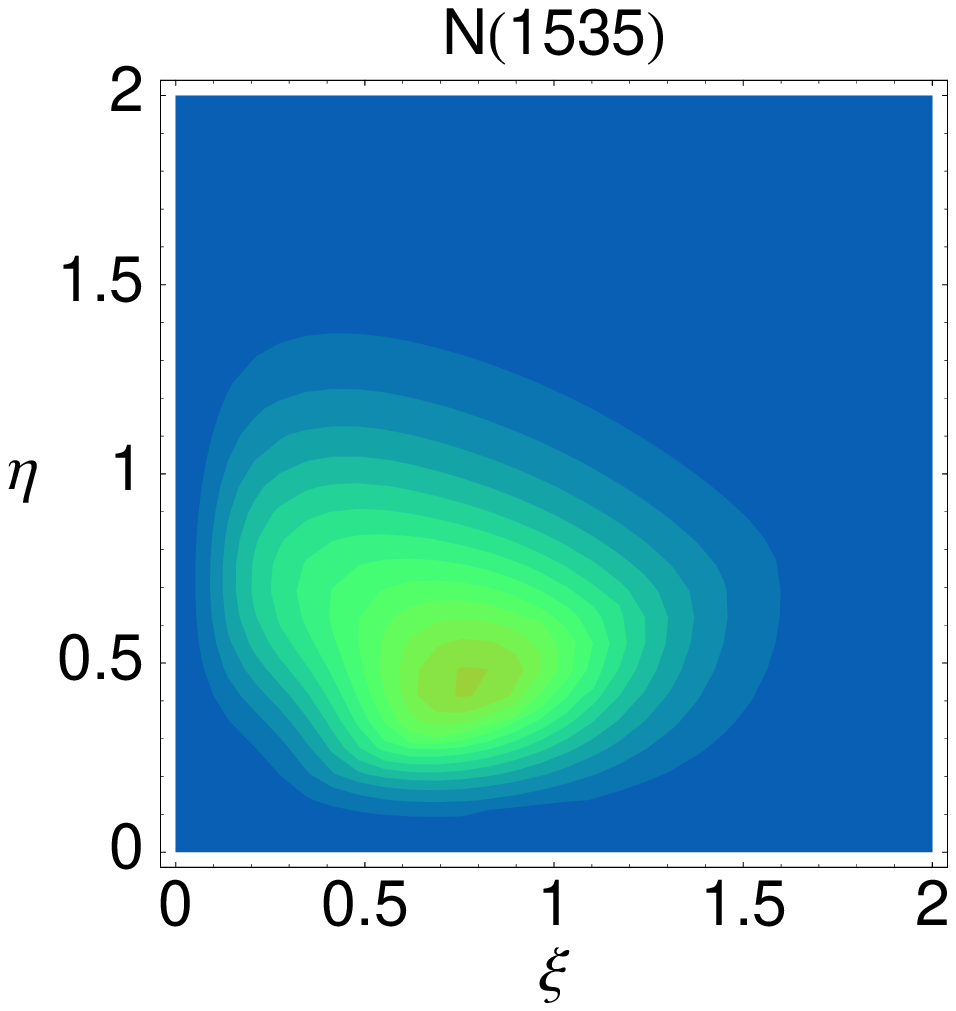}
\includegraphics[width=3.9cm]{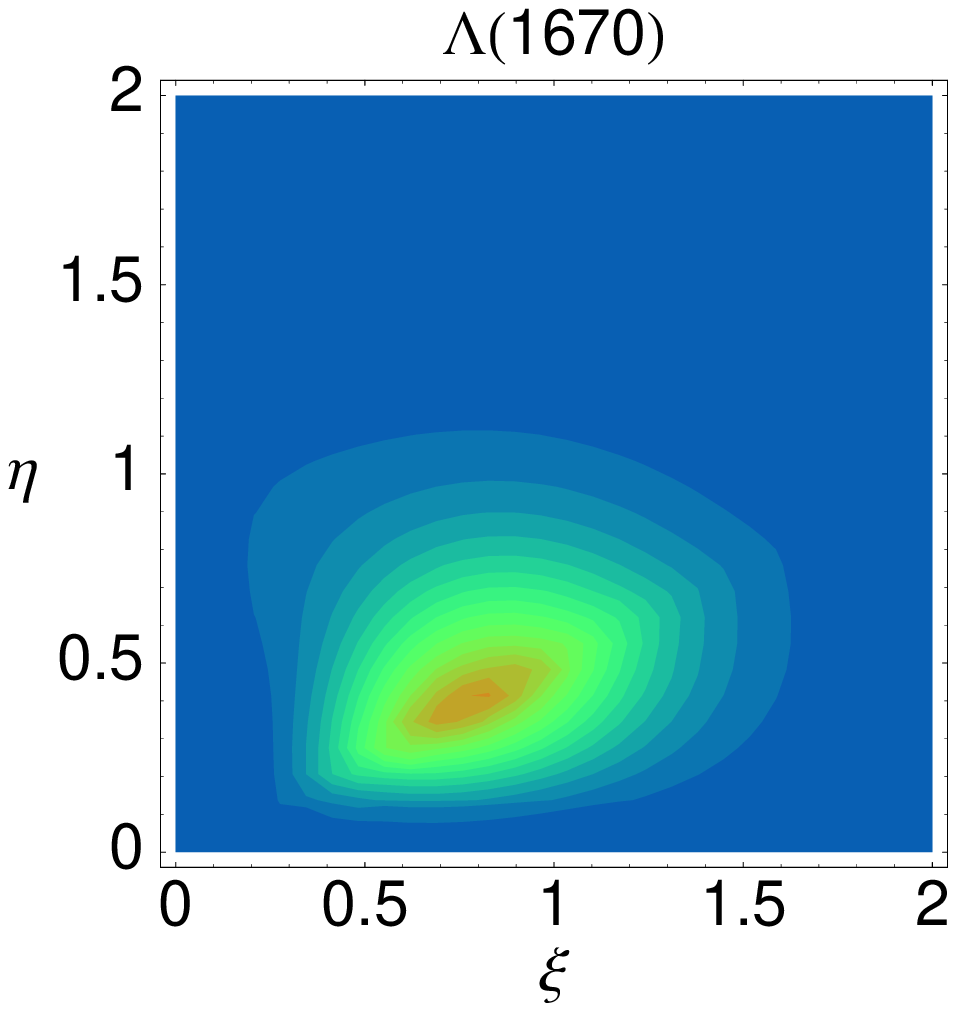}
\includegraphics[width=3.9cm]{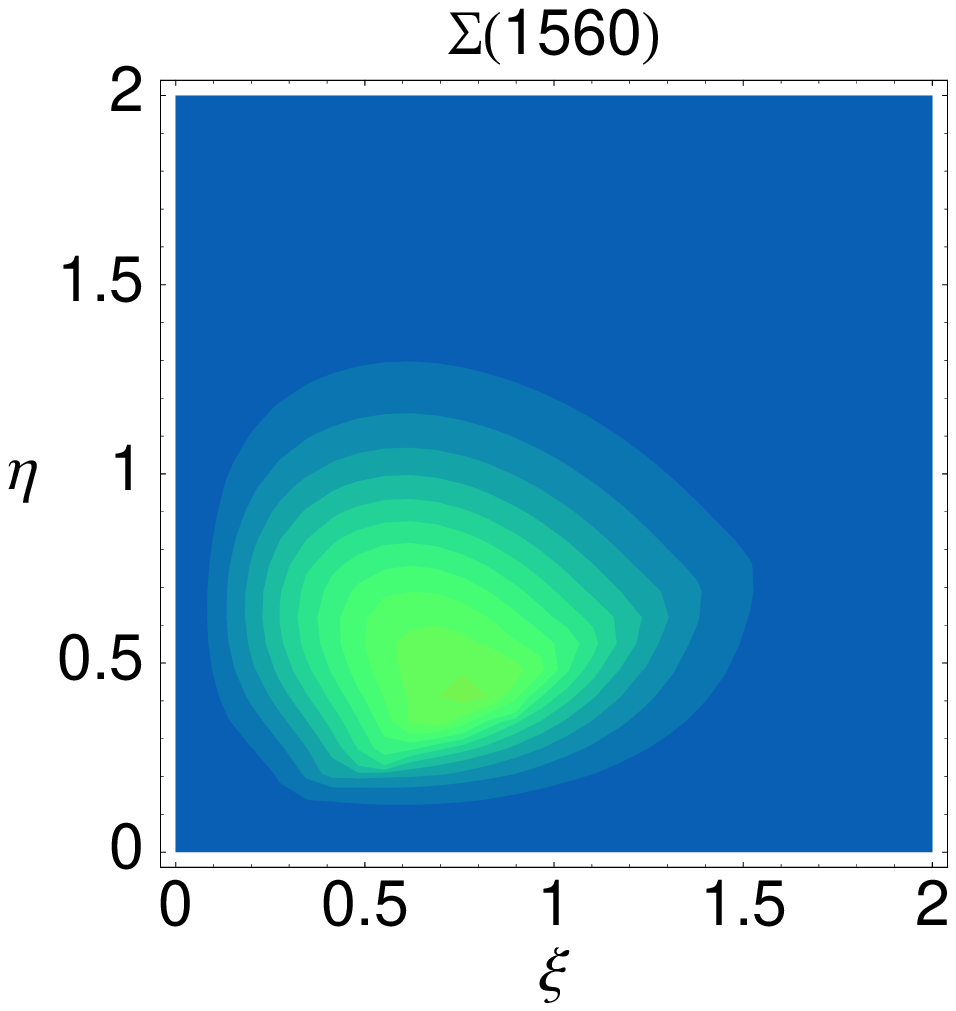}
\caption{Same as Fig.~\ref{fig:multi_1} for the $\frac{1}{2}^-$ octet baryon
states $N(1535)$, $\Lambda(1670)$, $\Sigma(1560)$.}
\label{fig:multi_9}
\end{figure*}

Into the next octet of excited states we assign the 
$N(1535)$, $\Lambda(1670)$, and $\Sigma(1560)$ resonances, which all have
$J^P=\frac{1}{2}^-$. Only for the $\Lambda(1670)$ we find a sizable flavor singlet
component, mixing this state with the $\Lambda(1405)$. The spatial probability
density distributions are shown in Fig.~\ref{fig:multi_9}.

For the $\Sigma$ member in this octet we advocate the experimentally measured
$\Sigma(1560)$, which has only two-star status and is not considered in the
PDG classification (cf. Table~\ref{tab:multiplet_PDG}). Our identification of the
lowest $J^P=\frac{1}{2}^-$ $\Sigma$ RCQM eigenstate, being a flavor octet, with
$\Sigma(1560)$ is substantiated mainly by its decay properties~\cite{Melde:2006yw}.
Also, if this state becomes better established from experiment at this low energy,
there is hardly another choice of placing it into a flavor multiplet.

The $\Xi$ assignments in this multiplet must again be left open, as the possible
remaining candidates seen in phenomenology, the $\Xi(1620)$ and maybe the $\Xi(2120)$,
bear only one-star status and their $J^P$ is not measured. 
\subsection*{Octet: $\bf N(1650)$, $\bf \Lambda(1800)$, $\bf \Sigma(1620)$} 
\begin{figure*}
\includegraphics[width=3.9cm]{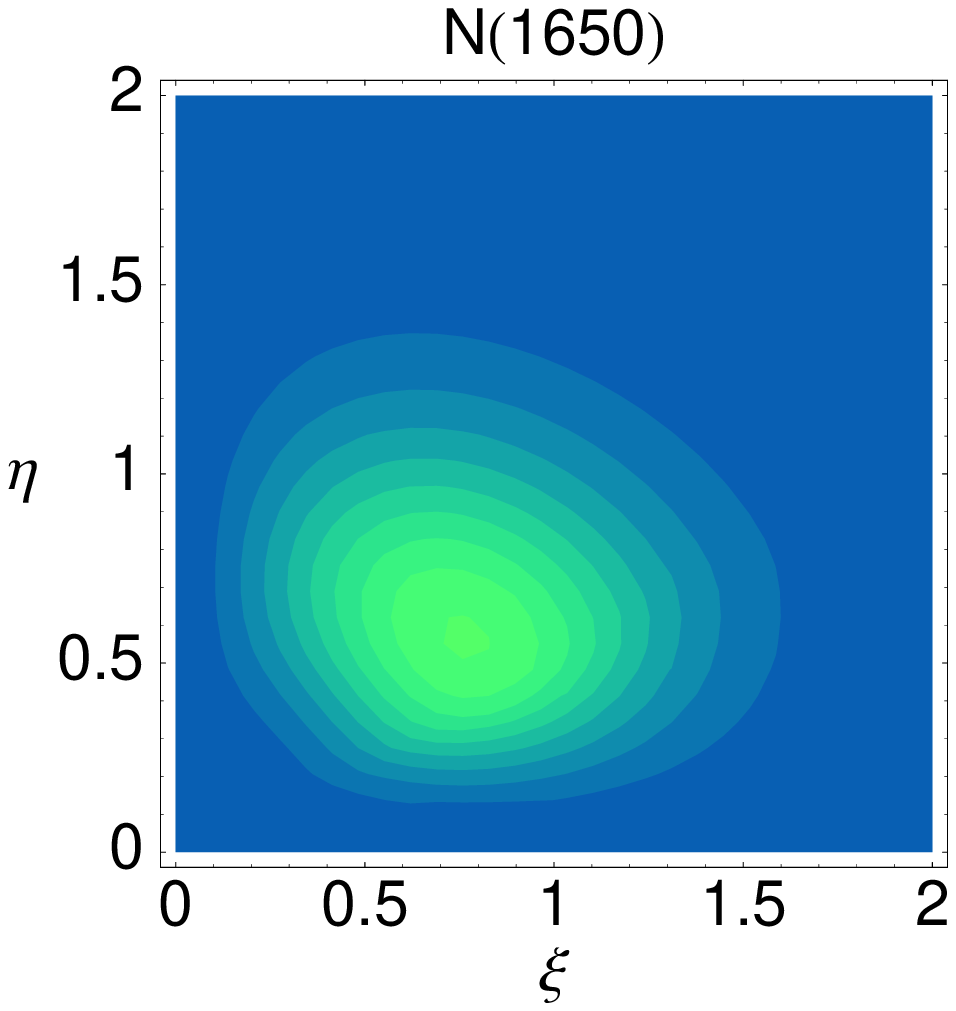}
\includegraphics[width=3.9cm]{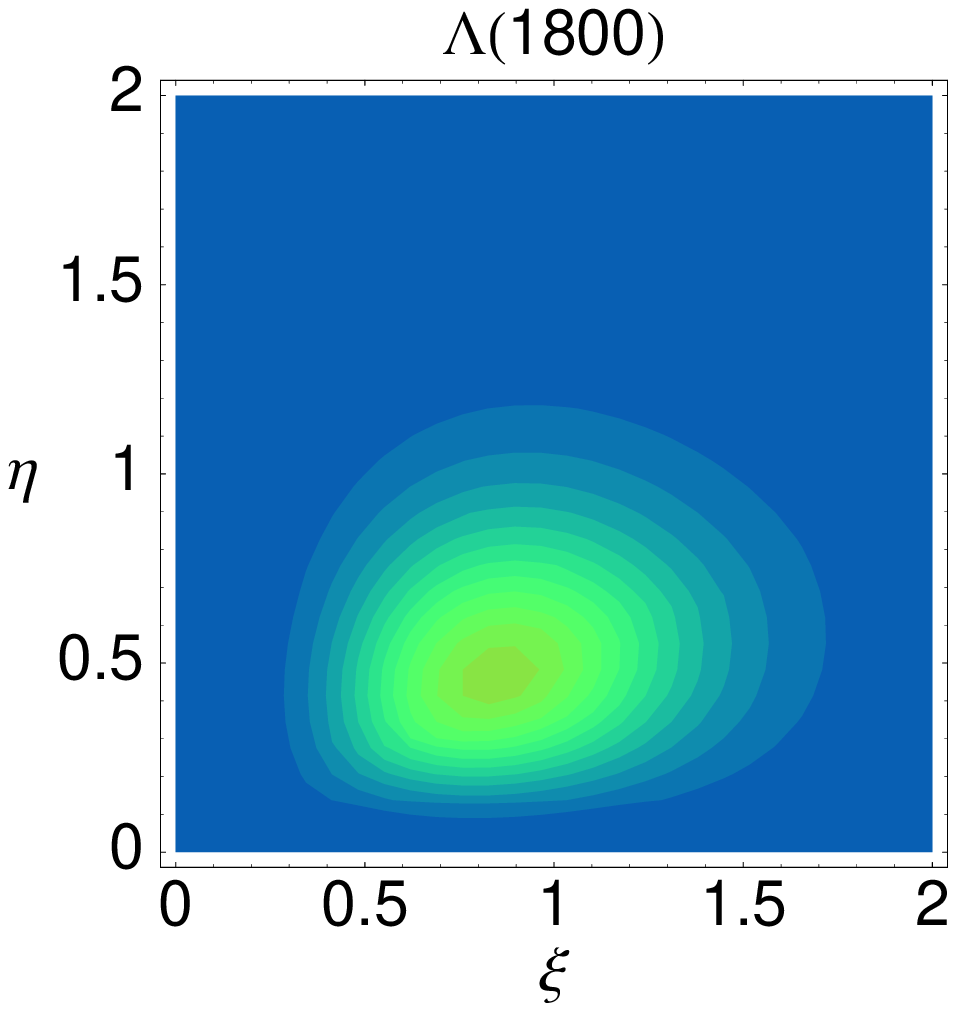}
\includegraphics[width=3.9cm]{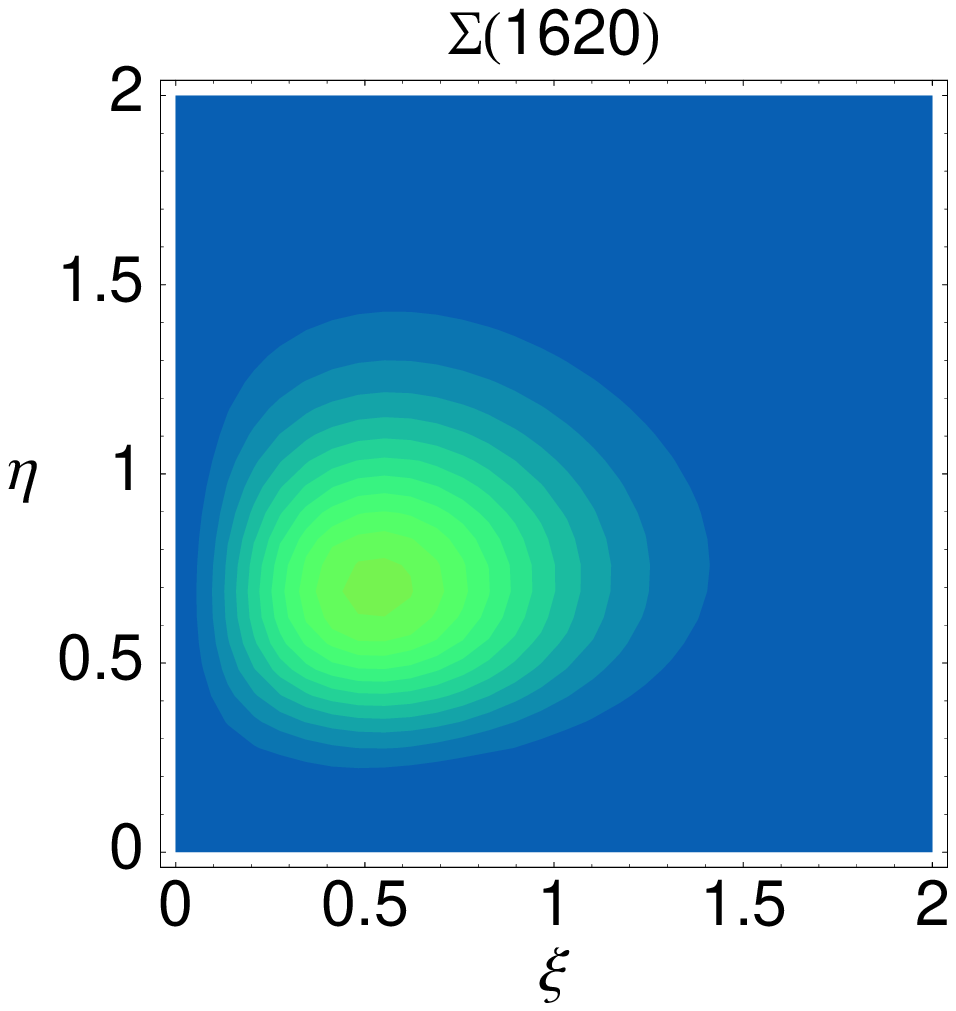}
\caption{Same as Fig.~\ref{fig:multi_1} for the $\frac{1}{2}^-$ octet baryon
states $N(1650)$, $\Lambda(1800)$, $\Sigma(1620)$.}
\label{fig:multi_14}
\end{figure*}

The next octet is the one with $J^P=\frac{1}{2}^-$ containing the $N(1650)$,
$\Lambda(1800)$, and $\Sigma(1620)$ composed of $L=1$ and $S=\frac{3}{2}$.
They all exhibit a pure flavor octet content with no admixtures at all. The
corresponding spatial probability density distributions are shown in
Fig~\ref{fig:multi_14}. 

For the $\Sigma$ member of this octet it is suggested to choose the
$\Sigma(1620)$, which has again only two-star status, however, with known
$J^P=\frac{1}{2}^-$. This identification is further motivated by the decay
properties of the second excited octet $\Sigma$ state with
$J^P=\frac{1}{2}^-$~\cite{Melde:2006yw}.
\subsection*{Octet: $\bf N(1520)$, $\bf \Lambda(1690)$,
$\bf \Sigma(1670)$, $\bf \Xi(1820)$}
\begin{figure*}
\includegraphics[width=3.9cm]{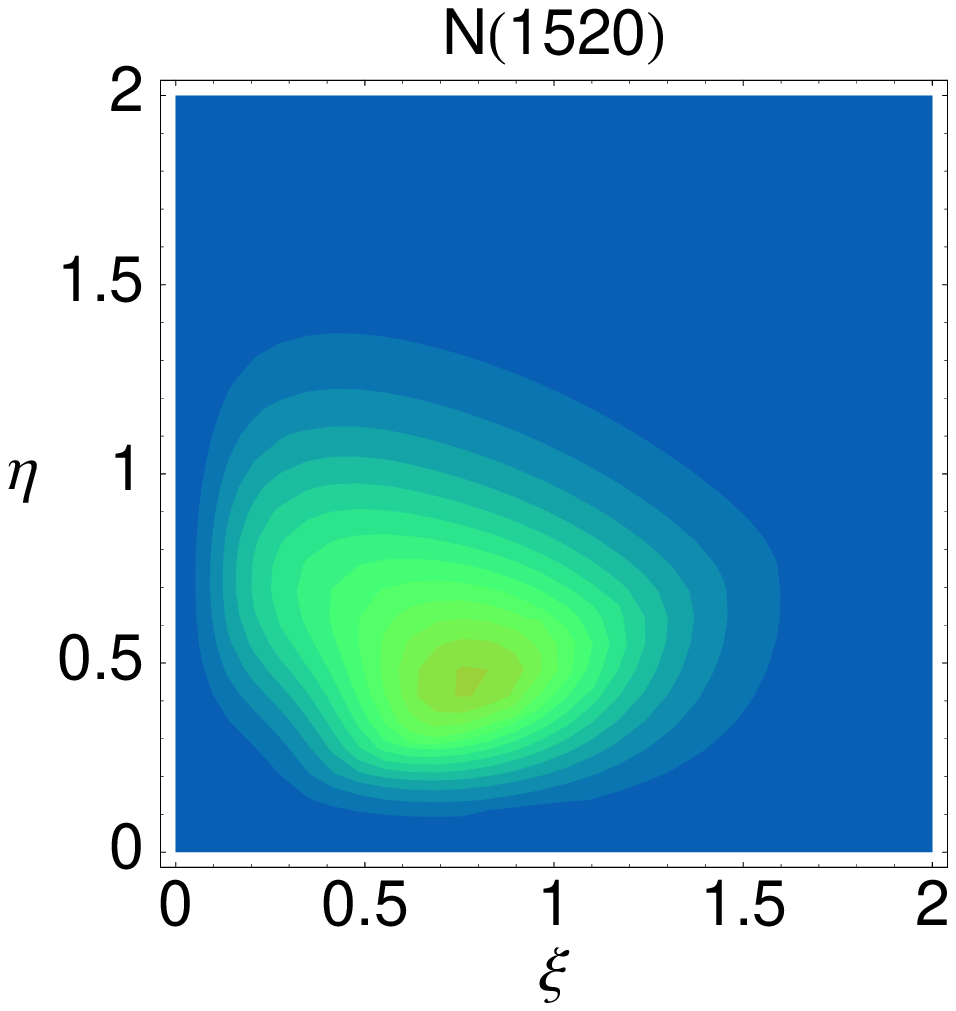}
\includegraphics[width=3.9cm]{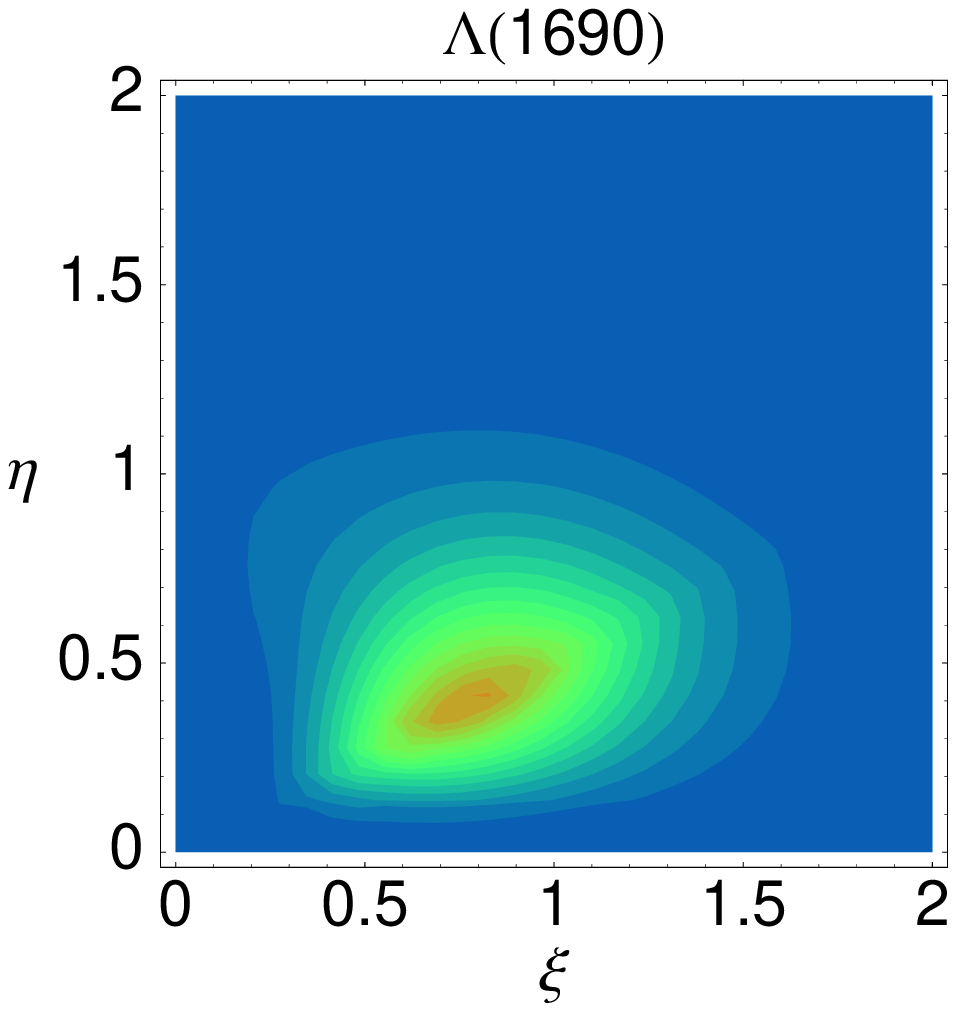}
\includegraphics[width=3.9cm]{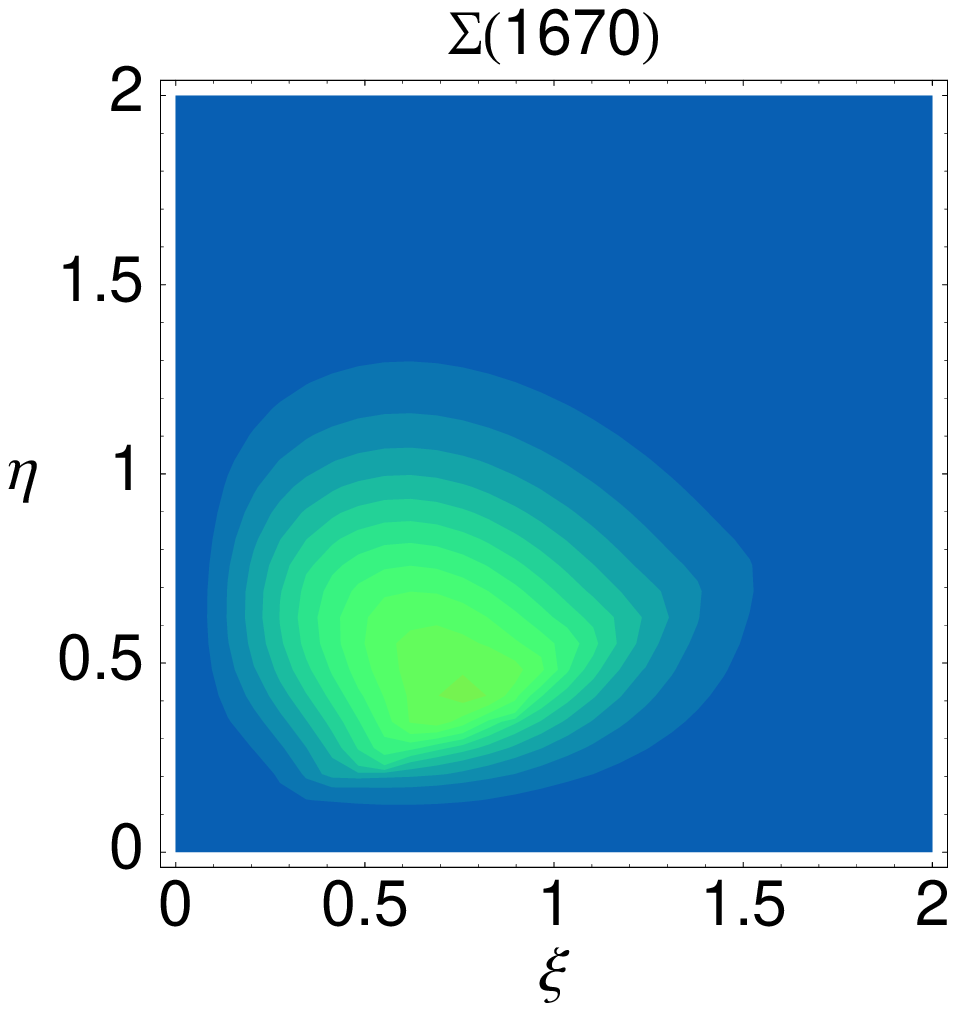}
\includegraphics[width=3.9cm]{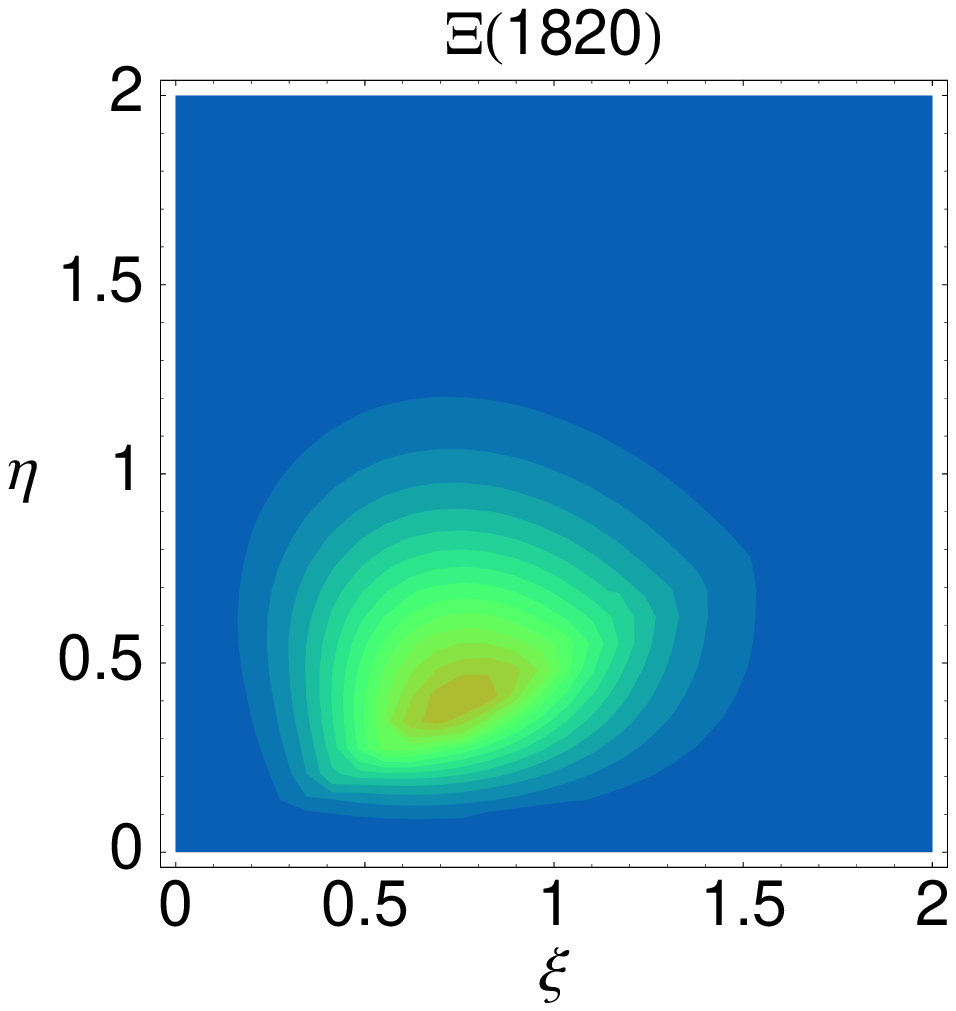}
\caption{Same as Fig.~\ref{fig:multi_1} for the $\frac{3}{2}^-$ octet baryon
states $N(1520)$, $\Lambda(1690)$, $\Sigma(1670)$, $\Xi(1820)$.}
\label{fig:multi_8}
\end{figure*}

The $J^P=\frac{3}{2}^-$ octet with $N(1520)$ is completely filled with resonances
experimentally established, namely with $\Lambda(1690)$,
$\Sigma(1670)$, and $\Xi(1820)$. The same classifications are suggested by the PDG
and also by GP. In our RCQMs the $\Lambda(1690)$ is degenerate with the
$\Lambda(1670)$ and thus also exhibits a sizable singlet admixture. The $N(1520)$
is a pure flavor octet, and the $\Sigma(1670)$ and $\Xi(1820)$ contain only
small decuplet components. The spatial probability density distributions of all
of these resonances are shown in Fig.~\ref{fig:multi_8}.  
\subsection*{Octet: $\bf N(1700)$, $\bf \Sigma(1940)$} 
\begin{figure*}
\includegraphics[width=3.9cm]{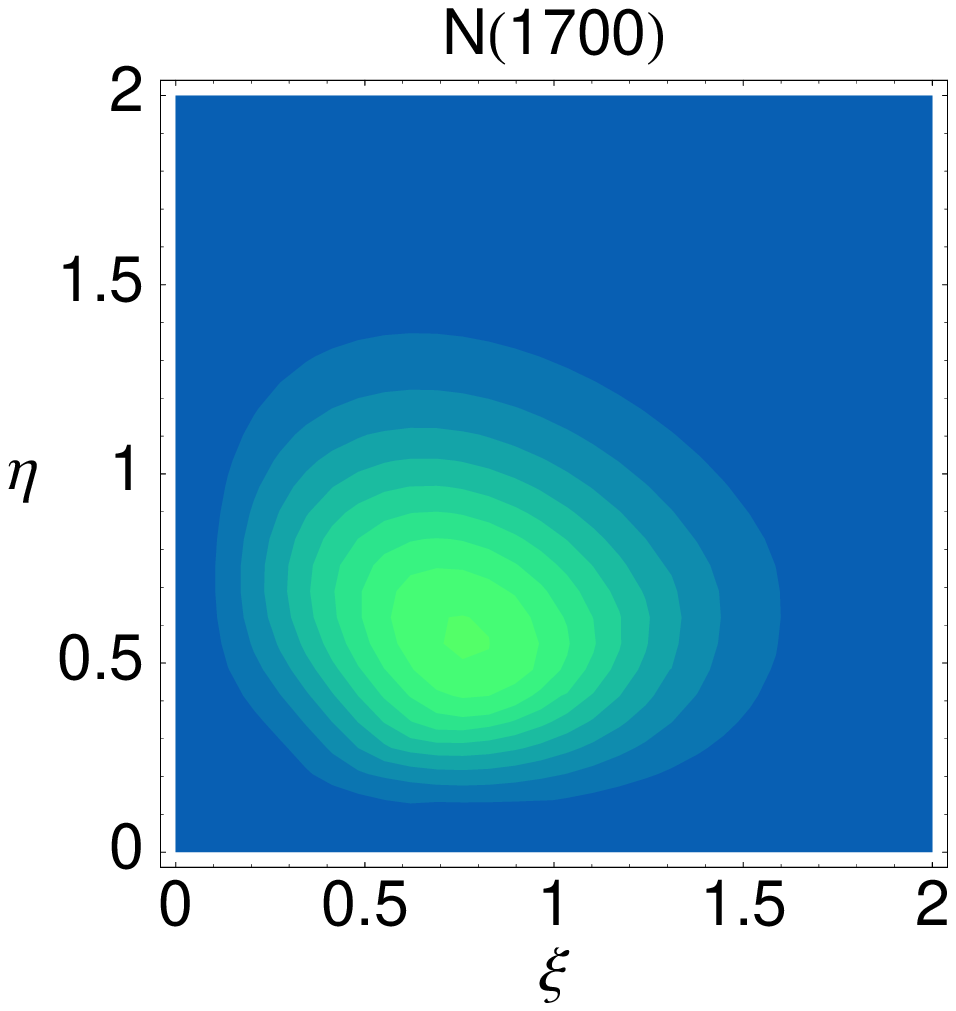}
\includegraphics[width=3.9cm]{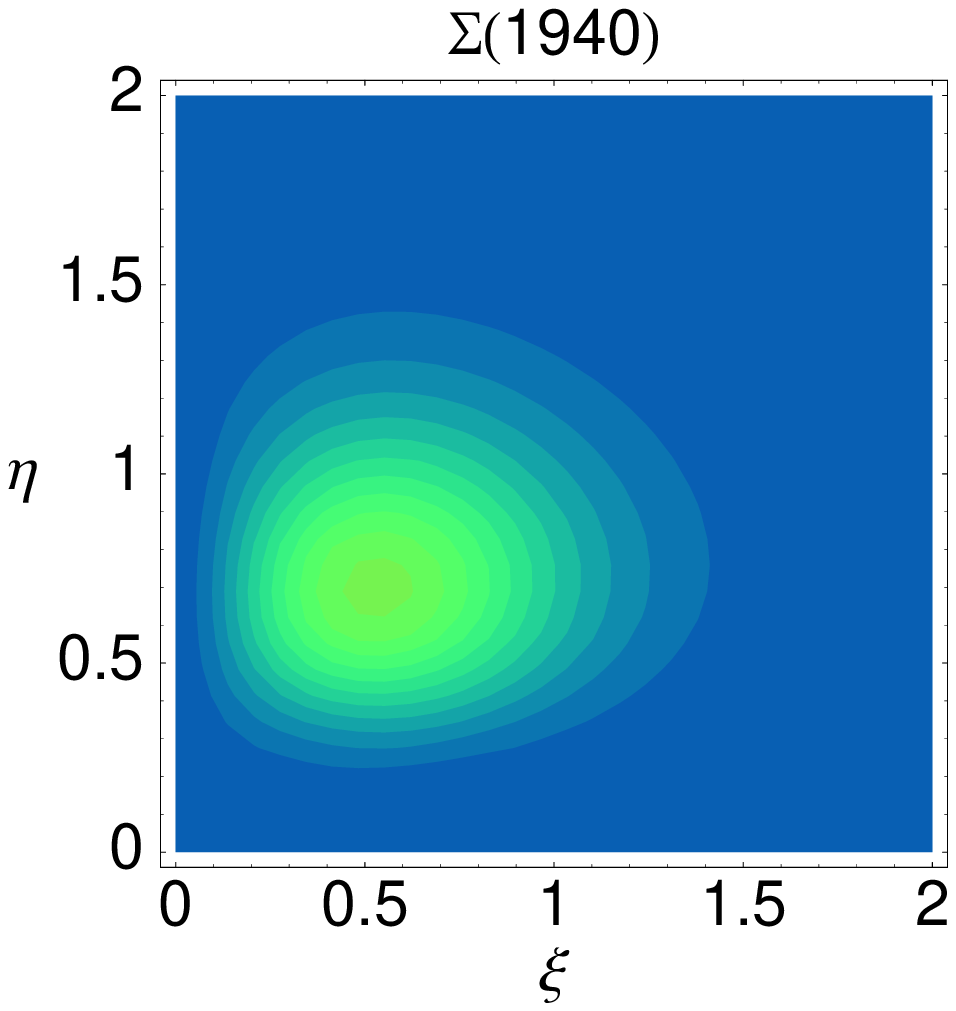}
\caption{Same as Fig.~\ref{fig:multi_1} for the $\frac{3}{2}^-$ octet baryon
states $N(1700)$, $\Sigma(1940)$.}
\label{fig:multi_11}
\end{figure*}

For the $J^P=\frac{3}{2}^-$ octet involving $N(1700)$ we can at most classify the
$\Sigma(1940)$ as a further member. This is in line with the classification by GP,
while the PDG reports $\Sigma(1940)$ with $J^P=\frac{3}{2}^-$ but does not
classify it into this octet nor into
the decuplet with $J^P=\frac{3}{2}^-$. The latter would be an alternative possibility,
and one must await further experimental results, especially on the mesonic decays,
for which no data have so far been reported by the PDG. For the $\Lambda$ member
of this octet there is no reliable experimental evidence, and the situation in the
$\Xi$ sector is similarly questionable. 
The spatial density distributions of the two states classified into this octet are
demonstrated in Fig.~\ref{fig:multi_11}.

\subsection*{Octet: $\bf N(1675)$, $\bf \Lambda(1830)$,
$\bf \Sigma(1775)$, $\bf \Xi(1950)$} 
\begin{figure*}
\includegraphics[width=3.9cm]{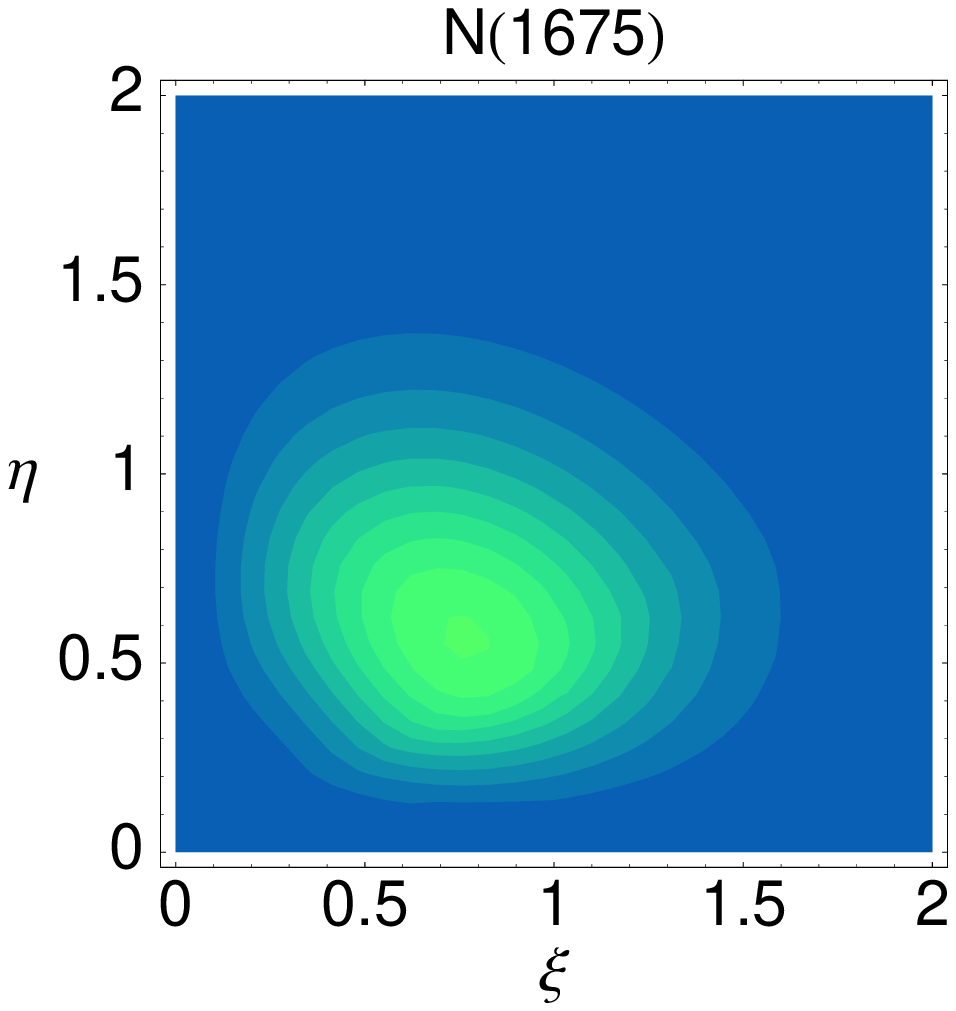}
\includegraphics[width=3.9cm]{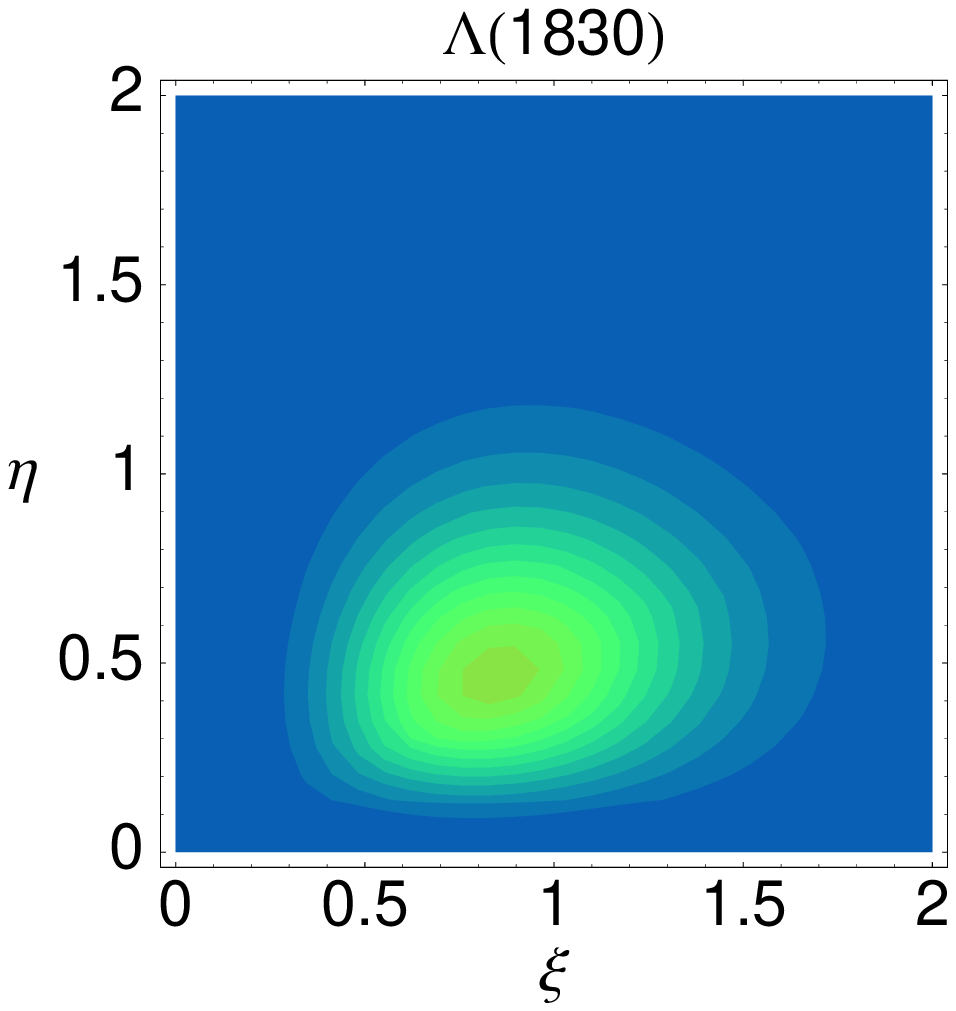}
\includegraphics[width=3.9cm]{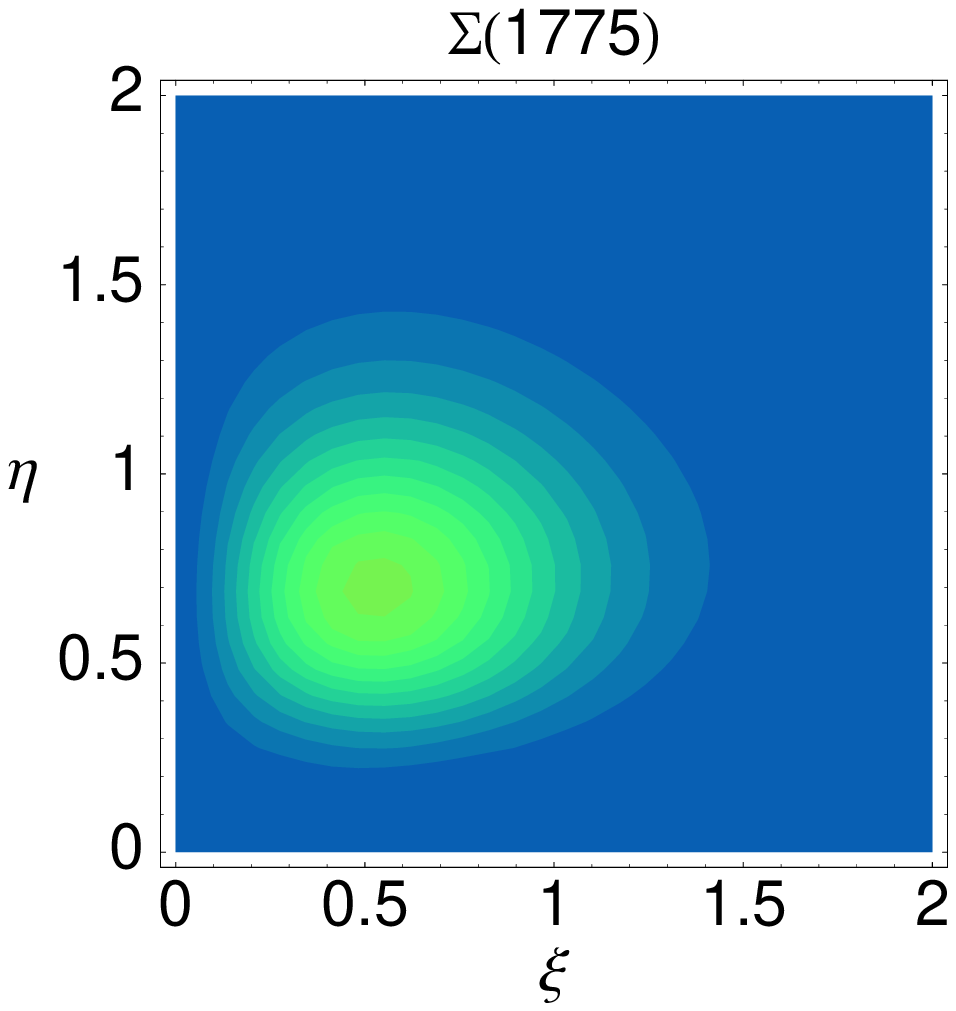}
\includegraphics[width=3.9cm]{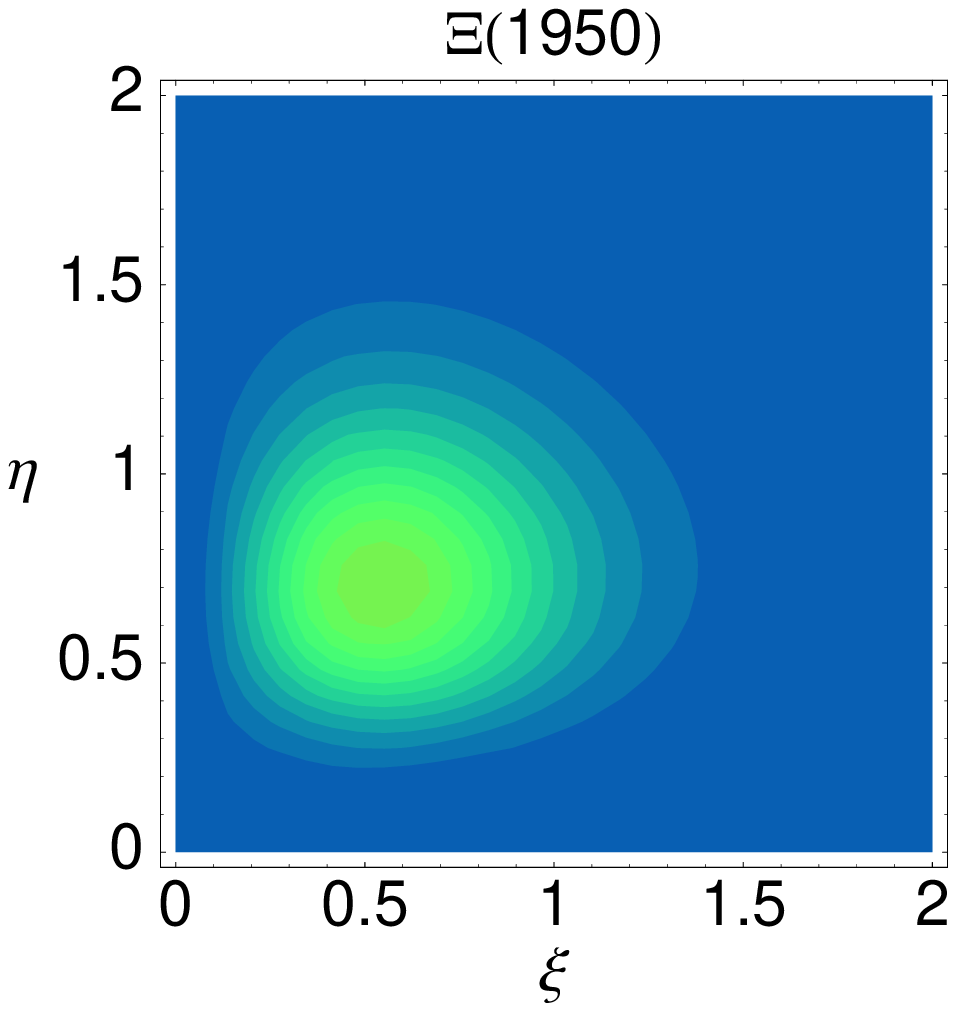}
\caption{Same as Fig.~\ref{fig:multi_1} for the $\frac{5}{2}^-$ octet baryon
states $N(1675)$, $\Lambda(1830)$, $\Sigma$(1775), $\Xi(1950)$.}
\label{fig:multi_12}
\end{figure*}

The $J^P=\frac{5}{2}^-$ octet with $L=1$, $S=\frac{3}{2}$ is filled with the
resonances $N(1675)$, $\Lambda(1830)$, $\Sigma(1775)$, and $\Xi(1950)$. All have
100 \% octet flavor content.
GP arrive at the same classification, while the PDG does not include the $\Xi(1950)$.
The latter is of unknown spin and parity, wherefore it could also be identified
with another flavor multiplet. Due to its decay properties (to be discussed below)
we suggest it to be a member of this octet. The spatial density distributions of
all of these octet states are shown in Fig.~\ref{fig:multi_12}.
\subsection*{Decuplet: $\bf \Delta(1600)$, $\bf \Sigma(1690)$}  
\begin{figure*}
\includegraphics[width=3.9cm]{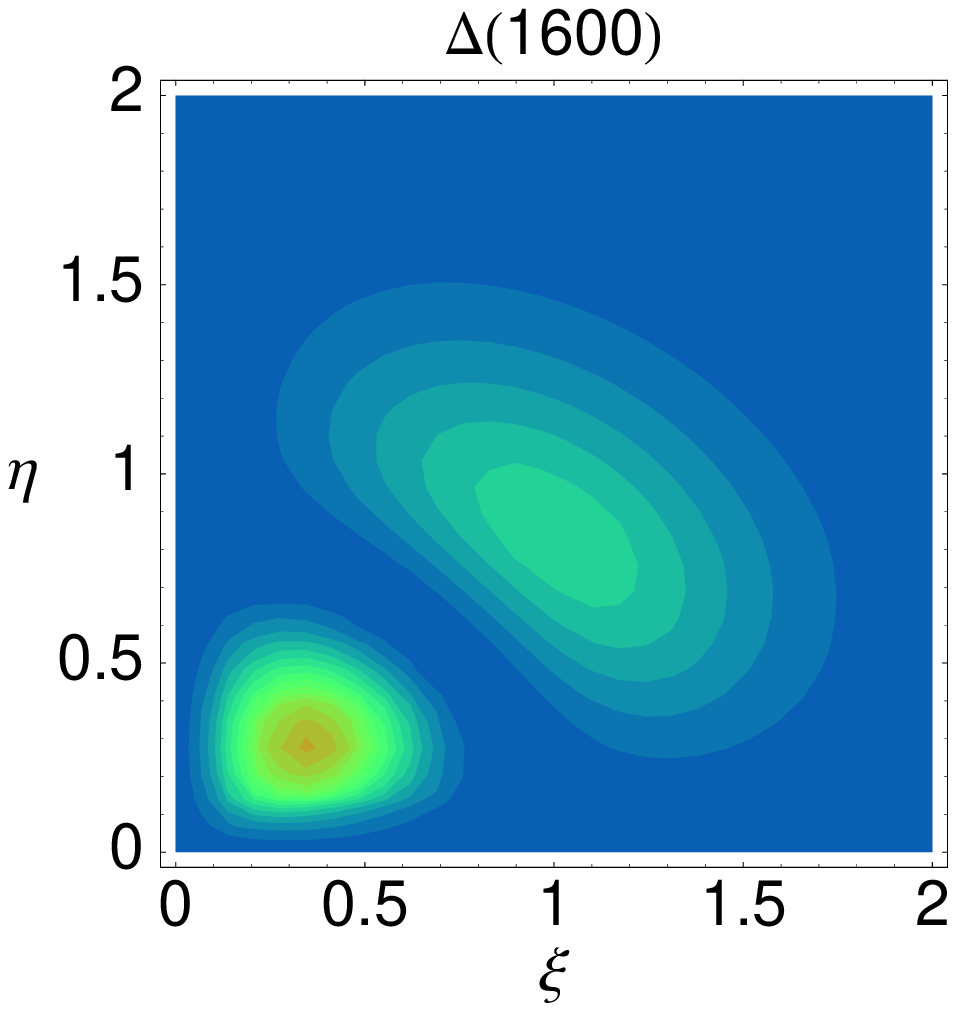}
\includegraphics[width=3.9cm]{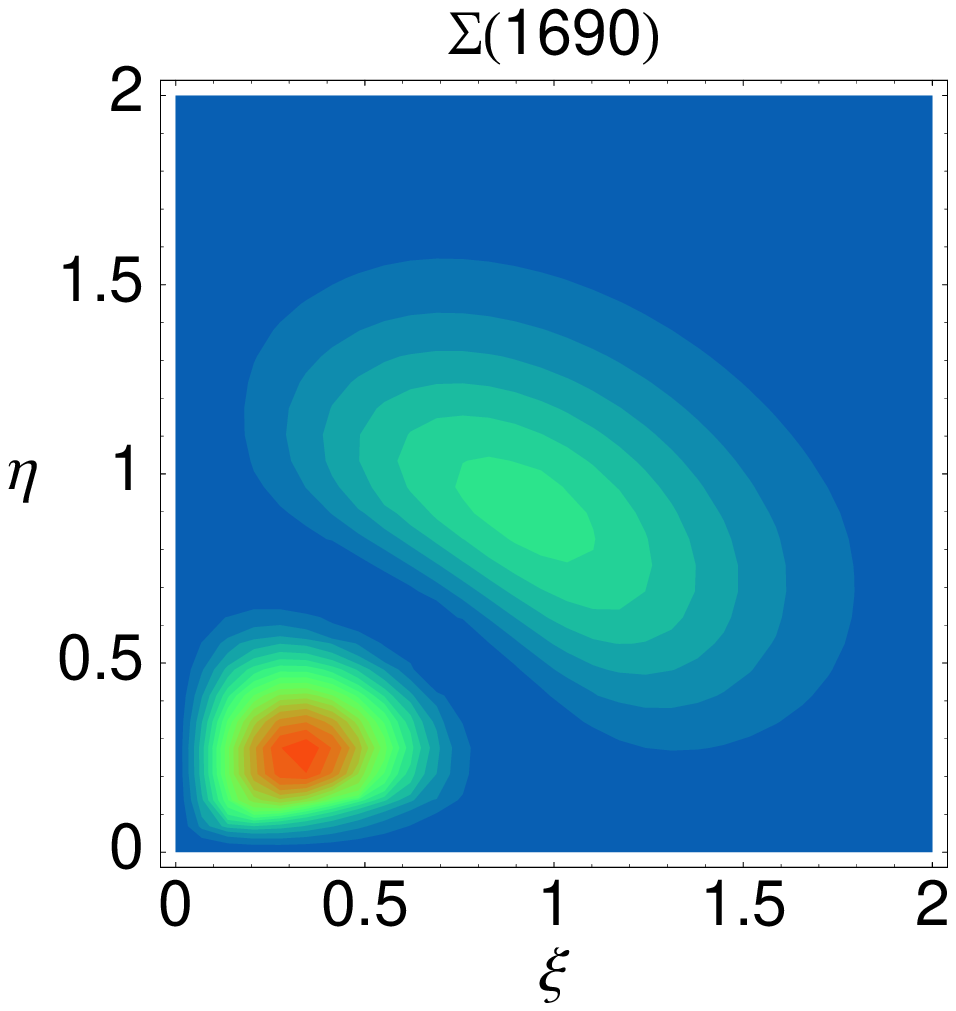}
\caption{Same as Fig.~\ref{fig:multi_1} for the $\frac{3}{2}^+$ decuplet baryon
states $\Delta(1600)$, $\Sigma(1690)$.}
\label{fig:multi_5}
\end{figure*}

The $\frac{3}{2}^+$ decuplet with the $\Delta(1600)$ contains the first
radial excitations above the decuplet ground states. We identify the  
$\Sigma(1690)$ to be a member of this decuplet. It has two-star status with unknown
$J^P$. While GP arrive at the same classification, the $\Sigma(1690)$ is not
considered by the PDG. Given the classification of $\Sigma$ resonances into octets
as resulting from Table~\ref{tab:multiplet_oct},
it appears most natural to identify the $\Sigma(1690)$
with the first radial excitation of decuplet states. Both the $\Delta(1600)$ and
$\Sigma(1690)$ have practically pure decuplet flavor content and their spatial
probability density distributions are shown
in Fig.~\ref{fig:multi_5}. They exhibit the typical nodal behaviour of first
radial excitations (cf. the analogous octet states in Fig.~\ref{fig:multi_3}). 
\subsection*{Decuplet: $\bf \Delta(1620)$, $\bf \Sigma(1750)$} 
\begin{figure*}
\includegraphics[width=3.9cm]{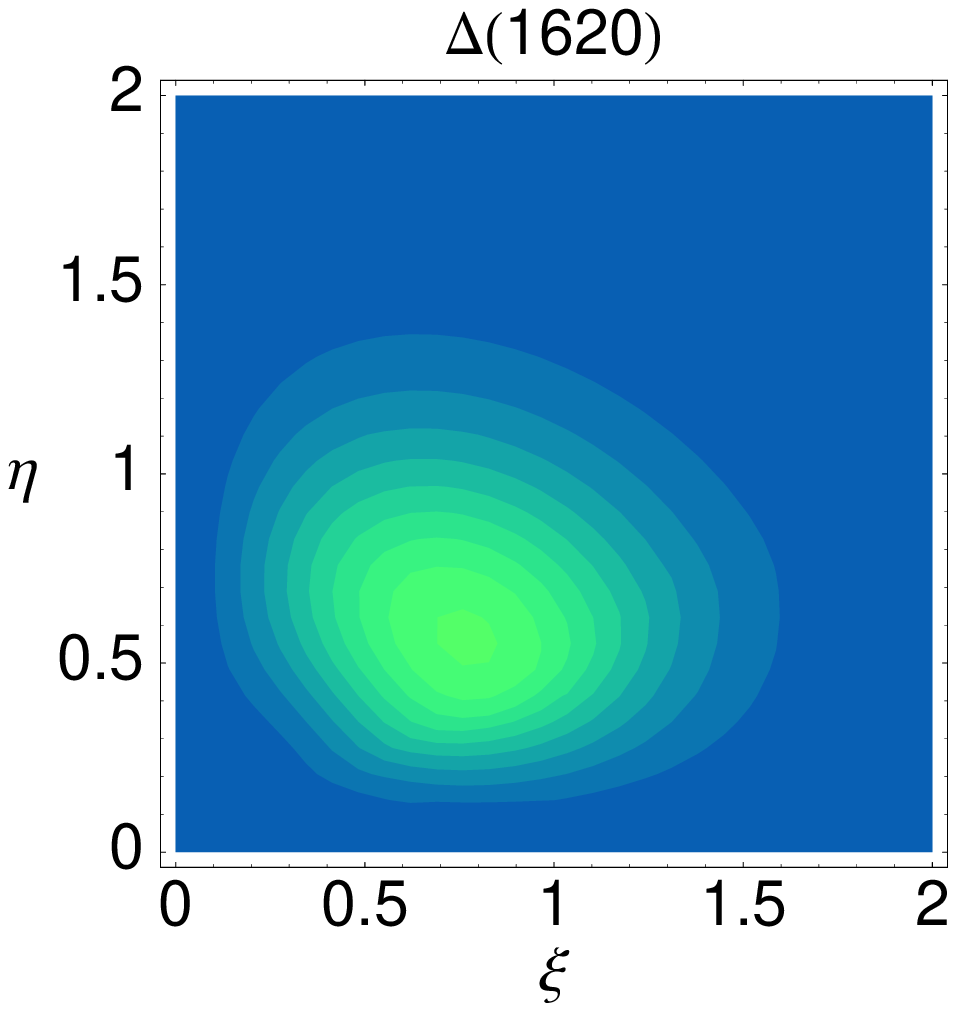}
\includegraphics[width=3.9cm]{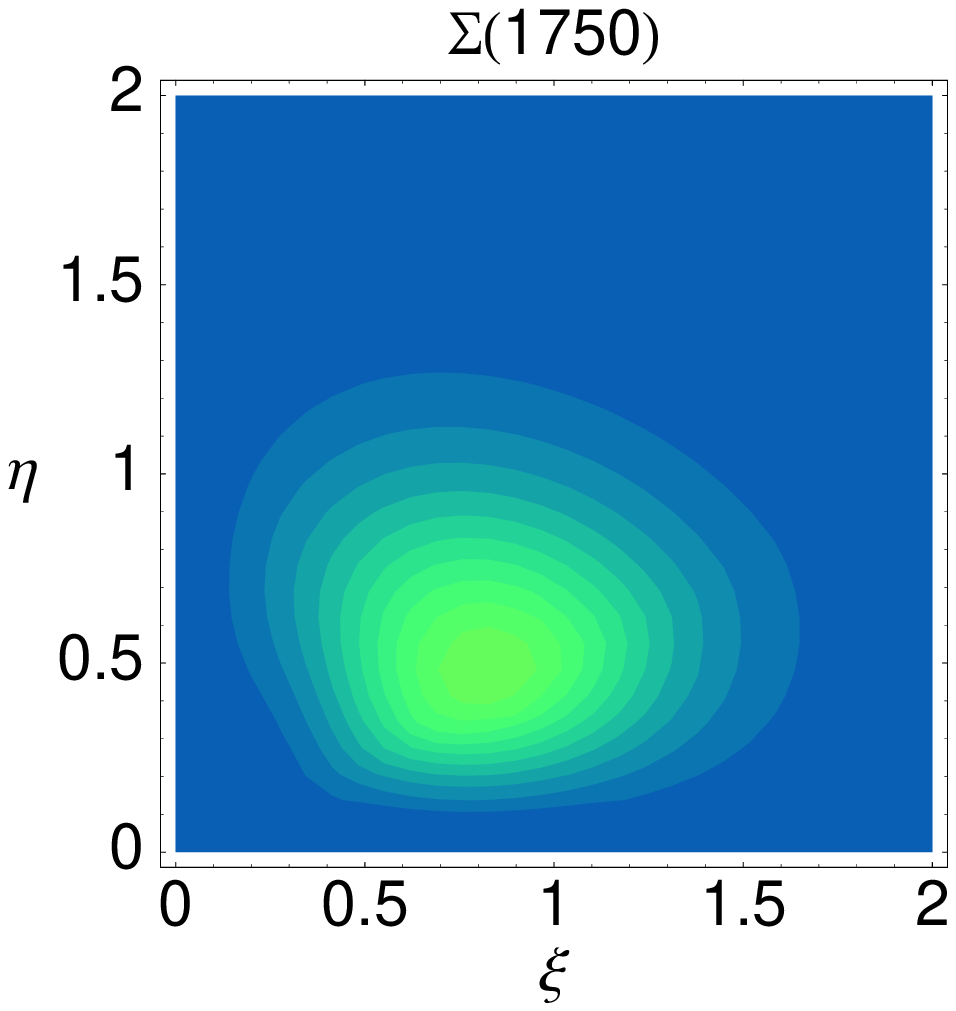}
\caption{Same as Fig.~\ref{fig:multi_1} for the $\frac{1}{2}^-$ decuplet baryon
states $\Delta(1620)$, $\Sigma(1750)$.}
\label{fig:multi_10}
\end{figure*}

In the $\frac{1}{2}^-$ decuplet we have the $\Delta(1620)$ and $\Sigma(1750)$
resonances. Whereas the classification of $\Delta(1620)$ is beyond doubt, as it
represents the lowest $\frac{1}{2}^-$ $\Delta$ excitation
coming with $L=1$ and $S=\frac{1}{2}$ (see Fig.~\ref{fig:NDel}), the identification
of $\Sigma(1750)$ as a decuplet member depends on the classification of $\Sigma(1560)$
and $\Sigma(1620)$ as octet states. In the GBE RCQM only the third state of the
$J^P=\frac{1}{2}^-$ excitations turns out to be a decuplet state. The mass eigenvalue
fits best with the $\Sigma(1750)$ and it produces a decay width that falls into
the general pattern established for relativistic results~\cite{Melde:2006yw}
(cf. also the discussion in the next section).
Furthermore the classification of $\Sigma(1750)$ into this decuplet agrees with
the one by GP. While the $\Delta(1620)$ is a pure flavor decuplet state, the $\Sigma(1750)$
bears a slight octet admixture. The spatial probability density distributions of
the two members of this decuplet are depicted in Fig.~\ref{fig:multi_10}.
\subsection*{Decuplet: $\Delta(1700)$}  
\begin{figure}
\includegraphics[width=3.9cm]{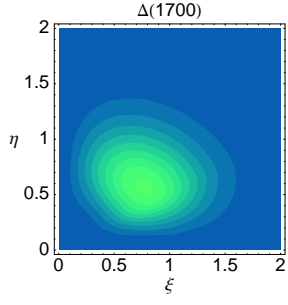}
\caption{Same as Fig.~\ref{fig:multi_1} for the $\frac{3}{2}^-$ decuplet baryon
state $\Delta(1700)$.}
\label{fig:multi_13}
\end{figure}

For the last decuplet one only has the $\Delta(1700)$ resonance with
$J^P=\frac{3}{2}^-$ and $L=1$, $S=\frac{1}{2}$. No other members of this decuplet are
experimentally established firmly enough. The $\Delta(1700)$ is a pure flavor
decuplet state and its spatial probability density distribution is plotted in
Fig.~\ref{fig:multi_13}.

\section{Mesonic Decays}
\label{sec:mesdecay}

\begin{figure*}
\includegraphics[angle=0,clip=,height=12cm]{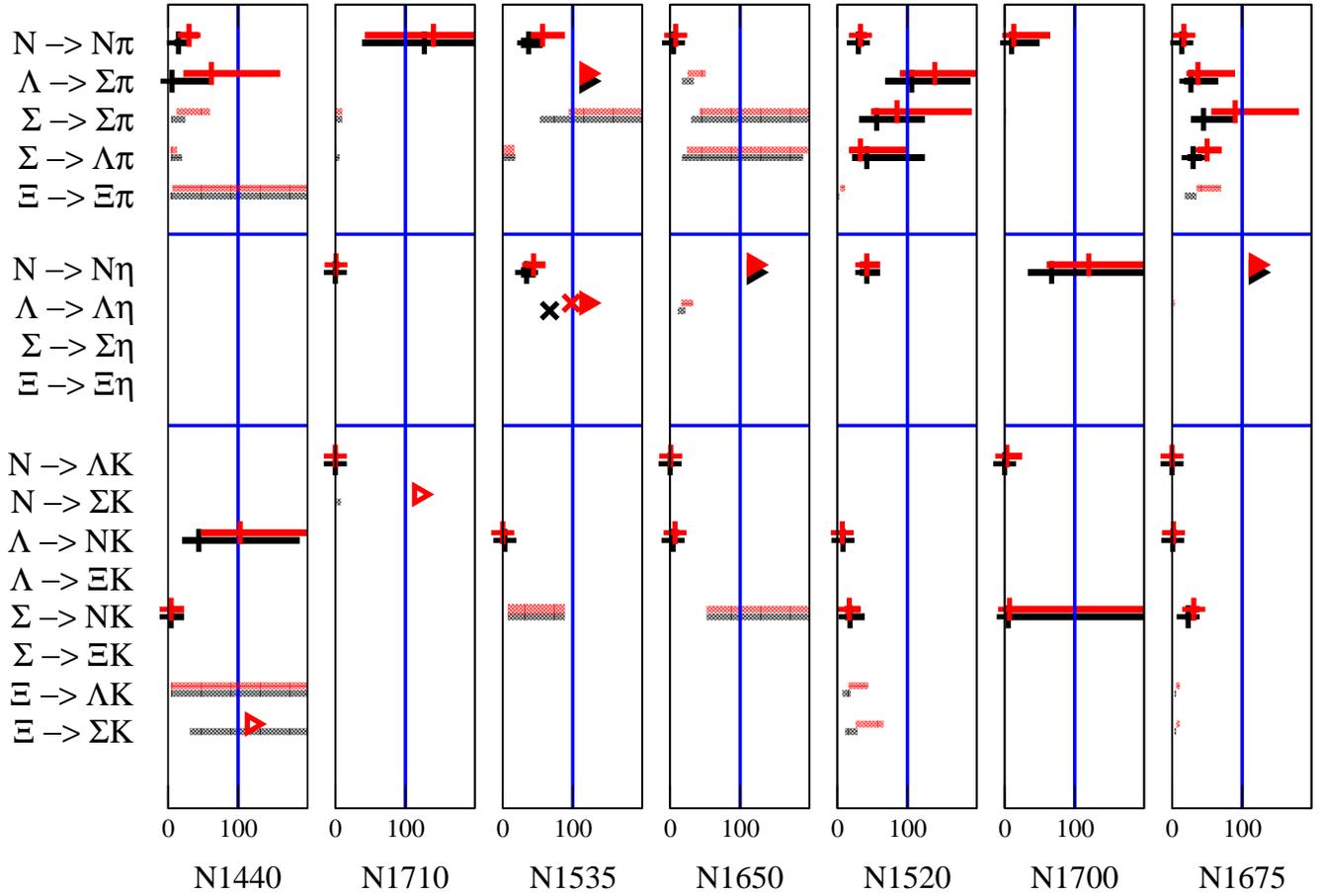}
\caption{Predictions for partial $\pi$, $\eta$, and $K$ decay widths of the GBE
(black/lower entries) and OGE (red/upper entries) RCQMs for the octets in
Table~\ref{tab:multiplet_oct} from the PFSM calculation.
The results shown by + crosses are presented as percentages of the best estimates
for experimental
data reported by the PDG~\cite{PDBook}, with the horizontal lines showing the
experimental uncertainties. In case of shaded lines without crosses the PDG gives only total
decay widths, and the theoretical results are represented relative to them. The
triangles point to results outside the plotted range. For the particular decay
$\Lambda(1670) \rightarrow \Lambda \eta$ in addition to the theoretical masses also
experimental ones were used, and the corresponding results are marked by $\times$ crosses.
For further explanations see the text.}
\label{fig:decay_graph_OGE_GBE_oct}
\end{figure*}

\begin{figure*}
\includegraphics[angle=0,clip=,height=12cm]{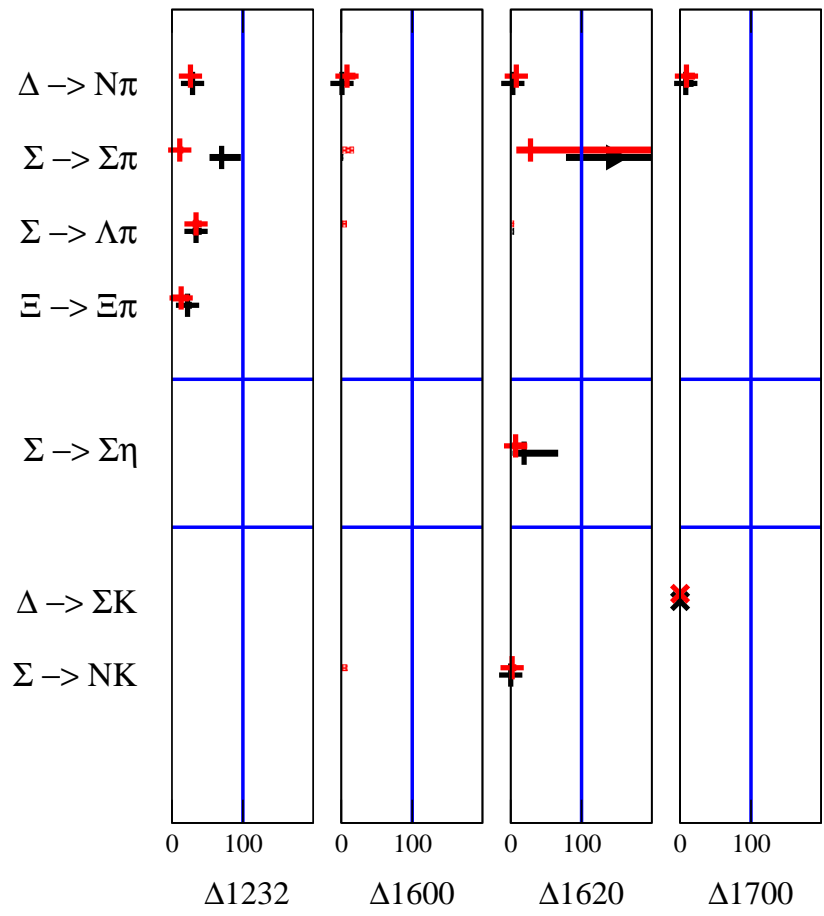}
\caption{Same as Fig.~\ref{fig:decay_graph_OGE_GBE_oct} but for the decuplets in
Table~\ref{tab:multiplet_decu}.}
\label{fig:decay_graph_OGE_GBE_decu}
\end{figure*}

\begin{figure*}
\includegraphics[angle=0,clip=,height=12cm]{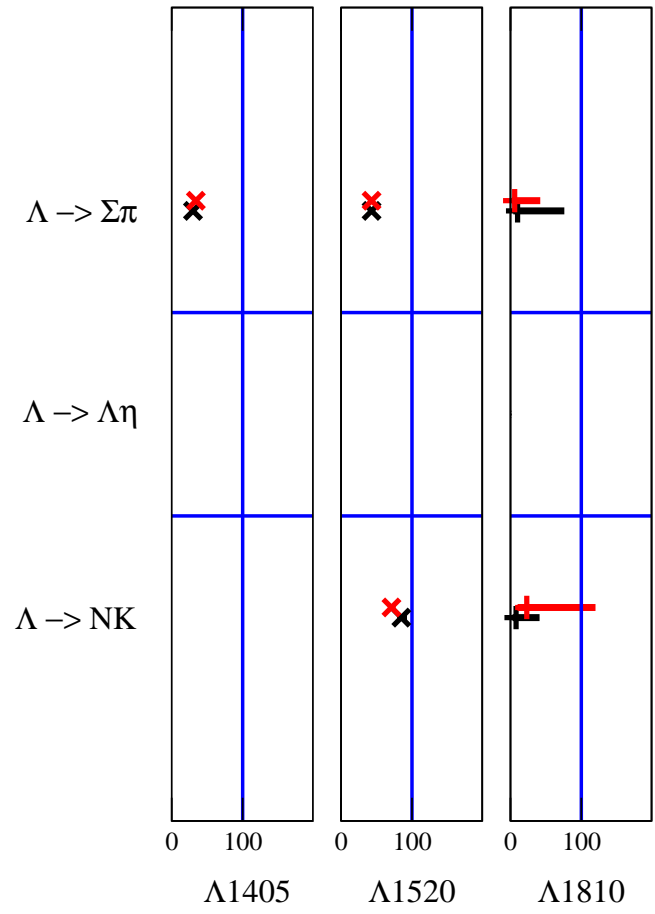}
\caption{Same as Fig.~\ref{fig:decay_graph_OGE_GBE_oct} but for the singlets in
Table~\ref{tab:multiplet_singl}.}
\label{fig:decay_graph_OGE_GBE_singl}
\end{figure*}

The decay properties of the baryon resonances occurring in
Tables~\ref{tab:multiplet_oct},~\ref{tab:multiplet_decu}, and~\ref{tab:multiplet_singl},
are presented in
Figs.~\ref{fig:decay_graph_OGE_GBE_oct},~\ref{fig:decay_graph_OGE_GBE_decu},
and~\ref{fig:decay_graph_OGE_GBE_singl}. The covariant predictions for partial
decay widths of all kind of decay modes are shown for both the GBE and OGE RCQMs.

We have chosen the same representation of the results as in
Fig.~\ref{fig:decay_graph_est} but here the experimental uncertainties of the
decay widths as reported by the PDG are included too. Partial decay widths are not
always available from experiment. In such cases we advocate the total decay widths
and present the theoretical results of partial decay widths relative to them
(shown by shaded lines without central values in the figures).

Beyond the relativistic results already published in
refs.~\cite{Melde:2005hy,Melde:2006yw,Sengl:2007yq} also the additional RCQM
predictions for the two-star resonances of Table~\ref{tab:multi3} are included
into Figs.~\ref{fig:decay_graph_OGE_GBE_oct} to~\ref{fig:decay_graph_OGE_GBE_singl}.
In most cases the predictions of the GBE and OGE RCQMs are quite similar. If differences
occur, they are in the first instance caused by resonance mass effects. The congruence
of the results from the GBE and OGE RCQMs becomes even more pronounced, if experimental
masses are employed instead of the theoretical ones. Only for certain
decays wave-function effects are responsible too. The typical pattern that emerges
is a general underestimation of the experimental data, with only a few exceptions.

The partial decay widths of the octet involving the $N(1440)$ resonance are
given in the first column of Fig.~\ref{fig:decay_graph_OGE_GBE_oct}. In the nonstrange
sector there are only the $\pi$ decays. Among them the most prominent decay
$N(1440) \rightarrow N\pi$ is grossly underestimated by both \mbox{RCQMs}. In this case also the difference between the GBE and OGE RCQMs is sizable, even though it gets reduced,
when mass effects are wiped out, i.e. when experimental masses are used instead
of the theoretical ones~\cite{Melde:2005hy}. The situation is quite similar for
the $\Lambda(1600) \rightarrow \Sigma\pi$ decay and to some extent also for the
$\Sigma(1660) \rightarrow \Sigma\pi$ and $\Sigma(1660) \rightarrow \Lambda\pi$
decays~\cite{Melde:2006yw}. The decay width for the $\Xi(1690) \rightarrow \Xi\pi$
turns out to be very small, and we can only relate it to an experimental total width. 
Also the $K$ decay widths follow the characteristic of remaining too small. The OGE RCQM
results for $\Lambda(1600) \rightarrow NK$ and $\Xi(1690) \rightarrow \Sigma K$,
which are relatively bigger, are dominated by mass effects. The corresponding decay
widths get much reduced if the experimental masses are employed (see
ref.~\cite{Sengl:2007yq} and cf. Table~\ref{tab:multi3}).

In the next octet we have only the decays of $N(1710)$ and $\Sigma(1880)$. It appears
that the $N(1710) \rightarrow N\pi$ result contradicts the pattern usually
found for the relativistic decay widths, as the theoretical prediction comes out
large and relatively close to the (central value) of the experimental datum. However,
it should be considered that the $N(1710)$, though being of three-star status,
is not so safely established experimentally, as the various partial wave analyses
do not agree
very well~\cite{PDBook}. In case of the strange decay the prediction of the OGE
RCQM for the $N(1710) \rightarrow \Sigma K$ width is mainly caused by a
threshold/mass effect. The large overshooting is essentially removed if the
experimental mass is used~\cite{Sengl:2007yq}. For the $\Sigma(1880)$ all the
$\pi$, $\eta$, and $K$ decay widths are extremely small often compatible with
zero (see Table~\ref{tab:multi3}) and thus not even visible in
Fig.~\ref{fig:decay_graph_OGE_GBE_oct}.

In the octet involving the $N(1535)$ the $N \rightarrow N\pi$ and $N \rightarrow N\eta$
decays follow the usual pattern. For the $\Lambda(1670)\rightarrow\Sigma\pi$ decay
both RCQMs predict a partial decay width much too large. The reason might be that
the $\Lambda(1670)$ has a relatively big flavor singlet admixture of 28 \%. We note
that a similar result is found for the $\Lambda(1690)$ resonance in the octet
involving the $N(1520)$. The $\Lambda(1670)\rightarrow \Lambda\eta$ decay is
rather sensitive to mass effects. In case of the GBE RCQM the channel is closed, whereas
for the OGE RCQM the prediction is too high. If the experimental masses are employed
instead of the theoretical ones, both RCQMs produce a result close or slightly
below the experimental width (cf. the $\times$ crosses in
Fig.~\ref{fig:decay_graph_OGE_GBE_oct}). The $\Lambda \rightarrow NK$ decay width
turns out to be extremely small. In this octet we have in addition the $\Sigma(1560)$
decays, for which no estimates of experimental widths are given by the PDG. While
the $\Sigma \rightarrow \Sigma\pi$ decay width might appear rather larger, the
$\Sigma \rightarrow \Lambda\pi$ and $\Sigma \rightarrow NK$ widths come out quite
small. This behaviour suggests that the identification of the $J^P=\frac{1}{2}^-$
$\Sigma$ eigenstate of this octet with the $\Sigma(1560)$ is the most reasonable
choice. Further experimental information on the decay properties of the $\Sigma(1560)$
would be highly welcome.

In the next octet the relativistic prediction for the $N(1650) \rightarrow N\pi$ decay
width is again too small, while the one of $N(1650) \rightarrow N\eta$ comes out
unusually large, as a notable exception~\cite{Melde:2005hy}.
The latter is true for both the GBE and OGE hyperfine interactions, and
even in the case when experimental masses are used the situation is not changed.
The $N(1650) \rightarrow \Lambda K$ decay width is again practically zero.
For the $\Lambda(1800)$ an experimental partial width is only available for the
strange $NK$ decay channel. The theoretical prediction grossly underestimates the
rather large $\Lambda \rightarrow NK$ width. The $\Lambda \rightarrow \Sigma\pi$
and $\Lambda \rightarrow \Lambda\eta$ widths can only be compared to the total
$\Lambda$ decay width and they result relatively small. Similarly, for the
$\Sigma(1620)$ the decay widths can only be related to the total width. In this
light the partial $\Sigma\pi$, $\Lambda\pi$, and $NK$ decay widths appear to
be relatively large, but we note that their sum still lies within the range of
the total width. It is remarkable that among the above three channels the
$\Sigma(1620) \rightarrow NK$ decay mode is the strongest one. In view of the
reported data being rather old it would be desirable to have new measurements
on the $\Sigma(1620)$.

For the octet with $N(1520)$ we have decays of all members including the $\Xi$.
The partial widths of the $\pi$ and $\eta$ decays of $N(1520)$ are both predicted
too small thus conforming to the typical pattern. The same appears to be true
for the $\pi$ and $\eta$ decays of the strange resonances. The only exception
is the $\Lambda(1690) \rightarrow \Sigma\pi$ decay width coming out relatively
large (and being in agreement with experiment). As in the case of
$\Lambda(1670)$ the possible reason is again the considerable singlet admixture of
28 \%. On the other hand, the $\Lambda(1690) \rightarrow \Lambda\eta$ decay
width is practically zero (and thus not visible in
Fig.~\ref{fig:decay_graph_OGE_GBE_oct}); there is no datum on that decay by the PDG.
All the $K$ decay widths are predicted too small by both RCQMs, where for the
$\Xi(1820)$ no partial widths are reported by the PDG.

In the $J^P=\frac{3}{2}^-$ octet with $N(1700)$ the $\pi$ decay widths are
rather small, where in case of the $\Sigma(1940)$ comparison is possible
only to the total experimental width; because of its magnitude the small
theoretical result is not visible in
Fig.~\ref{fig:decay_graph_OGE_GBE_oct}. Likewise the $K$ decay widths are
very small. For the $N(1700)$ only the $\Lambda K$ width is available from
phenomenology, while the $\Sigma K$ width is not given. In this octet, only the
$N(1700) \rightarrow N\eta$ width appears to be of
considerable size but this is again an artefact of the representation in
Fig.~\ref{fig:decay_graph_OGE_GBE_oct}, as the experimental width is reported
to be extremely small, practically consistent with zero.

In the last octet with $J^P=\frac{5}{2}^-$ we have all kind of decay modes. The
$\pi$ decay widths are all too small. In the $\eta$ channel only the decay width
of $N(1675)$ is known from phenomenology. However, it is reported to be consistent
with zero and thus the percentage of theoretical prediction falls outside the
range plotted in Fig.~\ref{fig:decay_graph_OGE_GBE_oct}. The $\eta$ decay widths of
$\Lambda(1830)$, $\Sigma(1775)$, and $\Xi(1950)$ are rather small, in some cases
practically zero (and thus not visible in Fig.~\ref{fig:decay_graph_OGE_GBE_oct});
the corresponding partial widths are not known from experiment.
Similarly the $K$ decay widths are all quite small, except for
$\Sigma(1775) \rightarrow NK$. For the latter the experimental partial decay
width is underestimated, like for the two other decays 
$N(1675) \rightarrow\Lambda K$ and $\Lambda(1830) \rightarrow NK$, for which
partial widths are reported from experiment.

For the lowest decuplet involving the $\Delta(1232)$ only $\pi$ decays are
possible. In all cases the theoretical results underestimate the experimental data.
Only in case of the GBE RCQM the prediction of the $\Sigma(1385) \rightarrow \Sigma\pi$
comes close to the experimental value. However, if experimental masses are used,
the theoretical width is again reduced~\cite{Melde:2006yw}. As a result the decay
widths of this decuplet are quite consistent with the general pattern found for the
relativistic quark model predictions.

In the next decuplet with $\Delta(1600)$ we also have the $\Sigma(1690)$. The partial
widths of all the possible decay modes of these radial-excitation states are found to
be extremely small. Only for the $\Delta(1600) \rightarrow N \pi$ decay a partial
width is reported from experiment. For the $\Sigma(1690) \rightarrow \Sigma \pi$,
 $\Sigma(1690) \rightarrow \Lambda \pi$,  and $\Sigma(1690) \rightarrow NK$ we
 can only compare to the total width (see Table~\ref{tab:multi3}).

The situation is quite similar for the decays in the next decuplet involving the
$\Delta(1620)$. Here, we have the $\Sigma(1750)$, which bears a three-star status,
and partial widths are available not only for the
$\Sigma(1750) \rightarrow \Sigma \pi$ but also for the 
$\Sigma(1750) \rightarrow \Sigma \eta$ as well as $\Sigma(1750) \rightarrow NK$
decays. An exception to the typical pattern appears to be only the prediction for
the partial width of the $\Sigma(1750) \rightarrow \Sigma \pi$ decay in case of the
GBE RCQM. The pertinent theoretical value falls within the error bars of the
phenomenological width, which is reported to be rather small and could
even be compatible with zero~\cite{PDBook}.

As a representative of the next decuplet with $J^P=\frac{3}{2}^-$ we only have the
$\Delta(1700)$. Its $\pi$ and $K$ decay widths are again extremely small. For the
latter we remark that the decay is only possible, if experimental masses are used.
Both RCQMs predict the $\Delta(1700)$ mass too low (see
Table~\ref{tab:masses} and/or Fig.~\ref{fig:NDel}).

Finally we consider the singlet decays in Fig.~\ref{fig:decay_graph_OGE_GBE_singl}.
As has already been  stated in the previous section,
the description of the $\Lambda(1405)$ poses a serious problem for all kind of
constituent quark models (relying on \{QQQ\} configurations only), as its mass
cannot be reproduced in accordance with the experimental observation. It comes
out at least 150-200 MeV too high for the RCQMs we consider here (see
Fig.~\ref{fig:LaSig}). These
shortcomings have a big influence on the predictions for the $\pi$ decay
width; they come out way too big~\cite{Melde:2006yw}. Therefore, we give in
Fig.~\ref{fig:decay_graph_OGE_GBE_singl} the predictions obtained with the
experimental masses. They fall below the experimental values and thus fit into
the typical pattern.

Essentially, the same observations hold true for the next singlet, the $J^P=\frac{3}{2}^-$
$\Lambda(1520)$. Only, we have here in addition the $\Lambda \rightarrow NK$
channel open with the corresponding partial decay width reported from
phenomenology. Both the $\pi$ and $K$ decay widths of this resonance are again
predicted too small.

Finally, we have the $J^P=\frac{1}{2}^+$ singlet $\Lambda(1810)$. The dominant
decays are to the $\pi$ and $K$ channels, which are of comparable strengths. The
theoretical predictions underestimate both of them. The
$\Lambda(1810) \rightarrow \Lambda \eta$ decay width comes out extremely small but
different from zero; due to its smallness the corresponding entry is not visible
in Fig.~\ref{fig:decay_graph_OGE_GBE_singl}. 

\section{Conclusions
\label{sec:summary}}

We have presented a comprehensive study of the light and strange baryon resonances
below $\approx 2$ GeV in the framework of relativistic constituent quark models.
In particular, we have considered their eigenvalue spectra and the properties
of the eigenstates with regard to their spin-flavor contents and spatial structures.
In addition we have explored the pattern of the predictions for the partial widths
of the various mesonic decay modes. We have combined these findings to identify the
theoretical eigenstates with experimental resonances and thus arrived
at a new classifications scheme for the known resonances into flavor multiplets.
It has turned out that consideration of all of these evidences is required in
order to produce a maximally reliable classification of the low-lying baryon
resonances.

\begin{figure*}
\includegraphics[width=7.6cm]{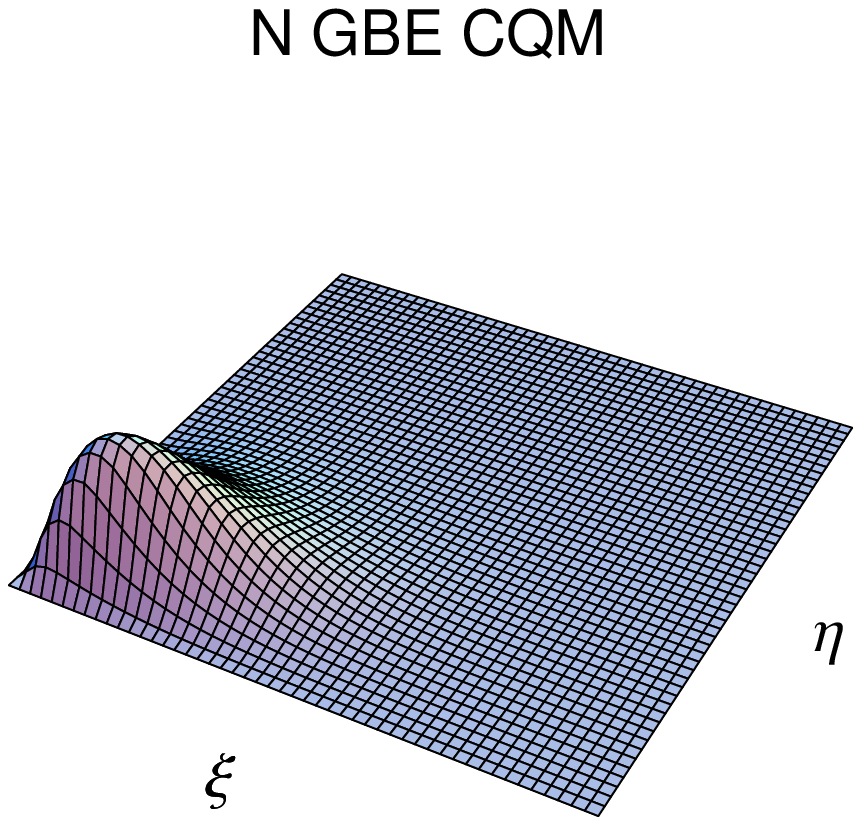}
\includegraphics[width=7.6cm]{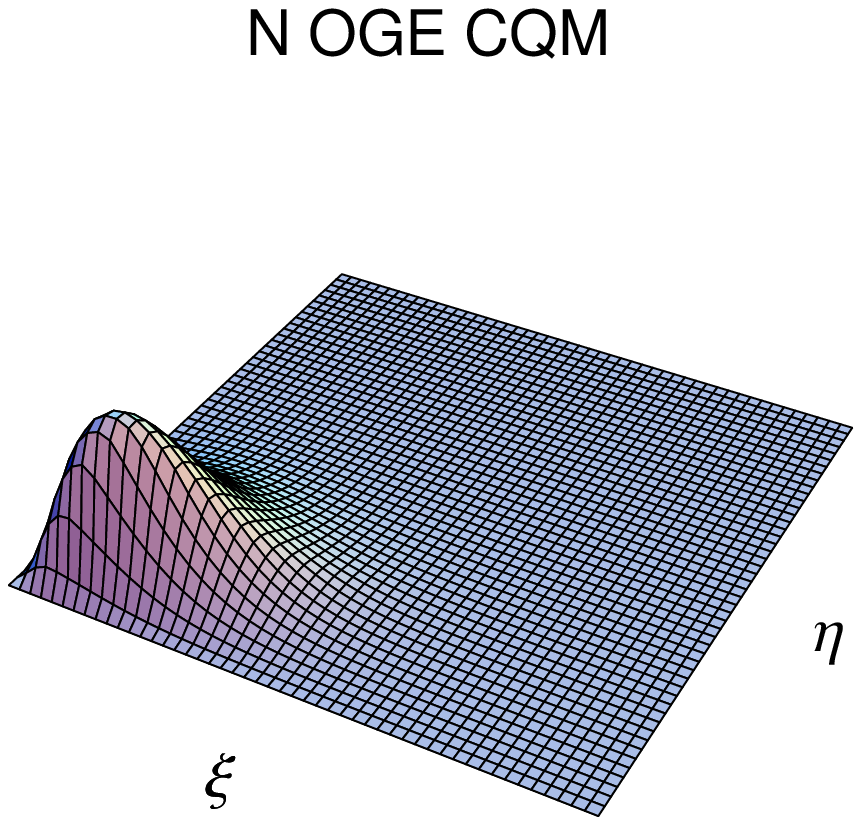} \\[1.6cm]
\includegraphics[width=7.6cm]{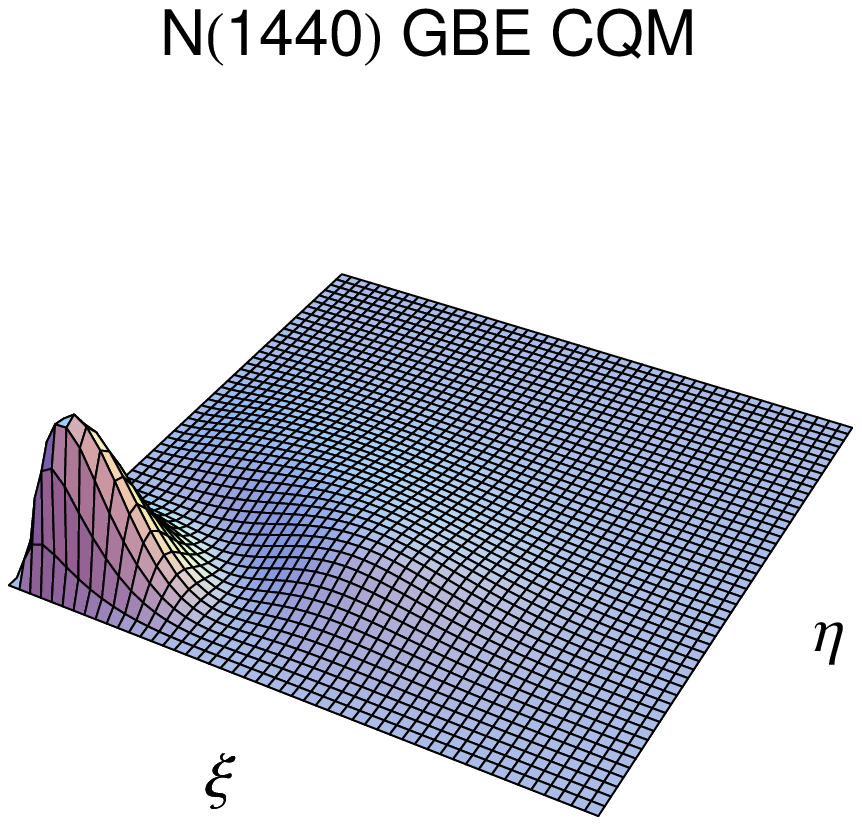}
\includegraphics[width=7.6cm]{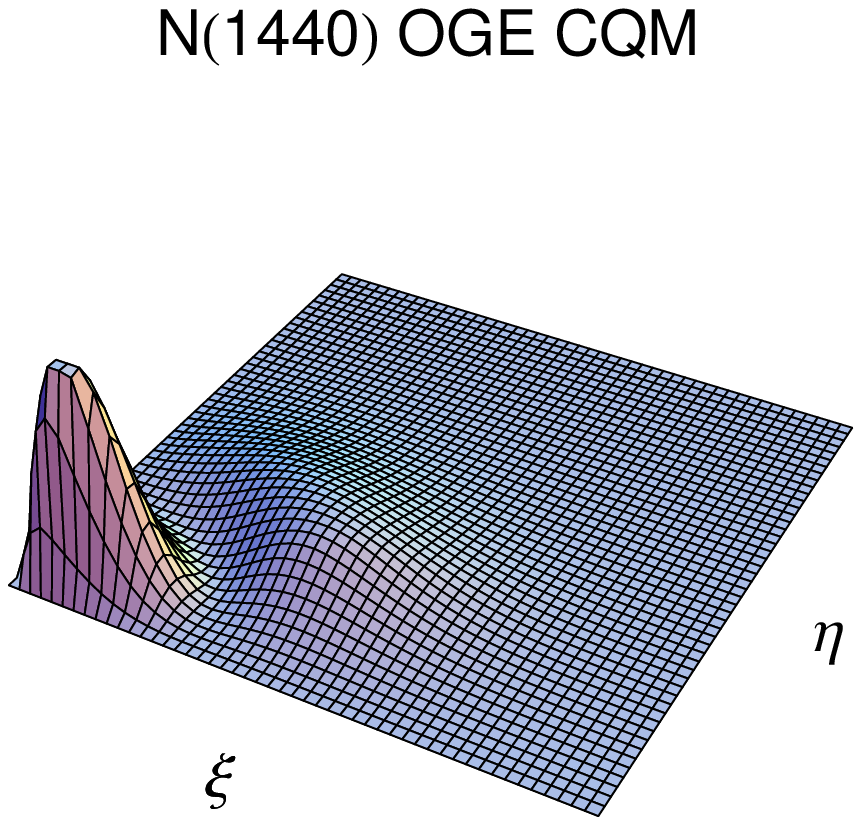}
\caption{Spatial probability density distributions $\rho(\xi,\eta)$ of the nucleon
ground state $N(939)$ and the Roper resonance $N$(1440) as a function of the radial
parts of the Jacobi coordinates $\xi$ and $\eta$ for the GBE RCQM (left plots) and
the OGE RCQM (right plots).
\label{fig:GBE-OGE-wf}
}
\end{figure*}

The relativistically invariant eigenvalue spectra have been obtained by the
solution of the eigenvalue problem of the baryon mass operator using a stochastic
variational method. Two different
types of hyperfine interactions for a \{QQQ\} system of confined constituent quarks
have been considered: the ones resulting from Goldstone-boson-exchange and from
one-gluon-exchange dynamics, respectively. Their characteristic excitation spectra
have been identified. By the solution of the eigenvalue problem,
at the same time, the baryon eigenstates have been obtained. Their configuration-space
representations in the baryon rest frame (i.e. the baryon wave functions) have been
analyzed considering their spin-flavor and spatial structures. Detailed evidences
have thus been gained on the properties of the resonance eigenstates within each
flavor multiplet with definite $(LS)J^P$ and certain radial excitation. The mixtures
between flavor octet and singlet as well as octet and decuplet have been determined,
and, for the first time, the characteristic spatial probability distributions for
the ground and excited states have been shown. For the RCQMs considered here the
spin-flavor and spatial structures of the baryon wave functions are qualitatively
rather similar (cf. the examples shown in Fig.~\ref{fig:GBE-OGE-wf}). In the
discussion of the baryon properties in section~\ref{sec:class} we have therefore contented
ourselves with presenting only the detailed results of the GBE RCQM. While the
main conclusions do not depend on the type of RCQM considered, the distinct
interactions present in either the GBE or OGE RCQMs do lead to notable differences
in the theoretical predictions.

Recently, first covariant predictions by RCQMs for mesonic decay widths of baryon
resonances have become
available~\cite{Melde:2005hy,Melde:2006yw,Sengl:2007yq,Metsch:2004qk,Migura:2007}.
The corresponding results provide additional
insight for the classification of baryon resonances into flavor multiplets. Here,
we have completed the theoretical results of decay widths for the GBE and OGE RCQMs
regarding all baryon resonances below $\approx 2$ GeV, with total orbital angular
momentum $L<2$, and at least two-star status. The evolving pattern shows that
the experimental data for partial decay widths are in general underestimated.
There are only a few notable exceptions, the $N(1710) \rightarrow N\pi$ decay, where
the experimental situation can be considered as unsettled, the
$\Lambda(1670) \rightarrow \Sigma \pi$ and the $\Lambda(1690) \rightarrow \Sigma \pi$
decays, where we have found considerable singlet admixtures, and the
$N(1650) \rightarrow N\eta$ decay, where no convincing explanation is readily at hand.
      
Using all of these evidences a new classification of baryon ground states and resonances
into flavor singlet, octet, and decuplet has been proposed. In most instances it is
in accordance with the classification by the PDG~\cite{PDBook} and also with the
one by Guzey and Polyakov~\cite{Guzey:2005vz}. Only for the $\Lambda(1810)$ and
for the $J^P=\frac{1}{2}^-$ octets and decuplets we find differences.
Regarding the $\Lambda(1810)$ our classification differs from both of these schemes,
as we identify this resonance as a flavor singlet. However, this is in agreement with
the assignment suggested by Matagne and Stancu~\cite{Matagne:2006zf}.
Contrary to the PDG but in accordance with Guzey and Polyakov, the $\Sigma(1560)$ and
$\Sigma(1620)$ are placed into the $J^P=\frac{1}{2}^-$ octets involving the
$N(1535)$ and the $N(1650)$, respectively. As a consequence,
the $\Sigma(1750)$ is assigned to the decuplet involving the $\Delta(1620)$.
Furthermore, we identify the $\Sigma(1940)$ as a member of the octet involving
$N(1700)$ and the $\Sigma(1690)$ as a member of the decuplet involving the
$\Delta(1600)$. In addition, we classify the $\Xi(1690)$ and the $\Xi(1950)$ into the
octets involving the $N(1440)$ and $N(1675)$, respectively. 

We remark that the suggested classification must also be considered with some caution.
What regards theory, it is based on the description of the baryons
as eigenstates of an invariant mass operator relying on \{QQQ\} degrees of freedom only.
As this might not be adequate for resonances, it nevertheless constitutes a limitation
for present methods of solving a relativistic few-quark problem. For the decay widths
a restricted decay operator has been employed. Future works towards improving the
relativistic description of baryons should primarily aim at extending the RCQMs to
include additional degrees of freedom, like explicit couplings to decay channels.
In such a framework, in particular the resonant states will be generated more
realistically.

Regarding phenomenology, sufficient experimental evidence is lacking for some
$\Sigma$ resonances and above all in the $\Xi$ sector. For a complete assignment
of states in the flavor octets additional information also on $\Lambda$ resonances
is urgently needed.
In particular, determinations of $J^P$ for the lesser known resonances and further
measurements of the various partial decay widths would be highly welcome.

\begin{acknowledgments}
This work was supported by the Austrian Science Fund (FWF-Project P19035).
B. Sengl acknowledges support through the Doktoratskolleg
'Hadrons in Vacuum, Nuclei, and Stars'
(FWF-Project W1203). The authors profited from valuable 
discussions with L.~Canton, A.~Krassnigg and R.~F. Wagenbrunn. 
They are also grateful to F.~Stancu for pointing out the classification
of $\Lambda$(1810) as flavor singlet as found in ref.~\cite{Matagne:2006zf}.
\end{acknowledgments}

\addcontentsline{toc}{chapter}{Bibliography}

\begin{thebibliography}{10}

\bibitem{Melde:2005hy}
T. Melde, W. Plessas, and R.~F. Wagenbrunn, Phys. Rev. C {\bf 72},  015207
  (2005); Erratum, Phys. Rev. C {\bf 74}, 069901 (2006).

\bibitem{Melde:2006yw}
T. Melde, W. Plessas, and B. Sengl, Phys. Rev. C {\bf 76},  025204  (2007).

\bibitem{Sengl:2007yq}
B. Sengl, T. Melde, and W. Plessas, Phys. Rev. D {\bf 76},  054008  (2007).

\bibitem{Stancu:1989iu}
F. Stancu and P. Stassart, Phys. Rev. D {\bf 39},  343  (1989).

\bibitem{Capstick:1993th}
S. Capstick and W. Roberts, Phys. Rev. D {\bf 47},  1994  (1993).

\bibitem{Geiger:1994kr}
P. Geiger and E.~S. Swanson, Phys. Rev. D {\bf 50},  6855  (1994).

\bibitem{Krassnigg:1999ky}
A. Krassnigg {\it et~al.}, Few Body Syst. Suppl. {\bf 10},  391  (1999).

\bibitem{Plessas:1999nb}
W. Plessas {\it et~al.}, Few Body Syst. Suppl. {\bf 11},  29  (1999).

\bibitem{Theussl:2000sj}
L. Theussl, R.~F. Wagenbrunn, B. Desplanques, and W. Plessas, Eur. Phys. J. A
  {\bf 12},  91  (2001).

\bibitem{Metsch:2004qk}
B. Metsch, AIP Conf. Proc. {\bf 717},  646  (2004).

\bibitem{Migura:2007}
S. Migura, Ph.D. Thesis, University of Bonn, Bonn, Germany, 2007.

\bibitem{Merten:2002nz}
D. Merten {\it et~al.}, Eur. Phys. J. {\bf A14},  477  (2002).

\bibitem{Metsch:2003ix}
B. Metsch, U. Loering, D. Merten, and H. Petry, Eur. Phys. J. {\bf A18},  189
  (2003).

\bibitem{Melde:2007iu}
T. Melde and W. Plessas, arXiv:0711.0881 [nucl-th]  (2007).

\bibitem{PDBook}
W.-M. {Yao} {\it et~al.}, {J. Phys. G} {\bf 33},  1+  (2006).

\bibitem{Cardarelli:1995dc}
F. Cardarelli, E. Pace, G. Salme, and S. Simula, Phys. Lett. {\bf B357},  267
  (1995).

\bibitem{Wagenbrunn:2000es}
R.~F. Wagenbrunn {\it et~al.}, Phys. Lett. {\bf B511},  33  (2001).

\bibitem{Glozman:2001zc}
L.~Y. Glozman {\it et~al.}, Phys. Lett. {\bf B516},  183  (2001).

\bibitem{Boffi:2001zb}
S. Boffi {\it et~al.}, Eur. Phys. J. {\bf A14},  17  (2002).

\bibitem{Melde:2007zz}
T. Melde {\it et~al.}, Phys. Rev. D {\bf 76},  074020  (2007).

\bibitem{Berger:2004yi}
K. Berger, R.~F. Wagenbrunn, and W. Plessas, Phys. Rev. D {\bf 70},  094027
  (2004).

\bibitem{Keister:1991sb}
B.~D. Keister and W.~N. Polyzou, Adv. Nucl. Phys. {\bf 20},  225  (1991).

\bibitem{Dirac:1949}
P. Dirac, Rev. Mod. Phys. {\bf 21},  392  (1949).

\bibitem{Leutwyler:1977vy}
H. Leutwyler and J. Stern, Ann. Phys. {\bf 112},  94  (1978).

\bibitem{Bakamjian:1953}
B. Bakamjian and L.~H. Thomas, Phys. Rev. {\bf 92},  1300  (1953).

\bibitem{Glozman:1998ag}
L.~Y. Glozman, W. Plessas, K. Varga, and R.~F. Wagenbrunn, Phys. Rev. D {\bf
  58},  094030  (1998).

\bibitem{Glozman:1998fs}
L.~Y. Glozman {\it et~al.}, Phys. Rev. C {\bf 57},  3406  (1998).

\bibitem{Bhaduri:1981pn}
R.~K. Bhaduri, L.~E. Cohler, and Y. Nogami, Nuovo Cim. A {\bf 65},  376
  (1981).

\bibitem{Suzuki:1998bn}
Y. Suzuki and K. Varga, {\em Stochastic Variational Approach to
  Quantum-Mechanical Few-Body Problems} (Springer Verlag, Berlin, 1998).

\bibitem{Samios:1974tw}
N.~P. Samios, M. Goldberg, and B.~T. Meadows, Rev. Mod. Phys. {\bf 46},  49
  (1974).

\bibitem{Guzey:2005vz}
V. Guzey and M.~V. Polyakov, hep-ph/0512355  (2005).

\bibitem{Melde:2004qu}
T. Melde, L. Canton, W. Plessas, and R.~F. Wagenbrunn, Eur. Phys. J. A {\bf
  25},  97  (2005).

\bibitem{LeYaouanc:1988aa}
A. LeYaouanc, L. Oliver, O. Pene, and J.C. Raynal, 
  {\em Hadron Transitions in the Quark Model}
  (Gordon and Breach, New York, 1988).
  
\bibitem{Sengl:2006}
B. Sengl, Ph.D. Thesis, University of Graz, Graz, 2006.

\bibitem{Aubert:2006ux}
B. Aubert {\it et~al.},   hep-ex/0607043 (2006).

\bibitem{Pervin:2007wa}
M. Pervin and W. Roberts,  arXiv:0709.4000 [nucl-th] (2007).

\bibitem{Matagne:2006zf}
N. Matagne and F. Stancu, Phys. Rev. D {\bf 74},  034014  (2006).

\bibitem{Papp:2000kp}
Z. Papp, A. Krassnigg, and W. Plessas, Phys. Rev. C {\bf 62},  044004  (2000).

\end{thebibliography}

%
%
%
%
\clearpage
\renewcommand{\arraystretch}{1.3}
\begin{table}[h!]
\begin{center}
\caption{Energy eigenvalues (in MeV) of the ground and resonance
states with total angular momentum and parity $J^P$ from the GBE and OGE RCQMs
in comparison to the experimental
masses according to the PDG~\cite{PDBook}. In each case the number in the parentheses
denotes the k-th excitation in the respective $J^P$
column starting with $k=0$. The resonances denoted by mass values in square
brackets represent states not definitely classified by the PDG. 
\label{tab:masses}
}
\begin{ruledtabular}
\begin{tabular}{ccrrc}
Baryon&$J^P$&\multicolumn{2}{c}{Theory}&Experiment\\
&&GBE&OGE& \\
\hline
$N(939)$
&$\frac{1}{2}^+$
&$939$ (0)
&$939$ (0)
&$938 -940$
\\
$N(1440)$
&$\frac{1}{2}^+$
&$1459$ (1)
&$1577$ (1)
&$1420-1470$
\\
$N(1520)$
&$\frac{3}{2}^-$
&$1519$ (0)
&$1521$ (0)
&$1515-1525$
\\
$N(1535)$
&$\frac{1}{2}^-$
&$1519$ (0)
&$1521$ (0)
&$1525-1545$
\\
$N(1650)$
&$\frac{1}{2}^-$
&$1647$ (1)
&$1690$ (1)
&$1645-1670$
\\
$N(1675)$
&$\frac{5}{2}^-$
&$1647$ (0)
&$1690$ (0)
&$1670-1680$
\\
$N(1700)$
&$\frac{3}{2}^-$
&$1647$ (1)
&$1690$ (1)
&$1650-1750$
\\
$N(1710)$
&$\frac{1}{2}^+$
&$1776$ (2)
&$1859$ (2)
&$1680-1740$
\\
\hline
$\Delta(1232)$
&$\frac{3}{2}^+$
&$1240$ (0)
&$1231$ (0)
&$1231-1233$
\\
$\Delta(1600)$
&$\frac{3}{2}^+$
&$1718$ (1)
&$1854$ (1)
&$1550-1700$
\\
$\Delta(1620)$
&$\frac{1}{2}^-$
&$1642$ (0)
&$1621$ (0)
&$1600-1660$
\\
$\Delta(1700)$
&$\frac{3}{2}^-$
&$1642$ (0)
&$1621$ (0)
&$1670-1750$
\\
\hline
$\Lambda(1116)$
&$\frac{1}{2}^+$
&$1136$ (0)
&$1113$ (0)
&$1116$
\\
$\Lambda(1405)$
&$\frac{1}{2}^-$
&$1556$ (0)
&$1628$ (0)
&$1402-1410$
\\
$\Lambda(1520)$
&$\frac{3}{2}^-$
&$1556$ (0)
&$1628$ (0)
&$1519-1521$
\\
$\Lambda(1600)$
&$\frac{1}{2}^+$
&$1625$ (1)
&$1747$ (1)
&$1560-1700$
\\
$\Lambda(1670)$
&$\frac{1}{2}^-$
&$1682$ (1)
&$1734$ (1)
&$1660-1680$
\\
$\Lambda(1690)$
&$\frac{3}{2}^-$
&$1682$ (1)
&$1734$ (1)
&$1685-1695$
\\
$\Lambda(1800)$
&$\frac{1}{2}^-$
&$1778$ (2)
&$1844$ (2)
&$1720-1850$
\\
$\Lambda(1810)$
&$\frac{1}{2}^+$
&$1799$ (2)
&$1957$ (2)
&$1750-1850$
\\
$\Lambda(1830)$
&$\frac{5}{2}^-$
&$1778$ (0)
&$1844$ (0)
&$1810-1830$
\\
\hline
$\Sigma(1193)$
&$\frac{1}{2}^+$
&$1180$ (0)
&$1213$ (0)
&$1189-1197$
\\
$\Sigma(1385)$
&$\frac{3}{2}^+$
&$1389$ (0)
&$1373$ (0)
&$1383-1387$
\\
$\Sigma[1560]$
&$\frac{1}{2}^-$
&$1677$ (0)
&$1732$ (0)
&$1546-1576$
\\
$\Sigma[1620]$
&$\frac{1}{2}^-$
&$1736$ (1)
&$1829$ (2)
&$1594-1643$
\\
$\Sigma(1660)$
&$\frac{1}{2}^+$
&$1616$ (1)
&$1845$ (1)
&$1630-1690$
\\
$\Sigma(1670)$
&$\frac{3}{2}^-$
&$1677$ (0)
&$1732$ (0)
&$1665-1685$
\\
$\Sigma[1690]$
&$\frac{3}{2}^+$
&$1865$ (1)
&$1991$ (1)
&$1670-1727$
\\
$\Sigma(1750)$
&$\frac{1}{2}^-$
&$1759$ (2)
&$1784$ (1)
&$1730-1800$
\\
$\Sigma(1775)$
&$\frac{5}{2}^-$
&$1736$ (0)
&$1829$ (0)
&$1770-1780$
\\
$\Sigma(1880)$
&$\frac{1}{2}^+$
&$1911$ (2)
&$2049$ (2)
&$1806-2025$
\\
$\Sigma[1940]$
&$\frac{3}{2}^-$
&$1736$ (1)
&$1829$ (2)
&$1900-1950$
\\
$\Sigma$
&$\frac{3}{2}^-$
&$1759$ (2)
&$1784$ (1)
&
\\

\hline
$\Xi(1318)$
&$\frac{1}{2}^+$
&$1348$ (0)
&$1346$ (0)
&$1315-1321$
\\
$\Xi(1530)$
&$\frac{3}{2}^+$
&$1528$ (0)
&$1516$ (0)
&$1532-1535$
\\
$\Xi[1690]$
&$\frac{1}{2}^+$
&$1805$ (1)
&$1975$ (1)
&$1680-1700$
\\
$\Xi(1820)$
&$\frac{3}{2}^-$
&$1792$ (0)
&$1894$ (0)
&$1818-1828$
\\
$\Xi[1950]$
&$\frac{5}{2}^-$
&$1881$ (0)
&$1993$ (0)
&$1935-1965$
\end{tabular}
\end{ruledtabular}
\end{center}
\end{table}

\renewcommand{\arraystretch}{1.3}
{\footnotesize
\begin{table}
\begin{center}
\caption{
Classification of baryon ground and resonance states into flavor
multiplets by the PDG~\cite{PDBook}. In addition to established states
also some two-star resonances are included, namely $\Sigma$(1620) and $\Sigma$(1880).
Entries with question marks have not yet received any assignments.
\label{tab:multiplet_PDG}
}
\vspace{0.2cm}
{\begin{tabular}{@{}crc llll@{}}
\hline\hline
&multiplet &$(LS)J^P$&&&&\\
\hline
& octet & $(0\frac{1}{2})\frac{1}{2}^+$
& $N(939)$ 
& $\Lambda(1116)$ 
& $\Sigma(1193)$ 
& $\Xi(1318)$ 
\\
& octet &$(0\frac{1}{2})\frac{1}{2}^+$ 
& $N(1440)$ 
& $\Lambda(1600)$ 
& $\Sigma(1660)$ 
& $\Xi(?)$
\\
& octet & $(0\frac{1}{2})\frac{1}{2}^+$
& $N(1710)$ 
& $\Lambda(1810)$
& $\Sigma(1880)$
&$\Xi(?)$
\\
& octet & $(1\frac{1}{2})\frac{1}{2}^-$
& $N(1535)$ 
& $\Lambda(1670)$ 
& $\Sigma(1620)$
& $\Xi(?)$
\\
& octet 
& $(1\frac{3}{2})\frac{1}{2}^-$
& $N(1650)$ 
& $\Lambda(1800)$ 
& $\Sigma(1750)$
& $\Xi(?)$ 
\\
& octet &$(1\frac{1}{2})\frac{3}{2}^-$ 
& $N(1520)$ 
& $\Lambda(1690)$ 
& $\Sigma(1670)$ 
& $\Xi(1820)$ 
\\
& octet & $(1\frac{3}{2})\frac{3}{2}^-$
& $N(1700)$ 
& $\Lambda(?)$ 
& $\Sigma(?)$
& $\Xi(?)$
\\
& octet &$(1\frac{3}{2})\frac{5}{2}^-$ 
& $N(1675)$ 
& $\Lambda(1830)$ 
& $\Sigma(1775)$ 
& 
$\Xi(?)$
\\
\hline
& singlet & $(1\frac{1}{2})\frac{1}{2}^-$
&-
& $\Lambda(1405)$ 
&-  
& - \\
& singlet &$(1\frac{1}{2})\frac{3}{2}^-$
&-  
& $\Lambda(1520)$ 
& - 
& -\\
\hline
& decuplet& $(0\frac{3}{2})\frac{3}2{}^+$
& $\Delta(1232)$ 
&-
& $\Sigma(1385)$ 
& $\Xi(1530)$ 
\\
& decuplet& $(0\frac{3}{2})\frac{3}2{}^+$
& $\Delta(1600)$ 
&-
& $\Sigma(?)$ 
& $\Xi(?)$ 
\\
& decuplet& $(1\frac{1}{2})\frac{1}{2}^-$
& $\Delta(1620)$ 
&- 
& $\Sigma(?)$
& $\Xi(?)$
\\
& decuplet &$(1\frac{1}{2})\frac{3}{2}^-$
& $\Delta(1700)$ 
&-
& $\Sigma(?)$ 
& $\Xi(?)$  
\\
\hline\hline
\end{tabular}}
\end{center}
\end{table}
}

\begin{table*}
\caption{Covariant predictions of partial widths for various decay modes of
hyperon resonances by the GBE CQM~\cite{Glozman:1998ag} and the
OGE CQM~\cite{Theussl:2000sj}. The denotation of the states follows the multiplet
assignments in the present work (see
Tables~\ref{tab:multiplet_oct},~\ref{tab:multiplet_decu}, and~\ref{tab:multiplet_singl}).
Regarding phenomenological data~\cite{PDBook} a comparison is possible only
to total decay widths.
\label{tab:multi3}
}
\begin{ruledtabular}
\begin{tabular}{@{}ccc cc cc@{}}
Decay&$J^P$&Experiment 
&\multicolumn{2}{c}{With Theoretical Masses}
&\multicolumn{2}{c}{With Experimental Masses}\\
&&[MeV]
&GBE & OGE  
&GBE & OGE \\
\hline
\small $\rightarrow \Sigma\pi $\\
$\Sigma(1560)$&
$\frac{1}{2}^-$&
$\Gamma(9-109)$
    &$58$ 
    &$102$ 
    &$44$
    &$70$
\\ 
$\Sigma(1620)$&
$\frac{1}{2}^-$&
$\Gamma(10-106)$ 
    &$32$ 
    &$44$ 
    &$21$
    &$26$
\\
$\Sigma$(1690)& 
$\frac{3}{2}^+$
& $\Gamma(15-300)$
    &$0.4$ 
    &$2.7$ 
    &$0.2$
    &$1.1$
\\
$\Sigma$(1880)& 
$\frac{1}{2}^+$
& $\Gamma(30-372)$
    &$3.0$ 
    &$3.0$ 
    &$1.8$
    &$0.4$
\\
\hline
\small $\rightarrow \Lambda\pi $\\
$\Sigma(1560)$&
$\frac{1}{2}^-$&
$\Gamma(9-109)$ 
    &$1.6$ 
    &$1.5$
    &$2.1$
    &$2.2$
\\ 
$\Sigma(1620)$&
$\frac{1}{2}^-$&
$\Gamma(10-106)$ 
    &$19$ 
    &$25$ 
    &$17$
    &$23$
    \\
$\Sigma$(1690)& 
$\frac{3}{2}^+$
& $\Gamma(15-300)$
    &$\approx 0$ 
    &$1.2$ 
    &$\approx 0$
    &$0.6$
\\
$\Sigma$(1880)& 
$\frac{1}{2}^+$
& $\Gamma(30-372)$
    &$1.7$ 
    &$0.5$ 
    &$1.6$
    &$0.3$
\\
\hline
\small $\rightarrow NK $\\
$\Sigma(1560)$&
$\frac{1}{2}^-$&
$\Gamma(9-109)$ 
    &$8$ 
    &$8$
    &$6$
    &$5$
\\ 
$\Sigma(1620)$&
$\frac{1}{2}^-$&
$\Gamma(10-106)$ 
    &$55$ 
    &$55$ 
    &$57$
    &$58$
   \\
$\Sigma$(1690)& 
$\frac{3}{2}^+$ 
& $\Gamma(15-300)$
    &$\approx 0$ 
    &$1.4$ 
    &$\approx 0$
    &$0.8$
\\
$\Sigma$(1880)& 
$\frac{1}{2}^+$
& $\Gamma(30-372)$
    &$\approx 0$ 
    &$\approx 0$ 
    &$\approx 0$
    &$\approx 0$
\\
\hline
\small $\rightarrow \Xi\pi $\\
$\Xi(1690)$&
$\frac{1}{2}^+$&
$\Gamma(< 30)$ 
    &$0.8$ 
    &$1.8$
    &$0.5$
    &$0.5$
\\ 
$\Xi(1950)$
&
$\frac{5}{2}^-$&
$\Gamma(40-80)$ 
    &$14$ 
    &$28$
    &$25$
    &$26$
\\
\hline
\small $\rightarrow \Lambda K $\\
$\Xi(1690)$&
$\frac{1}{2}^+$&
$\Gamma(<30)$ 
    &$1.1$ 
    &$1.3$ 
    &$0.5$
    &$0.4$
\\ 
$\Xi(1950)$
&
$\frac{5}{2}^-$&
$\Gamma(40-80)$ 
    &$2.5$ 
    &$4.4$
    &$4.3$
    &$3.6$
\\
\hline
\small $\rightarrow \Sigma K $\\
$\Xi(1690)$&
$\frac{1}{2}^+$&
$\Gamma(<30)$ 
    &$9.2$ 
    &$55$ 
    &$0.1$
    &$0.2$
\\ 
$\Xi(1950)$
&
$\frac{5}{2}^-$&
$\Gamma(40-80)$ 
    &$2.3$ 
    &$4.3$
    &$3.7$
    &$3.6$
\\
\hline 
\small $\rightarrow \Xi\eta $\\
$\Xi(1950)$
&
$\frac{5}{2}^-$&
$\Gamma(40-80)$ 
    &$$ 
    &$\approx 0$
    &$0.1$
    &$0.1$
\\
\end{tabular}
\end{ruledtabular}
\end{table*}

\renewcommand{\arraystretch}{1.3}
\begin{table}
\begin{center}
\caption{Classification of flavor octet baryons. The denotation of the mass eigenstates is
made according to the nomenclature of baryon states seen in experiment. The superscripts denote the percentages of octet content as calculated with the GBE
CQM~\cite{Glozman:1998ag}. States in bold face have either not been assigned by the PDG
or differ from their assignment.
\label{tab:multiplet_oct}
}
\vspace{0.2cm}
{\begin{tabular}{@{}l llll  @{}}
\hline\hline
$(LS)J^P$& &&&\\
\hline
 $(0\frac{1}{2})\frac{1}{2}^+$
& $N(939)^{100}$  
& $\Lambda(1116)^{100}$
& $\Sigma(1193)^{100}$ &
$\Xi(1318)^{100}$
 \\
$(0\frac{1}{2})\frac{1}{2}^+$ 
& $N(1440)^{100}$  
& $\Lambda(1600)^{96}$
& $\Sigma(1660)^{100}$ 
& $\mbox{\boldmath$\Xi$}{\bf (1690)}^{100}$
\\
$(0\frac{1}{2})\frac{1}{2}^+$
& $N(1710)^{100}$  
&
& $\Sigma(1880)^{99}$ 
& 
 \\
$(1\frac{1}{2})\frac{1}{2}^-$
& $N(1535)^{100}$  
& $\Lambda(1670)^{72}$
& $\mbox{\boldmath$\Sigma$}{\bf (1560)}^{94}$
&  
 \\
$(1\frac{3}{2})\frac{1}{2}^-$
& $N(1650)^{100}$ 
& $\Lambda(1800)^{100}$ 
&$\mbox{\boldmath$\Sigma$}{\bf (1620)}^{100}$
& 
 \\
$(1\frac{1}{2})\frac{3}{2}^-$ 
& $N(1520)^{100}$
& $\Lambda(1690)^{72}$ 
& $\Sigma(1670)^{94}$ 
& $\Xi(1820)^{97}$
 \\
$(1\frac{3}{2})\frac{3}{2}^-$
& $N(1700)^{100}$ 
&
& ${\bf \Sigma(1940)}^{100}$ 
& 
 \\
$(1\frac{3}{2})\frac{5}{2}^-$ 
&$N(1675)^{100}$  
& $\Lambda(1830)^{100}$
& $\Sigma(1775)^{100}$ 
& $\mbox{\boldmath$\Xi$}{\bf (1950)}^{100}$
 \\
\hline\hline
\end{tabular}}
\end{center}
\end{table}

\renewcommand{\arraystretch}{1.3}
\begin{table}
\begin{center}
\caption{
Classification of flavor decuplet baryons. Analogous notation as in
Table~\ref{tab:multiplet_oct}.
\label{tab:multiplet_decu}
}
\vspace{0.2cm}
{\begin{tabular}{@{}l llll @{}}
\hline\hline
 $(LS)J^P$&&&&\\
\hline
$(0\frac{3}{2})\frac{3}2{}^+$
& $\Delta(1232)^{100}$ 
& $\Sigma(1385)^{100}$ 
& $\Xi(1530)^{100}$ 
& $\Omega(1672)^{100}$
 \\
$(0\frac{3}{2})\frac{3}{2}^+$
& $\Delta(1600)^{100}$ 
& $\mbox{\boldmath$\Sigma$}{\bf (1690)}^{99}$
& & 
  \\
$(1\frac{1}{2})\frac{1}{2}^-$
& $\Delta(1620)^{100}$  
&  $\mbox{\boldmath$\Sigma$}{\bf (1750)}^{94}$
& &
\\
$(1\frac{1}{2})\frac{3}{2}^-$
& $\Delta(1700)^{100}$ 
& & & 
 \\
\hline\hline
\end{tabular}}
\end{center}
\end{table}

\renewcommand{\arraystretch}{1.3}
\begin{table}
\begin{center}
\caption{Classification of flavor singlet baryons. Analogous notation as in
Table~\ref{tab:multiplet_oct}.
\label{tab:multiplet_singl}
}
\vspace{0.2cm}
{\begin{tabular}{@{}l l @{}}
\hline\hline
$(LS)J^P$&\\
\hline
$(1\frac{1}{2})\frac{1}{2}^-$
& $\Lambda(1405)^{71}$ 
 \\
$(1\frac{1}{2})\frac{3}{2}^-$  
& $\Lambda(1520)^{71}$
 \\
$(0\frac{1}{2})\frac{1}{2}^+$ 
&  $\mbox{\boldmath$\Lambda$}{\bf (1810)}^{92}$
\\
\hline\hline
\end{tabular}}
\end{center}
\end{table}

%
%
%
%
%

\clearpage


\end{document}